%% file: main.tex
\documentclass[preprint]{elsarticle} % options: preprint or review

\makeatletter
	\def\ps@pprintTitle{%
 	\let\@oddhead\@empty
	\let\@evenhead\@empty
	\def\@oddfoot{\centerline{\thepage}}%
	\let\@evenfoot\@oddfoot}
\makeatother

\usepackage{etoolbox}
\patchcmd{\MaketitleBox}{\footnotesize\itshape\elsaddress\par\vskip36pt}{\footnotesize\itshape\elsaddress\par\parbox[b][36pt]{\linewidth}{\vfill\hfill\textnormal{\today}\hfill\null\vfill}}{}{}%
\patchcmd{\pprintMaketitle}{\footnotesize\itshape\elsaddress\par\vskip36pt}{\footnotesize\itshape\elsaddress\par\parbox[b][36pt]{\linewidth}{\vfill\hfill\textnormal{\today}\hfill\null\vfill}}{}{}%

\usepackage[%pdfmark,%
            colorlinks=true,%
            breaklinks=true,%
            linkcolor=blue,
            urlcolor=blue,%
            citecolor=blue,%
            pdftitle={Predicting Solar Wind Streams from the Inner-Heliosphere to Earth via Shifted Operator Inference},  	% Give the title of paper
            pdfkeywords={solar wind modeling; space weather prediction; magnetohydrodynamics; data-driven model reduction; scientific machine learning; operator inference},	 % give keywords (the more the better, usually 4-6)
            pdfauthor={Opal Issan and Boris Kramer},		 %
            bookmarksopen=false,%
            pdfpagemode=None]{hyperref}

\input{ListPackages}
\input{mymacros.tex}

\graphicspath{{./figures/}}

\begin{document}
	
	% Here is a frontmatter of an article I wrote, please just edit
	\begin{frontmatter}
		
		\title{Predicting Solar Wind Streams from the Inner-Heliosphere to Earth \\ via Shifted Operator Inference}
		
		%\tnotetext[mytitlenote]{}

		\author[ucsd]{Opal Issan\corref{cor1}}
		\ead{oissan@ucsd.edu}
		\cortext[cor1]{Corresponding author}
		\author[ucsd]{Boris Kramer}
		
		\address[ucsd]{Department of Mechanical and Aerospace Engineering, University of California San Diego, CA, United States}
		
\begin{abstract}
    Solar wind conditions are predominantly predicted via three-dimensional numerical magnetohydrodynamic (MHD) models. Despite their ability to produce highly accurate predictions, MHD models require computationally intensive high-dimensional simulations. This renders them inadequate for making time-sensitive predictions and for large-ensemble analysis required in uncertainty quantification. 
    This paper presents a new data-driven reduced-order model (ROM) capability for forecasting heliospheric solar wind speeds. 
    Traditional model reduction methods based on Galerkin projection have difficulties with advection-dominated systems---such as solar winds---since they require a large number of basis functions and can become unstable.
    A core contribution of this work addresses this challenge by extending the non-intrusive operator inference ROM framework to exploit the translational symmetries present in the solar wind caused by the Sun's rotation.
    The numerical results show that our method can adequately emulate the MHD simulations and is more accurate than a reduced-physics surrogate model, the Heliospheric Upwind Extrapolation model. 
\end{abstract}	
		
\end{frontmatter}

%%%%%%%%%%%%%%%%%%%%%%%%%%%%%%%%%%%%%%%%
%%%%%%%%%%%%%%%%%%%%%%%%%%%%%%%%%%%%%%%
\section{Introduction} \label{sec:intro}
%%%%%%%%%%%%%%%%%%%%%%%%%%%%%%%%%%%%%%%
Magnetohydrodynamic (MHD) modeling of coronal and interplanetary solar wind can significantly improve the prediction of catastrophic space weather events. Such space weather geomagnetic storms can have detrimental effects on spacecrafts, cause electric power outage, satellite collisions, telecommunication interruption, and expose astronauts to harmful radiation. 
Very recently, 40 out of 49 of SpaceX's starlink satellites failed to reach their low-Earth orbits presumably due to the effects of a geomagnetic storm around Feb. 2, 2022; the geomagnetic storm increased Earth's upper atmosphere density causing orbital drag \cite{andrews_2022_starlink}. This event further emphasises the need for real-time modeling of solar storms.
The most substantial source of space weather events are coronal mass ejections, corotating interaction regions, and high-speed solar wind streams that reach the Earth's magnetosphere. Three-dimensional MHD solar wind models, such as the Magnetohydrodynamics Around a Sphere \cite{riley01a} model, Enlil \cite{odstrcil_2003_enlil}, and the Space Weather Modeling Framework \cite{toth_2005_swmf}, can provide high-fidelity predictions. Apart from providing a global assessment of coronal and heliospheric properties, MHD modeling can connect \textit{in situ} magnetic and plasma observations from one spacecraft to the other, providing crucial support to interplanetary missions \cite{riley2021_psp_mas}. 
Although MHD models are an important tool in understanding observed coronal and heliospheric dynamics, they require computationally intensive high-dimensional simulations. This renders them infeasible for time-sensitive predictions and large-ensemble methods such as quantifying forecast uncertainty and performing parameter sensitivity analysis.
Thus, there is a need for computationally efficient surrogate models that are capable of reproducing MHD results with sufficient fidelity.

The present work addresses this challenge by proposing a new method to derive a data-driven reduced-order model (ROM) for solar wind predictions that particularly focuses on issues that arise from the advection-dominated nature of the problem. The proposed method efficiently learns predictive ROMs from high-fidelity simulations (or other data) of solar wind models, while simultaneously producing an interpretable model and physical model form. We subsequently review the related literature, both focusing on the mathematical aspects of surrogate modeling for advection-dominated systems and the application domain of efficient heliospheric modeling.

Solar wind predictions are produced by a chain of coupled models in different parts of the Sun-Earth domain, i.e., the solar surface, corona, and heliosphere. In the heliospheric domain, there are mainly two classes of surrogate solar wind models: reduced-physics (white-box) and data-driven (black-box) models. The first approach is based on physical simplifications of the MHD equations. The simplest reduced-physics model is the ballistic approximation which assumes that each solar wind parcel maintains a constant radial speed as it propagates in the heliosphere. This approximation is mainly used to map solar wind streams for short radial distances \cite{snyder66a}. An improved kinematic model that bridges the gap between the ballistic mapping and global three-dimensional MHD modeling is the Heliospheric Upwind Extrapolation model, where each parcel speed is dependent on its adjacent parcel speed \cite{riley_HUXP1_2011,riley_HUXP2_2021,owens2020computationally, reiss_thux_2020}. 
A second approach is to build surrogate solar wind models via data-driven and statistical techniques \cite{camporeale_sw_ml_2019}. Such methods mainly aim to forecast the solar wind at Earth's vicinity without computing the solar wind dynamics on the full heliospheric domain. Examples of data-driven models include an artificial neural network model \cite{yang_2018_ann}, a gradient-boosting regression-based model \cite{bailey_2021_gbr}, and a probability distribution function model based on past rotation solar wind observations \cite{bussy_2014_pdf}. 
As we will see later, our proposed reduced-order modeling methodology is data-driven, yet accounts for the physical properties of the solar wind rotation. It therefore serves as a hybrid gray-box approach that leverages available physical information while remaining computationally efficient.

From a methodological perspective, several ROM strategies have targeted advection-dominated scenarios. For background, traditional ROMs are derived via (Petrov-) Galerkin projection, where the full-order model (FOM) is projected onto a low-dimensional subspace, see~\cite{holmes_lumley_berkooz_1996,antoulas05,hesthaven2016certified}. This class of ROMs aims to identify a small set of basis functions that minimize a certain error metric. 
However, a well-reported issue with linear-subspace ROMs is that they fail to model advection-dominated problems due to poorly decaying Kolmogorov N-width~\cite{greif2019decay, OR16, Wel17} which results in a slow decay of the singular values, see, e.g., \cite{rowley_2000translation_symmetry,iollo2014advection_modes,mojgani2017lagrangian,reiss2018shifted,lu2020lagrangian}. An accurate ROM would require a large number of basis functions, rendering it inefficient from a computational perspective. Additionally, a large number of global basis functions can lead to numerical instabilities.
The efforts to address the challenge posed by advection-dominated systems can be roughly categorized into Lagrangian-based approaches and methods that leverage a transport-invariant coordinate frame. 
The first line of research leverages Lagrangian coordinate grids to build a ROM that propagates both the wave physics \textit{and} the coordinate grid in time~\cite{mojgani2017lagrangian,lu2020lagrangian,lu_2021_hodograph}. These methods work extremely well, yet require knowledge of the underlying equations to solve for the Lagrangian grid, limiting their range of applicability.  
The second line of research, which our work builds upon, is based on transforming the dynamics to a moving coordinate frame via a time-dependent shift that is added to the spatial coordinates. In the moving frame, the system dynamics are absent of advection. The shift function can be numerically learned in various ways. For instance, the shifted proper orthogonal decomposition method~\cite{reiss2018shifted} proposes to detect the shift either through tracking peaks of the solution, or through an expansive SVD algorithm, where different candidate shifts are applied to the data matrix and then the SVD is computed. The shifts leading to the best singular value decay are then selected and a corresponding ROM is built via Galerkin projection. This strategy is computationally quite expensive due to the need for many SVDs of a large (yet rectangular) data matrix.
The authors in \cite{mirhoseini_zahr2021ROM-IFT} propose an implicit feature-tracking algorithm that is based on a minimal-residual ROM. This algorithm works well on complex geometries, however, finding the domain mapping can be quite expensive, too. 
Very recent work by~\cite{papapicco2022neural} builds on the trend of machine learning to derive two separate neural networks, one for detecting nonlinear shifts in the transport velocity, and a second for interpolating a shifted solution back to the reference frame. The method is fully data-driven, yet does not propose a predictive ROM that integrates the shift detection with a projection framework. 
Most closely related to our work is \cite{mendible2020dimensionality}, which proposes an unsupervised learning method to aid the identification and low-dimensional modeling of systems with translational symmetries. The data-based method uses sparse regression and ridge detection to identify models with non-constant wave speeds inherent to the data. The method performs well on several examples, including multiple waves travelling in opposite directions. While the examples all demonstrate interesting wave phenomena, the governing equations were always known, allowing to derive solid intuition about the wave-speed library. 
Other approaches that have been developed that do not fit precisely into these categories include the work by \cite{pehersto2020transportROM} that updates the ROM basis online to avoid slowly decaying Kolmogorov N-width, and the work by \cite{iollo2014advection_modes} that applies two separate mechanisms to deal with advection-diffusion systems, namely a representation of translational features via advection modes, and then the subsequent residual (features that are not purely advective) via global modes.  
Building on the philosophy of inducing time-dependent shifts into the ROM framework, we extend the non-intrusive projection-based operator inference ROM framework~\cite{peherstorfer2016data} towards advection-dominated systems by transforming the dynamics to a moving coordinate frame. Standard operator inference has successfully been applied to diverse applications such as combustion \cite{SKHW2020_learning_ROMs_combustor,mcquarrie2021data}, chemical reactors~\cite{BGKPW2020_OpInf_nonpoly}, ocean flows~\cite{yildiz21OPINF_RSWE}, Hamiltonian systems~\cite{SWK21-HOPINF}, and general reaction systems in the presence of incomplete data~\cite{UP21NonMarkovian}. Since operator inference learns the operators that would be obtained through intrusive Galerkin projection (which can be done exactly with additional data pre-processing, see~\cite{peherstorfer2020sampling}) it inherits problems that intrusive Galerkin ROMs face in the presence of strong advection.

We propose a new strategy for efficient data-driven heliospheric solar wind modeling. The method, \textit{shifted operator inference (sOpInf)}, builds on standard operator inference and extends it towards the challenges faced in solar wind predictions. Our proposed method first determines a moving coordinate frame where the dynamics are absent of translation and rotation and subsequently transforms the system into the new time-dependent coordinates. Two methods for predicting the shift are proposed. It then performs model learning in the shifted coordinate systems and subsequently makes predictions with the sOpInf-ROM via interpretable ODE simulation. Our hypothesis aligns with the previously cited references, in that simple translational patterns in the data and model can (and should) be exploited in the ROM approach.
Our proposed approach (1) speeds up the MHD simulation by several orders of magnitude, (2) preserves the solar wind spiral pattern created by the Sun's rotation, and (3) uncovers macroscopic coherent structures present in the evolution of solar wind streams by analyzing the velocity field modal decomposition. We present computationally-efficient data-driven ROMs for two heliospheric solar wind models: the MAS (Magnetohydrodynamics Around a Sphere) model and the HUX (Heliospheric Upwinding eXtrapolation) model. 

This paper is organized as follows. Section~\ref{sec:sw-model} describes the MAS and HUX heliospheric solar wind models and in Section~\ref{sec:ROM} we present the proposed method, shifted operator inference. In Section~\ref{sec:Numerical-Results} we demonstrate the performance of sOpInf on the MAS and HUX solar wind speed simulated data. Section~\ref{sec:conclusion} then offers conclusions and an outlook to future work.

%%%%%%%%%%%%%%%%%%%%%%%%%%%%%%%%%%%%%%%
\section{Solar Wind Models} \label{sec:sw-model}
%%%%%%%%%%%%%%%%%%%%%%%%%%%%%%%%%%%%%%%
This section introduces the solar wind models considered in this study. Section~\ref{sec:MAS-Model} presents the MAS  model. Section~\ref{sec:HUX} discusses an approximation to that model with similar physical attributes, the HUX model.

%%%%%%%%%%%%%%%%%%%%%%%%%%%%%%%%%%%%%%%
\subsection{Spherical Magnetohydrodynamics: The MAS Model} \label{sec:MAS-Model}
%%%%%%%%%%%%%%%%%%%%%%%%%%%%%%%%%%%%%%%
The MAS (Magnetohydrodynamics Around a Sphere) model is the primary MHD model in the CORHEL (CORona-HELiosphere) software and is publicly available at NASA's community-coordinated modeling center~\cite{CCMC}. The MAS model solves the time-dependent resistive MHD equations and has been used to study coronal mass ejections \cite{Lionello_2013}, coronal dynamics \cite{riley01a}, solar wind structure \cite{riley12e}, and connect \textit{in-situ} spacecraft observations \cite{riley2021_psp_mas}.
Herein, we focus our effort on analyzing the MAS solar wind radial velocity results and therefore exclude the discussion of other plasma components such as the magnetic field, plasma temperature, density, pressure, etc. This is because many space weather operational forecast models for satellite control and Earth-based infrastructure are particularly interested in near-Earth solar wind speed. Predicting the solar wind speed is pivotal for assessing the risk of geomagnetic storms because (1) coronal mass ejections, which are the most fundamental source of space weather events, are modeled as perturbations to the ambient solar wind; (2) interaction regions between fast and slow solar wind, known as co-rotating interaction regions, mainly present during solar minimum, are a driver of moderate geomagnetic activity \cite{riley12e}; and (3) high-speed solar wind streams cause an additional acceleration of energetic electrons in the radiation belts \cite{cranmer_2017}.

%%%%%%%%%%%%%%%%%%%%%%%%%%%%%%%%%%%%%%%
\subsubsection{Governing Equation} \label{MAS-governing-equations}
%%%%%%%%%%%%%%%%%%%%%%%%%%%%%%%%%%%%%%%
The MAS model solves a system of three-dimensional time-dependent resistive MHD equations in spherical coordinates ($r,\theta, \phi$), where $r$ is the radial distance from the Sun, $\theta$ is Carrington latitude in heliographic (rotating) coordinate system (HG), $\phi$ is the Carrington longitude in the HG coordinate system. The governing equations are 
\begin{align} 
                 \nabla \times \bB & = \frac{4\pi}{c} \bJ, \label{MAS-EQ1} \\
                 \nabla \times \bE & = -\frac{1}{c} \frac{\partial \bB}{\partial t}, \label{MAS-EQ2}\\
   \bE + \frac{1}{c}\boldsymbol{v}\times \bB & = \eta \bJ, \label{MAS-EQ3} \\
       \prhodt + \diverg(\rho \boldsymbol{v}) & = 0, \label{MAS-EQ4-Continuity} \\
\rho \left (\frac{\partial \boldsymbol{v}}{\partial t} + \boldsymbol{v} \cdot \nabla \boldsymbol{v} \right ) & = \frac{1}{c} \bJ \times \bB - \nabla p - \nabla p_{w} + \rho \bg + \diverg(\nu \rho \nabla \boldsymbol{v}), \label{MAS-EQ5-Momentum} \\
\frac{1}{\gamma -1} \left (\pTdt + \boldsymbol{v} \cdot \nabla T \right ) &= -T \diverg \boldsymbol{v} + S  \label{MAS-EQ6-Energy}.
\end{align}
and the initial and boundary condition are described in \cite{riley12e,linker_mhd_1999,MAS_user_guide}. Here, $\bB(r, \theta, \phi, t)$ is the magnetic field, $\bJ(r, \theta, \phi, t)$ is the current density, $\bE(r, \theta, \phi, t)$ is the electric field, $\boldsymbol{v}(r, \theta, \phi, t)$ is the plasma velocity, $T(r, \theta, \phi, t)$ is the plasma temperature, $\rho(r, \theta, \phi, t)$ is the plasma mass density, and $p(r, \theta, \phi, t)$ is the plasma pressure, and $p_{w}(r, \theta, \phi, t)$ is the Alfv\'en wave pressure. 
The constant $c$
%$=\SI{299792458}{\meter^{1}\second^{-1}}$ 
denotes the speed of light in a vacuum and $\bg(r) = - \frac{G M_{s}}{r^2} \hat{\be}_{r}$ is the gravitational acceleration, where $\hat{\be}_{r}$ is the unit vector in the radial direction, $G$
%$=\SI{6.6743e-11}{\meter^3\kilogram^{-1}\second^{-2}}$ 
is the universal gravitational constant, and $M_{s}= \SI{1.99e30}{\kg}$ is the solar mass. 
For the simulations used in this study, the constant resistivity is set to $\eta = \SI{4.6779e-5}{\second}$ and the kinematic viscosity $\nu = \SI{3.3503e16}{\meter^{2} \second^{-1}}$.
In the energy equation described in Eq.~\eqref{MAS-EQ6-Energy}, the thermodynamic approximation sets the ratio of specific heats to $\gamma = 5/3$. Moreover, the energy source terms are denoted by $S=S(T)$; for more details about the thermodynamic energy source term see the MAS user guide \cite{MAS_user_guide}. 
The MAS boundary conditions exploit photospheric magnetic field observations (e.g. data from the Wilcox Solar Observatory, the Global Oscillation Network Group, and the Solar Dynamics Observatory spacecraft), see~\cite{MAS,MAS_user_guide}. Here, we choose to analyze the scenario where the thermodynamic MAS results are driven by a synoptic map of the photospheric magnetic field as it reaches a dynamic steady state \cite{riley01a}. Most MHD models, e.g. MAS, solve for the ambient solar wind via time-dependent simulations and allow the solution to relax to steady state. While customary, this results in a large run time to compute the steady state. There are other MHD models, such as \cite{pizzo_3paper}, that solve directly for the steady solar wind, yet they have their own set of numerical challenges as it is no longer straightforward to extract the physical variables from the flux variables.

%%%%%%%%%%%%%%%%%%%%%%%%%%%%%%%%%%%%%%%
\subsubsection{Numerical Solver}\label{MAS-numerical-solver}
%%%%%%%%%%%%%%%%%%%%%%%%%%%%%%%%%%%%%%%
The MAS model equations are numerically solved on a nonuniform logically-rectangular staggered grid using finite differences. The nonuniform mesh allows for adjustment of the grid point concentration based on transition and active regions. For more details about the numerical methods and their stability, see \cite{Lionello_stability_MHD_1999,Caplan_2017}. 
The MAS model divides its computational domain to two distinct regions: the corona and heliosphere. The corona is the region between $1R_{S}$ to $30R_{S}$ and the heliosphere is the region between $30R_{S}$ to $1.1 \text{AU}$. The value $R_{S}$ denotes solar radii unit of distance which is $695,700 \text{km}$, and 1/215th of an astronomical unit ($\text{AU}$), which is equal to the distance from the Sun to Earth. 
Numerical results and implementation details are discussed in Section~\ref{sec:data-computational-enviroment-etc}.

%%%%%%%%%%%%%%%%%%%%%%%%%%%%%%%%%%%%%%%
\subsection{Solar Wind Speed: The HUX Model} \label{sec:HUX}
%%%%%%%%%%%%%%%%%%%%%%%%%%%%%%%%%%%%%%%
The HUX (Heliospheric Upwinding eXtrapolation) model developed by \cite{riley_HUXP1_2011, riley_HUXP2_2021} is a two-dimensional time-stationary model that predicts the heliospheric solar wind speed. The HUX model has been incorporated into operational and ensemble-based space weather programs \cite{kumar_2020_hux_operational,Amerstorfer_2021_HUX_operational} as an MHD surrogate model to study retrospective time periods as well as real-time predictions. It has also been used to map streams directly from \textit{in-situ} spacecraft observations (e.g. Helios A/B) to Earth \cite{issan_HUXP3_2022}. The HUX model is based on simplified physical assumptions of the fluid momentum equation. In contrast to the MAS model, where the velocity field is solved via the MHD equations, the HUX model constructs a kinematic mapping where each plasma parcel speed is governed by its adjacent parcel's speed. 

We introduce the HUX model as a reduced-physics surrogate solar wind model that is capable of capturing the solar wind speed as it propagates away from the Sun. The HUX model shares many similarities with the MAS model, such as advection-dominated solutions, and can therefore be used to test our proposed method, shifted operator inference. Moreover, since both sOpInf and HUX are surrogate models to the MAS model, we will compare their capability to approximate the MAS results.  
This section describes the HUX governing equations along with their spatial discretization and implementation.

%%%%%%%%%%%%%%%%%%%%%%%%%%%%%%%%%%%%%%%
\subsubsection{Reduced-Physics Equation} \label{sec:hux-pde-section}
%%%%%%%%%%%%%%%%%%%%%%%%%%%%%%%%%%%%%%%
The HUX model~\cite{riley_HUXP1_2011, snyder66a} is derived from the fluid momentum equation in the corotating frame of reference with the Sun by considering Eq.~\eqref{MAS-EQ5-Momentum} in the absence of magnetic and viscous effects describing steady flow by replacing the time derivative ($\frac{\partial}{\partial t}$) with a spatial derivative ($-\Omega_{\text{rot}}\frac{\partial}{\partial \phi}$), i.e. the governing equations are 
\begin{equation} \label{momentum-equation}
-\Omega_{\text{rot}}(\theta) \frac{\partial \boldsymbol{v}(r, \theta, \phi)}{\partial \phi} + \left[\boldsymbol{v}(r, \theta, \phi) \cdot \nabla \right]\boldsymbol{v}(r, \theta, \phi)  = -\frac{1}{\rho(r, \theta, \phi)} \nabla p(r, \theta, \phi) + \bg(r),
\end{equation}
where $\boldsymbol{v} = [v_{r}(r, \theta, \phi), v_{\theta}(r, \theta, \phi), v_{\phi}(r, \theta, \phi)]$ is the solar wind proton velocity, $\rho(r, \theta, \phi)$ is the plasma density, and $\bg(r)$ is the gravitational acceleration specified in Section \ref{MAS-governing-equations}. The term $\Omega_{\text{rot}} (\theta) = \frac{2\pi}{25.38} - \frac{2.77\pi}{180} \cos(\theta - \frac{\pi}{2})^2$ is the angular frequency of the Sun's rotation, i.e., a function of latitude \cite{riley_HUXP2_2021}.
The analysis in \cite{riley_HUXP1_2011, riley_HUXP2_2021, issan_HUXP3_2022} justifies neglecting the pressure gradient and gravity terms in Eq.~\eqref{momentum-equation} and only taking into account variations of the velocity in the radial direction. As a result, Eq.~\eqref{momentum-equation} reduces to the two-dimensional nonlinear scalar homogeneous time-stationary equation
\begin{equation} \label{hux-underlying-equation-PDE}
-\Omega_{\text{rot}}(\theta = \hat{\theta}) \frac{\partial v_{r}(r, \phi)}{\partial \phi} + v_{r}(r, \phi)\frac{ \partial v_{r}(r, \phi)}{\partial r}=0,
\end{equation}
where the independent variables are $r$ and  $\phi$ and the dependent variable is the velocity in the radial direction $v_{r}(r, \phi)$. The angular frequency of the Sun's rotation is evaluated at a constant Carrington latitude $\hat{\theta}$; here we consider the Sun's equatorial plane ($\hat{\theta} = 0$) so that $\Omega_{\text{rot}}(0) = \frac{2 \pi}{25.38} \text{1/days}$ at the solar equator.  The initial-boundary value problem (IBVP) is subject to the initial condition $v_{r}(r_{0}, \phi) = v_{r_{0}} (\phi)$ and is defined on the periodic domain $0 \leq \phi \leq 2 \pi$ and $r \geq 30 R_{S}$, where beyond $30 R_{S}$, the solar wind travels along roughly radial trajectories, justifying the assumption of only considering the velocity in the radial direction.
Additionally, to account for the residual acceleration present in the inner heliosphere, the authors in \cite{riley_HUXP1_2011} suggested adding an acceleration boost to the initial velocity profile described by %
\begin{equation} \label{hux-acceleration-boost-term}
v_{\text{acc}}(r_{0}, v_{r_{0}} (\phi)) = \alpha [v_{r_{0}} (\phi)]( 1 - e^{-r_{0}/r_{h}}),
\end{equation}
where $v_{r_{0}} (\phi)$ is the initial radial velocity, $\alpha=0.15$ is the acceleration factor, and $r_{h}=50R_{S}$ is the radial location at which the acceleration ends. Hence, the acceleration boost, $v_{\text{acc}}(r_{0}, v_{r_{0}} (\phi))$, is added to the initial velocity profile $v_{r_{0}} (\phi)$ prior to solving the HUX equation. 

%%%%%%%%%%%%%%%%%%%%%%%%%%%%%%%%%%%%%%%
\subsubsection{Discretization via the Upwind Scheme} \label{sec:hux-pde}
%%%%%%%%%%%%%%%%%%%%%%%%%%%%%%%%%%%%%%%
This section describes the semi-discretization of Eq.~\eqref{hux-underlying-equation-PDE} in longitude, which then results in a set of ODEs. To begin, we rewrite Eq.~\eqref{hux-underlying-equation-PDE} in the hyperbolic conservation form
\begin{equation} \label{conservative-form-hux}
    \frac{\partial}{\partial r} v_{r}(r, \phi) + \frac{\partial}{\partial \phi} f[v_{r}(r, \phi)]= 0 ,
\end{equation}
where the physical flux function is $f[v_{r}(r, \phi)] = -\Omega_{\text{rot}}(\hat{\theta}) \ln[v_{r}(r, \phi)]$. We use the first-order conservative upwind method from \cite{issan_HUXP3_2022}, so to approximate the partial derivative of the flux function $f$ with respect to $\phi$ by
\begin{equation} \label{conservative-upwinding-derivative}
    \frac{\partial}{\partial \phi} f[v_{r}(r, \phi^{(j)})] \approx \frac {-\Omega_{\text{rot}}(\hat{\theta})}{\Delta \phi}  \left ( \ln[v_{r}(r, \phi^{(j+1)})] - \ln[v_{r}(r, \phi^{(j)})] \right ),
\end{equation}
where $n_{\phi}$ is the number of mesh points in longitude and $j = 1, 2, \ldots ,n_{\phi}$ denotes the longitude grid index. We discretize the longitudinal direction uniformly with $\Delta \phi$ mesh spacing and denote the discretized state vector as $\bv(r) = [v_{r}(r, \phi^{(1)}), v_{r}(r, \phi^{(2)}), \ldots, v_{r}(r, \phi^{(n_{\phi})})]^{\top} \in \real^{n_{\phi}}$. From here, we obtain the semi-discrete system of ordinary differential equations
\begin{equation} \label{ode-form-hux}
    \frac{\text{d}}{\text{d} r}\bv(r) = \bD\ln[\bv(r)]
\end{equation}
with the sparse matrix   
\begin{equation} \label{A-matrix}
    \bD = \frac{\Omega_{\text{rot}}(\hat{\theta})}{\Delta \phi}
    \left[
    \begin{array}{cccccc}
        -1 & 1 & 0 & & &\\
        0 & -1 & 1 & & & \\
        & \ddots & \ddots & \ddots & \ddots &\\
        & & & -1& 1 & 0 \\
        & & & & -1 & 1 \\
        1 & & & & 0 & -1 \\
    \end{array}
    \right] \in\real^{n_{\phi} \times n_{\phi}}.
\end{equation}
The initial condition $\mathbf{v}(r_{0}) \in \real^{n_{\phi}}$ is set as the MAS coronal solution, $\bv^{\text{MAS}} \in \mathbb{R}^{n_{\phi}}$, evaluated at $r_{0} = 30R_{S}$ along with adding the ad hoc acceleration boost described in Eq.~\eqref{hux-acceleration-boost-term}, i.e., 
\begin{equation}
    \bv(r_0) = \bv^{\text{MAS}}(r_0) \left[ 1 + \alpha( 1 - e^{-{r_{0}}/r_{h}})\right]. 
\end{equation}

%%%%%%%%%%%%%%%%%%%%%%%%%%%%%%%%%%%%%%%
\section{Shifted Operator Inference: A Non-intrusive Reduced-Order Model Approach for Advection-Dominated Systems} \label{sec:ROM}
%%%%%%%%%%%%%%%%%%%%%%%%%%%%%%%%%%%%%%%
This section proposes \textit{shifted operator inference (sOpInf)}, a non-intrusive data-driven modeling framework that expands standard operator inference~\cite{pehersto2020transportROM} towards the challenge of modeling solar wind, and more generally, advection-dominated systems. The proposed method ensures that the learned ROM is able to (i) capture the dynamics of the translational systems with only a few modes, (ii) retain the translation and rotation properties of the physical system, and (iii) accurately predict the shift velocity in the testing regime. This is done by transforming the original coordinates to a moving coordinate frame where the dynamics are absent of translation and rotation. 
Section~\ref{sec:ShiftedOperatorInference} describes the new method that leverages a coordinate shift to first transform the data and subsequently learn in the transformed coordinates. Section~\ref{sec:DeterminationOfShiftFunction} proposes two alternative strategies for deriving this coordinate transformation. Lastly, Section~\ref{sec:Burgers} illustrates the sOpInf methodology on an introductory example: the one-dimensional inviscid Burgers' equation.

%%%%%%%%%%%%%%%%%%%%%%%%%%%%%%%%%%%%%%%
\subsection{Shifted Operator Inference for Advection-Dominated Systems} \label{sec:ShiftedOperatorInference}
%%%%%%%%%%%%%%%%%%%%%%%%%%%%%%%%%%%%%%%
To illustrate the proposed methodology, we consider a generic $k-$dimensional time-dependent partial differential equation (PDE) for the scalar function $u(x_1, x_2, \ldots, x_k, t)$ of the form
\begin{equation} \label{eq:generic-first-order-pde}
    F\left (u,x_{1},\ldots, x_{k}, \frac{\partial u}{\partial t}, \frac{\partial u}{\partial x_{1}}, \ldots , \frac{\partial u}{\partial x_{k}}, t \right ) = 0,
\end{equation}
where $x_{1}, x_{2},\ldots, x_{k}\in \real$ denote the spatial coordinates and $t\in \real^+$. One may think of $t$ as time, or, as described in the previous section, we will also consider the independent variable to be the radial distance from the Sun, $r$. The function $F$ defines the equations of motion of the system which include advective terms. For simplicity, we focused the illustration on a  purely convective case, but our method can account for higher derivatives (e.g. diffusion terms) as well, see Section \ref{sec:Numerical-Results} where we consider a model with viscous forces. 

Our goal is to derive a data-driven ROM that can accurately predict the solutions to advection-dominated systems, i.e., that uses data of a semi-discretization of \eqref{eq:generic-first-order-pde} and produces a predictive and efficient low-dimensional model. The proposed method proceeds in four steps as outlined next.

%%%%%%%%%%%%%%%%%%%%%%%%%%%%%%%%%%%%%%%%%%%%%%%
\paragraph{(I) Data collection and translation}
%%%%%%%%%%%%%%%%%%%%%%%%%%%%%%%%%%%%%%%%%%%%%%%
The system~\eqref{eq:generic-first-order-pde} is in $k$ dimensions and is typically solved via a spatial discretization scheme at fixed spatial locations 
$\bx_{i} = [x_{i}^{(1)}, x_{i}^{(2)}, \ldots, x_{i}^{(n_{i})}] \in \real^{n_{i}}$, $i=1,2,\ldots,k$, which we refer to as the \textit{original coordinates}. These are assembled into the following spatial grid
\begin{equation*}
    \bX = \begin{bmatrix}
    x_{1}^{(1)} & \ldots & x_{1}^{(n_{x})} \\
    \vdots & \ddots & \vdots \\
    x_{k}^{(1)} & \ldots & x_{k}^{(n_{x})} 
    \end{bmatrix}\in \real^{k \times n_{x}}
    \qquad \text{and} \qquad \bx = \text{vec}(\bX) \in \real^n,
\end{equation*} 
where $n_{x}=\Pi_{i=1}^{k} n_{i}$ is the number of spatial grid points and $n=k\cdot n_x$. Depending on the context, using $\bX$ in matrix form or $\bx$ in vector form may be more preferred. We collect data (for instance, solar wind speed data) from the numerical solver in the original coordinates at instances $t_i$, i.e., $\bu_i \approx u(\bx,t_i)\in \real^n$ with $0 = t_0 < t_1< \dots < t_K = T$.  To account for the translational element in the data, we next shift each snapshot to a moving coordinate frame 
\begin{equation}
    \bu_{i} \approx u(\bx, t_{i}) \mapsto \tilde{u}(\tilde{\bx}(\bx, t_{i}), t_{i}) \approx \tilde{\bu}_{i} \qquad \text{with} \qquad \tilde{\bx}(\bx, t) = \bx + \bc(t)
\end{equation}
where $\tilde{\bx}(\bx, t_{i})$ denotes the moving coordinate frame and $\bc(t) \in \real^{n}$ represents the traveling wave speed. We evaluate $\tilde{\bu}_i$ via piecewise linear interpolation, i.e.,
\begin{equation*} 
    \tilde{\bu}_{i}= \mathcal{P}^{k}_{i} \left[ \bu_{i}, \bx, \tilde{\bx}(\bx, t_{i}) \right]
\end{equation*}
where $\mathcal{P}^{k}_{i}$ denotes the $k >1$ dimensional piecewise linear interpolant of $\bu_{i}$ in the original grid $\bx$ evaluated on the moving coordinate frame grid $\tilde{\bx}(\bx, t_{i})$. The piecewise linear interpolation is implemented in the Python package \texttt{scipy} under the function \texttt{scipy.interpolate.LinearNDInterpolator()}. Our method is not restricted to the type of interpolation, so higher order interpolation methods can be used as well (e.g. piecewise cubic interpolation). In the moving coordinate frame, the system no longer exhibits translational properties. Section~\ref{sec:DeterminationOfShiftFunction} proposes two techniques to determine $\bc(t)$. 
We then store the transformed data in the matrix
\begin{equation*}
\tilde{\bU} = [\tilde{\bu}_1 \quad \dots \quad  \tilde{\bu}_K] \in \real^{n \times K}, 
\label{eq:snapshot_matrix}
\end{equation*}
where in the applications that we consider, $n \gg K$, so the matrix $\tilde{\bU}$ is tall and skinny.

%%%%%%%%%%%%%%%%%%%%%%%%%%%%%%%%%%%%%%%%%%%%%%%
\paragraph{(II) Data reduction via projection}
%%%%%%%%%%%%%%%%%%%%%%%%%%%%%%%%%%%%%%%%%%%%%%%
Given high-dimensional data, we first identify the low-dimensional subspace in which to learn a ROM. In this work, we use the subspace spanned by the proper orthogonal decomposition (POD) modes~\cite{holmes_lumley_berkooz_1996}, which is obtained by computing the economy-sized singular value decomposition of the snapshot matrix, i.e., 
\begin{equation} \label{svd}
\tilde{\bU}  = \bV\boldsymbol{\Sigma}\mathbf{W}^{\top},
\end{equation}
where $\bV\in \real^{n \times K}$, $\boldsymbol{\Sigma} \in \real^{K \times K}$ and $\mathbf{W}  \in \real^{K \times K}$. The $\ell \ll n$  dimensional POD basis, $\bV_\ell = [\bv_1,\ldots,\bv_\ell]$, is given by the first $\ell$ columns of $\bV$. The basis dimensions can be chosen based on the cumulative energy criteria, e.g., 
\begin{equation} \label{energy-criteria}
\ell = \argmin_{\hat{\ell}}\frac{\sum_{i=1}^{\hat{\ell}}\sigma_{i}}{\sum_{i=1}^{K}\sigma_{i}} > \epsilon_{\text{tol}},
\end{equation}
where $\sigma_{i}=\boldsymbol{\Sigma}_{ii}$ are the singular values, and $\epsilon_{\text{tol}}$ is commonly chosen to be $\epsilon_{\text{tol}} =0.95$ or $\epsilon_{\text{tol}} =0.99$ which encapsulate 95\% and 99\% of the energy in the data, respectively. 
Next, we project the state snapshot data onto the POD subspace spanned by the columns of $\bV_\ell$ and obtain the reduced snapshot matrices
\begin{equation} \label{data-matrix-and-derivative}
\widehat{\bU} = \bV^{\top}_\ell \tilde{\bU} = [\widehat{\bu}_1 \quad \dots \quad \widehat{\bu}_K] \in \real^{\ell \times K},
\qquad \text{and} \qquad 
\dot{\widehat{\bU}} = [ \dot{\widehat{\bu}}_1\quad \dots \quad  \dot{\widehat{\bu}}_K]\in \real^{\ell \times K},
\end{equation}
where the columns of $\dot{\widehat{\bU}}$ are computed from $\widehat{\bU}$ using any time derivative approximation (see, e.g.,~\cite{martins2013review,knowles2014methodsDifferentiation,chartrand2017numericalDifferentiation}), or can be obtained---if available---by evaluating the right-hand-side of the governing equation (the residual) and projecting the resulting data.

%%%%%%%%%%%%%%%%%%%%%%%%%%%%%%%%%%%%%%%%%%%%%%%
\paragraph{(III) Model learning and prediction via operator inference}
%%%%%%%%%%%%%%%%%%%%%%%%%%%%%%%%%%%%%%%%%%%%%%%
In this section we start with the assumption that the finite-dimensional data-generating model (a spatial discretization of Eq.~\eqref{eq:generic-first-order-pde} in the moving coordinate frame) is of the form of a polynomial nonlinear system of ODEs, written as
\begin{equation}
\frac{\text{d}\tilde{\bu}}{\text{d}t} = {\bA} \tilde{\bu} + {\bH} (\tilde{\bu} \otimes \tilde{\bu}) + {\bC} (\tilde{\bu}\otimes \tilde{\bu} \otimes \tilde{\bu}) + {\bB} + \textrm{HOT}, \qquad  \tilde{\bu} \in \real^n\label{eq:poly_FOM}
\end{equation}
with matrices ${\bA}\in \real^{n\times n}$, ${\bH} \in \real^{n \times n^2}$ and ${\mathbf{C}} \in \real^{n \times n^3}$. The column-wise Kronecker product is denoted by $\otimes$. Boundary conditions can either be represented via the constant $\bB\in \real^n$ or time-dependent BCs as $\bB \eta(t)$.
The abbreviation ``HOT'' in Eq.~\eqref{eq:poly_FOM} denotes higher-order terms, and represents terms that are quartic and higher order. For example, in the case of Burgers' equation in Section~\ref{sec:Burgers} the term $u\frac{\partial u}{\partial x}$ is quadratic in the PDE state $u(x,t)$ and would therefore yield a discretized component ${\bH} (\tilde{\bu} \otimes \tilde{\bu})$. 

Approximating the high-dimensional state $\tilde{\bu}$ in a low-dimensional basis $\bV_\ell \in \real^{n \times \ell}$, with $\ell \ll n$, we write $\tilde{\bu} \approx \bV_\ell\hat{\bu}$. Using a Galerkin projection, this yields the ROM of Eq.~\eqref{eq:poly_FOM} as
\begin{equation}
\frac{\text{d}\hat{\bu}}{\text{d}t} = \widehat{\bA} \hat{\bu}+ \widehat{\bH} (\hat{\bu}\otimes^{\prime} \hat{\bu}) + \widehat{\mathbf{C}} (\hat{\bu}\otimes^{\prime} \hat{\bu}\otimes^{\prime} \hat{\bu})+ \widehat{\bB} + \textrm{HOT}, \qquad \hat{\bu} \in \real^\ell
\label{eq:poly_ROM}
\end{equation}
where $\otimes^{\prime}$ is the compact Kronecker product, which removes redundant terms in the standard Kronecker product $\otimes$. For example, for $\hat{\bu} = [\hat{u}_{1}, \hat{u}_{2}]^{\top}$ the standard Kronecker product yields $\hat{\bu} \otimes \hat{\bu} = [\hat{u}_{1}^2, \hat{u}_{1}\hat{u}_{2}, \hat{u}_{2}\hat{u}_{1}, \hat{u}_{2}^2]^{\top}$ and the compact Kronecker product yields $\hat{\bu} \otimes^{\prime} \hat{\bu} = [\hat{u}_{1}^2, \hat{u}_{1}\hat{u}_{2}, \hat{u}_{2}^2]^{\top}$, which uses only unique terms. For here on, we only use the compact Kronecker product for learned sOpInf ROMs. Consequently, the ROM operators and their dimensions are

$\widehat{\bA}=\bV_\ell^\top \bA \bV_\ell \in \real^{\ell \times \ell} $,  $\widehat{\bH} = \bV_\ell^\top \bH (\bV_\ell \otimes^{\prime} \bV_\ell) \in \real^{\ell \times {\frac{1}{2}\ell(\ell+1)}}$,
$\widehat{\bC} = \bV_\ell^\top \bC (\bV_\ell \otimes^{\prime} \bV_\ell \otimes^{\prime} \bV_\ell) \in \real^{\ell \times {\frac{1}{6} \ell (\ell + 1) (\ell + 2)}}$, and $\widehat{\bB} = \bV_\ell^\top \bB \in \real^\ell$ is the reduced constant vector $\bB$. 
We note again that projection preserves polynomial structure, that is, Eq.~\eqref{eq:poly_ROM} has the same polynomial form as Eq.~\eqref{eq:poly_FOM}, but in the reduced subspace defined by $\bV_\ell$. 

To simplify notation, we continue from now on with a quadratic system, but note that all results carry over directly to cubic, quartic and all higher-order polynomial terms. Nevertheless, we note that the number of elements in the ROM operators scales with $\ell^4$ for the cubic operator, $\ell^5$ for the quartic operator, etc., yet higher-order terms often exhibit significant block-sparsity that can be exploited in numerical implementations which limits the growth of computational cost to solve the ROM.
For terms in the governing equations that are not in polynomial form, the introduction of variable transformations and auxiliary variables via the process of lifting \cite{gu2011qlmor,KW18nonlinearMORliftingPOD,SKHW2020_learning_ROMs_combustor,QKMW_2019_Transform_and_Learn} can convert these terms to polynomial form.

The goal at this stage is to learn a ROM that evolves the shifted and projected snapshots in time. Operator inference solves a least-squares problem to find the reduced operators that yield the ROM that best matches the projected snapshot data in a minimum residual sense. For a quadratic ROM (with $\widehat{\bC}$ and HOT set to zero in Eq.~\eqref{eq:poly_ROM}) operator inference solves the least-squares problem
\begin{equation*}
\min_{ \widehat{\bA} \in \real^{\ell \times \ell} , \widehat{\bH} \in \real^{\ell \times {\frac{1}{2}\ell(\ell+1)}}, \widehat{\bB} \in \real^\ell}
\left \Vert  \left [ \widehat{\bA}\widehat{\bU}+ \widehat{\bH}(\widehat{\bU} \otimes^{\prime} \widehat{\bU})  + \widehat{\bB}\ \mathbf{1}_K - \dot{\widehat{\bU}}  \right ]^\top \right \Vert^2_{\text{F}},
\end{equation*}
where $\mathbf{1}_K \in \real^{K}$ is the length $K$ row vector with all entries set to one.
Note that this least-squares problem is linear in the coefficients of the unknown ROM operators
$\widehat{\bA}$, $\widehat{\bH}$, and $\widehat{\bB}$. The appeal of the operator inference approach comes from the ability to compute the ROM operators  $\widehat{\bA}$, $\widehat{\bH}$, and $\widehat{\bB}$ directly from data without needing explicit access to the original high-dimensional operators ${\bA}$, ${\bH}$, and ${\bB}$.
The unknown operators and known low-dimensional data are combined in the matrices
\begin{equation*}
\mathbf{O} = [\widehat{\bA}\quad \widehat{\bH}\quad \widehat{\bB}] \in \real^{\ell \times(\ell+{\frac{1}{2}\ell(\ell+1)} + 1)} \qquad \text{and} \qquad \mathbf{D} = \begin{bmatrix} \widehat{\bU}^{\top} \quad (\widehat{\bU} \otimes^{\prime} \widehat{\bU})^{\top} \quad {\mathbf{1}_K}\end{bmatrix} \in \real^{K \times (\ell + {\frac{1}{2}\ell(\ell+1)} + 1)},
\end{equation*}
respectively. 
The unknown operators are then obtained as a solution to the minimization problem
\begin{align}
\min_{\mathbf{O}\in \real^{\ell \times (\ell+{\frac{1}{2}\ell(\ell+1)}+ 1)} } \left \Vert \mathbf{D} \mathbf{O}^{\top} - \dot{\widehat{\bU}}^{\top} \right \Vert^2_{\text{F}}.
\label{eq:min_OD}
\end{align}
For $K>\ell+{\frac{1}{2}\ell(\ell+1)}+1$ this overdetermined linear least-squares problem has a unique solution~\cite[Sec. 5.3]{golub96matrix}. It follows from linear algebra (and is noted in \cite{peherstorfer2016data}) that Eq.~\eqref{eq:min_OD} can be written as $\ell$ independent least-squares problems, each of the form 
\begin{equation*}
    \min_{\mathbf{o}_i \in \real^{\ell+{\frac{1}{2}\ell(\ell+1)}+1}} \left \Vert\mathbf{D} \mathbf{o}_i - \mathbf{r}_i \right \Vert_2^2,
\end{equation*}
for $i = 1,\dots, \ell$, where $\mathbf{o}_i$ is a column of $\mathbf{O}^{\top}$ (row of $\mathbf{O}$) and $\mathbf{r}_i$ is a column of $\dot{\widehat{\bU}}^{\top}$. This makes the operator inference approach efficient and scalable.

To avoid overfitting and prevent potential instability of the learned ROMs, regularization becomes necessary, see~\cite{mcquarrie2021data} for a detailed regularization study of operator inference. In this work, we use an Tikhonov regularization penalty so that the least-squares problem becomes
\begin{equation}
    \min_{\mathbf{O}\in \real^{\ell \times (\ell+{\frac{1}{2}\ell(\ell+1)}+ 1)} } \left \Vert \mathbf{D} \mathbf{O}^{\top} - \dot{\widehat{\bU}}^{\top} \right \Vert^2_{\text{F}} + \left \Vert \boldsymbol{\Gamma} \mathbf{O}^{\top} \right \Vert^2_{\text{F}}
\label{eq:regularization-added}
\end{equation}
where $\boldsymbol{\Gamma} = \text{diag}(\lambda_{1} \bI_{(\ell)}, \lambda_{2} \bI_{({\frac{1}{2}\ell(\ell+1)})}, \lambda_{1}) \in \real^{(\ell + {\frac{1}{2}\ell(\ell+1)} + 1) \times (\ell + {\frac{1}{2}\ell(\ell+1)} + 1)}$ is the diagonal matrix used for regularization. The parameter $\lambda_{1}$ is the regularization parameter of the operators $\widehat{\bB} \in \real^{\ell}$ and $\widehat{\bA} \in \real^{\ell \times \ell}$ and $\lambda_{2}$ regularizes the operator $\widehat{\bH} \in \real^{\ell \times {\frac{1}{2}\ell(\ell+1)}}$. The regularization parameters $\lambda_{1}$ and  $\lambda_{2}$ are problem specific and should be chosen accordingly. We provide details in Section~\ref{sec:Numerical-Results} and refer to \cite[Sec.~IV.B]{SKHW2020_learning_ROMs_combustor} for more implementation details of operator inference. 

Having learned the ROM in Eq.~\eqref{eq:poly_ROM} from the shifted data, we can make efficient predictions in that low-dimensional subspace that go beyond the training data into the fully predictive regime. We thus simulate Eq.~\eqref{eq:poly_ROM} to obtain a solution $\hat{\bu}(t)$, which we then lift to $n$ dimensions to get the approximate ROM solution $\tilde{\bu}^{\text{ROM}}(t) = \bV_\ell\hat{\bu}(t) \approx \tilde{\bu}(t)$. 
We use the Operator Inference Python package version 1.2.1~\cite{OpInfGitHub} to implement the model learning and prediction step of sOpInf. 

%%%%%%%%%%%%%%%%%%%%%%%%%%%%%%%%%%%%%%%%%%%%%%%
\paragraph{(IV) Re-shifting predicted ROM data}
%%%%%%%%%%%%%%%%%%%%%%%%%%%%%%%%%%%%%%%%%%%%%%%
The predicted ROM solutions $\tilde{\bu}^{\text{ROM}}(t)$ will be in the moving coordinate frame and require reverse translation to the original coordinates. 
We shift the ROM-predicted solutions back to the original coordinate system via interpolation
\begin{equation*}
%\tilde{\bu}^{\text{ROM}}_{i}(\tilde{\bx}(\bx, t_{i}), t_{i}) \mapsto 
{\bu^{\text{ROM}}_{i}}(\bx, t_{i}) = \mathcal{P}^{k}_{i} \left[ \tilde{\bu}^{\text{ROM}}_{i}, \tilde{\bx}(\bx, t_{i}), \bx \right]
\end{equation*}
where $\mathcal{P}^{k}_{i}$ denotes the $k >1$ dimensional (here: piecewise linear) interpolant of $\tilde{\bu}^{\text{ROM}}_{i}$ in the moving coordinate frame grid $\tilde{\bx}(\bx, t_{i})$ evaluated on the original grid $\bx$; see part (\textit{I}) in Section~\ref{sec:ShiftedOperatorInference} for more details about the interpolation implementation.
After shifting back to the original coordinates, the predicted and reconstructed sOpInf snapshots are columns of the matrix~$\bU^{\text{ROM}}\in~\real^{n \times (K+m)}$, such that $t_{K+m} > t_{K}$, which is the final output of the algorithm.

The previous steps (\textit{I})--(\textit{IV}) are summarized in Algorithm~\ref{alg:sOpInf-Pseudocode}, which is written for a quadratic system; yet extensions to cubic, quartic, and other polynomial systems are straightforward. 
%

%%%%%%%%%%%%%%%%%%%%%%%%%%%%%%%%%%%%%%%%%%%%%%%%%%%%%%%%%%%%%%%%%%%%%%%%%%%%%%%%%%%%%%%%%%
\begin{algorithm}
\caption{Shifted operator inference (sOpInf) \label{alg:sOpInf-Pseudocode}}
\textbf{Input}: $\mathbf{U} =[\bu_1, \bu_2, \ldots, \bu_K] \in \real^{n \times K}$ such that each column, $\mathbf{\bu}_{i} \in \real^{n}$, is a snapshot observed at $t_{i}$, SVD cumulative energy threshold $\epsilon_{\text{tol}}>0$, and regularization coefficients $\{\lambda_{1}, \lambda_{2}\}$.\newline
\textbf{Output}: ${\bU}^{\text{ROM}} \in \real^{n \times (K+m )}$ sOpInf reconstructed and predicted snapshots, where $t_{K+m} > t_{K}$.  \newline 
\textbf{Begin}: 
\begin{algorithmic}[1]
\State Learn shift function $\bc(t)$. \Comment{Section \ref{sec:DeterminationOfShiftFunction}}
\State Shift snapshots to moving coordinate system \newline $ \bU(\bx, t_i) \mapsto \tilde{\bU}(\tilde{\bx}(\bx, t_i), t_i)$ with $\tilde{\bx}(\bx, t)= \bx + \bc(t)$.  \Comment{Section~ \ref{sec:ShiftedOperatorInference}(\textit{I})}
\State Determine low-dimensional subspace  matrix $\bV_\ell$ by using threshold $\epsilon_{\text{tol}}$. \Comment{Section \ref{sec:ShiftedOperatorInference}(\textit{II})}
\State Project to low-dimensional subspace  $\widehat{\bU} = \bV^{\top}_\ell \tilde{\bU}$ and compute $\dot{\widehat{\bU}}$. \Comment{Section \ref{sec:ShiftedOperatorInference}(\textit{II})}
\State Solve the linear least-squares problem in Eq.~\eqref{eq:regularization-added} with regularization coefficients $\{\lambda_{1}, \lambda_{2}\}$ to obtain the sought ROM operators; here $\widehat{\bB}, \widehat{\bA}, \widehat{\bH}$. \Comment{Section \ref{sec:ShiftedOperatorInference}(\textit{III})}
\State Simulate the ROM in Eq.~\eqref{eq:poly_ROM} to get $\hat{\bu}(t)$ and lift to $\tilde{\bu}^{\text{ROM}}(t) = \bV_\ell \hat{\bu}(t) $.\Comment{Section \ref{sec:ShiftedOperatorInference}(\textit{III})}
\State Shift ROM results to original coordinates $ \tilde{\bu}^{\text{ROM}}(\tilde{\bx}(\bx, t_i), t_i) \mapsto {\bu}^{\text{ROM}}(\bx, t_i) $.  \Comment{Section \ref{sec:ShiftedOperatorInference}(\textit{IV})}
\end{algorithmic}
\end{algorithm}
%%%%%%%%%%%%%%%%%%%%%%%%%%%%%%%%%%%%%%%%%%%%%%%%%%%%%%%%%%%%%%%%%%%%%%%%%%%%%%%%%%%%%%%%%%

%%%%%%%%%%%%%%%%%%%%%%%%%%%%%%%%%%%%%%%
\subsection{Determination of Spatial Shift Velocity} \label{sec:DeterminationOfShiftFunction}
%%%%%%%%%%%%%%%%%%%%%%%%%%%%%%%%%%%%%%%
There are various ways in which the traveling wave speed, $\bc(t)$, can be discovered. For instance, the authors in~\cite{mendible2020dimensionality} use sparse regression and spectral clustering to uncover the function $\bc(t)$.  We present two methods: the method of characteristics discussed in Section \ref{sec:MethodOfCharacteristics}, an analytic approach that requires knowledge of the underlying equations (but not the discretization or computer code), and the cross-correlation extrapolation method described in Section \ref{sec:cross-correlation}, a purely data-driven approach. In both cases, the shift function $\bc(t)$ is learned from the batch of training data and extrapolated in the testing regime.

%%%%%%%%%%%%%%%%%%%%%%%%%%%%%%%%%%%%%%%
\subsubsection{Method of Characteristics} \label{sec:MethodOfCharacteristics}
%%%%%%%%%%%%%%%%%%%%%%%%%%%%%%%%%%%%%%%
The method of characteristics can be applied to quasi-linear partial differential equations, in which along the characteristic curves the PDE can be transformed to a set of coupled ODEs. 
We exploit the method of characteristics to find the scalar shift function $c(t)$ for first-order PDEs describing the scalar quantity $u(x_{1}, x_{2}, \ldots, x_{k}, t):~[a_{1}, b_{1}] \times \ldots \times [a_{k}, b_{k}] \times [t_{0}, t_{f}] \mapsto \real^{+}$,  whose dynamics are described by the following quasi-linear PDE:
\begin{equation}\label{general-form-pde-moc}
    \frac{\partial u(x_{1}, \ldots, x_{k}, t)}{\partial t} + \sum_{i = 1}^{k} f_{i}\left[u(x_{1}, \ldots, x_{k}, t), t\right] \frac{\partial u(x_{1}, \ldots, x_{k}, t)}{\partial x_{i}}  = g\left[u(x_{1}, \ldots, x_{k}, t), t\right],
\end{equation}
subject to the initial condition $u(x_{1}, \ldots, x_{k}, t=0) = u_{0}(x_{1},  \ldots, x_{k})$ with periodic boundary conditions at each spatial coordinate boundaries $x_{i} = a_{i}$ and $x_{i} = b_{i}$ for $i = 1,2, \ldots, k$. 
The transport speed $f_{i}\left[u(x_{1}, \ldots, x_{k}, t), t\right]\in\mathbb{R}$ must be strictly positive or negative $\forall t \in [t_{0}, t_{f}]$ and $ \forall x_{i} \in [a_{i}, b_{i}]$ to enforce uni-directional characteristics. The function $g\left[u(x_{1}, \ldots, x_{k}, t), t\right]\in \real$ is the source term.
Then, by the method of characteristics, Eq.~\eqref{general-form-pde-moc} can be written as a system of $k+1$ ODEs, i.e.
\begin{subequations}\label{moc-general}
\begin{align}
\begin{split} \label{moc-general-u}
\frac{\text{d} u(x_{1}(t), \ldots, x_{k}(t), t)}{\text{d} t} &= g\left[u(x_{1}(t), \ldots, x_{k}(t),  t), t\right],
\end{split}\\
\begin{split} \label{moc-general-x}
\frac{\text{d} x_{i}(t)}{\text{d} t} &= f_{i}\left[u(x_{1}(t), \ldots, x_{k}(t),  t), t\right], \ \ i=1,\ldots, k.
\end{split}
\end{align}
\end{subequations}
We solve Eq.~\eqref{moc-general-u} first, which results in 
\begin{equation}\label{G-equation}
    u(x_{1}(t), \ldots, x_{k}(t), t) = G\left[u_{0}(x_{1}(0), \ldots, x_{k}(0)), t\right],
\end{equation}
where $G\left[u_{0}(x_{1}(0), \ldots, x_{k}(0)), t\right] \in \mathbb{R}$ describes the scalar quantity $u$ along the characteristic curves. 
Given this ansatz, we proceed to obtain the characteristic curves by solving Eq.~\eqref{moc-general-x} via separation of variables, such that
\begin{equation}\label{char-curves-gen}
    x_{i}(t) = x_{i}(0) + \int_{t_{0}}^{t} f_{i}\left[G\left[u_{0}(x_{1}(0), \ldots, x_{k}(0)), t\right], t\right] \text{d} t.
\end{equation}
For most equations that arise from conservation laws, the ODEs in Eq.~\eqref{moc-general-u}--\eqref{moc-general-x} can be solved analytically via Eq.~\eqref{G-equation}--\eqref{char-curves-gen}. In the case when Eq.~\eqref{moc-general-u} can not be solved analytically, the coupled system of ODEs in Eq.~\eqref{moc-general-u}--\eqref{moc-general-x} can be solved numerically on a discrete grid; or in the case when $f_{i}[G, t]$ is a nonelementary antiderivative, we can approximate the function $f_{i}$ using Taylor series, and integrate term-by-term. 

By following the characteristic curves described in Eq.~\eqref{char-curves-gen} we are able to discover a moving coordinate frame absent of advection. In fluid dynamics, the characteristic paths are referred to as the Lagrangian specification of the flow field, whereby the fluid motion is observed following an individual fluid parcel. We construct a moving coordinate frame, $\bx + \bc(t)$, that resembles the main direction of the Lagrangian frame of reference. 
The shift function $c_{i}(t)$ corresponding to the spatial coordinate $x_{i}$ can be obtained by computing the \textit{mean characteristic} emerging from a certain spatial domain. Let the spatial domain of $x_{i}$, for $i=1,2,\ldots,k$ be discretized on a grid with $n_{i}$ points such that $\bx_{i} = [x_{i}^{(1)}, x_{i}^{(2)},\ldots, x_{i}^{(n_{i})}] \in \real^{n_{i}}$.
The mean characteristic (and hence the shift function $c_{i}(t) \in \real$ for $i=1, 2,\ldots k$) emerging in the interval $[x_{i}^{(p)}, x_{i}^{(q)}]$ with $1\leq p < q \leq n_{i}$ is given as 
\begin{align}\label{characteristic-curves-shift-function}
c_{i}(t) &= \frac{1}{|x_{i}^{(p)} - x_{i}^{(q)}|} \int_{x_{i}^{(p)}}^{x_{i}^{(q)}} \int_{t_{0}}^{t} f_{i}\left[G\left[u_{0}(x_{1}(0), \ldots, x_{k}(0)), t\right], t\right] \text{d} t \ \text{d} x_{i} \\
&\approx  \frac{1}{(q-p)} \sum_{j=p}^{q}  \int_{t_{0}}^{t} f_{i}\left[G\left[u_{0}(x_{1}(0), \ldots, x_{i}^{(j)}(0), \ldots, x_{k}(0)), t\right], t\right] \text{d}t.
\end{align}
The spatial interval $[x_{i}^{(p)}, x_{i}^{(q)}]$ can be set to the entire spatial domain, yet it is usually set to be a specific region of interest. 

For problems with shock formation, the shift function $\bc(t)$ is computed via the mean characteristic curve only before shock formation, i.e. before the characteristic lines first intersect, and approximated via the shock curve after shock formation, i.e.
\begin{equation}\label{shock-piecewise-c-of-t}
c_{i}(t) = \left.
    \begin{cases}
        \frac{1}{|x_{i}^{(p)} - x_{i}^{(q)}|} \int_{x_{i}^{(p)}}^{x_{i}^{(q)}} \int_{t_{0}}^{t} f_{i}\left[G\left[u_{0}(x_{1}(0), \ldots, x_{k}(0)), t\right], t\right] \text{d} t \ \text{d} x_{i} & \text{if } t_{0} < t < t_{s} \\
        s_{i}(t) - s_{i}(t_{s}) + a & \text{if } t > t_{s} 
    \end{cases}
    \right\} 
\end{equation}
where $a = \frac{1}{|x_{i}^{(p)} - x_{i}^{(q)}|} \int_{x_{i}^{(p)}}^{x_{i}^{(q)}} \int_{t_{0}}^{t_{s}} f_{i}\left[G\left[u_{0}(x_{1}(0), \ldots, x_{k}(0)), t\right], t\right] \text{d} t \ \text{d} x_{i}$. The time of shock formation is denoted by $t_{s}$, and $s_{i}(t)$ is the shock trajectory in $x_{i}$ coordinate, see Section~\ref{sec:Burgers} for a one-dimensional inviscid Burgers' equation example. For one-dimensional problems where $g=0$, the shock trajectory can be computed via the entropy (Rankine-Hugoniot) condition or the Whitham's geometric equal area rule in which multi-valued regions of the solution are replaced with a discontinuity that satisfies conservation~\cite{whitham_1976}. The shock location is approximated by locating a vertical line that splits the multi-valued curve into two regions with equal area. 

In problems where there are multiple shock curves, such as shown in Section~\ref{sec:hux-numerical-results}, we suggest to approximate $\bc(t)$ by tracking only one shock curve. The selection of the shock is problem dependent, e.g. in Section~\ref{sec:hux-numerical-results} we choose to follow the first shock emerging. Although, if the choice of shock curve or spatial interval $[x^{(p)}, x^{(q)}]$ is ambiguous, computing the shock curve is unfeasible or computationally expensive, or the initial condition is noisy, we recommend to use the cross-correlation extrapolation technique (Section~\ref{sec:cross-correlation}) to find $\bc(t)$.
Additionally, it can be non-trivial to find such characteristics in the case of more complex hyperbolic PDEs that are not in the form of Eq.~\eqref{general-form-pde-moc}, e.g. coupled systems such as Eqs.~\eqref{MAS-EQ1}--\eqref{MAS-EQ6-Energy} with multiple dependent variables resulting in more than one characteristic curve emanating from a single spatial point, in which we recommend employing the data-driven cross-correlation extrapolation technique, which we present next.

%%%%%%%%%%%%%%%%%%%%%%%%%%%%%%%%%%%%%%%
\subsubsection{Cross-Correlation Extrapolation Method} \label{sec:cross-correlation}
%%%%%%%%%%%%%%%%%%%%%%%%%%%%%%%%%%%%%%%
Cross-correlation is a mathematical operation that is commonly used in signal processing and pattern recognition to measure similarity of two signals. For two finite discrete signals $\bf, \bg \in \mathbb{C}^{n}$, the univariate discrete \textit{circular cross-correlation} is defined as 
\begin{equation}\label{cc-dis-circular-definition}
    (\bf \star \bg)[\tau] \coloneqq \sum_{j=1}^{n} \overline{\bf[j]} \bg[(j + \tau)_{\text{mod$_n$}}],
\end{equation}
where $\overline{\bf}$ denotes the complex conjugate of $\bf$, the bracket $[j]$ denotes the $j$th element of the signal, and $\tau \in \mathbb{Z}$ is the discrete displacement. The discrete circular cross-correlation can be extended to the multi-variate case, for snapshots with $k \in \mathbb{Z}$ variables and tensor-valued $\bf, \bg \in \mathbb{C}^{n_{1} \times n_{2} \times \ldots \times n_{k}}$:
\begin{equation}\label{cc-dis-definition-multi-variate}
(\bf \star \overset {k}{\cdots } \star \bg)[\boldsymbol{\tau}] \coloneqq \sum_{j_{1}=1}^{n_{1}} \ldots \sum_{j_{k}=1}^{n_{k}}
\overline{\bf[j_{1}, \ldots, j_{k}]} \bg[(j_{1}+ \tau_{1})_{\text{mod}_{n_{1}}}, \ldots, (j_{k} + \tau_{k})_{\text{mod}_{n_{k}}}],
\end{equation}
where $\boldsymbol{\tau}= [\tau_{1}, \tau_{2},\ldots ,\tau_{k}] \in \mathbb{Z}^{k}$ is the multi-variate discrete displacement. When the signals correlate, the value of $\bf \star \bg$ is maximized. 

\begin{figure}
    \centering
    \begin{subfigure}[b]{0.32\textwidth}
        \centering
        \caption{$V_{30 R_{S}}$}
        \includegraphics[width=\textwidth]{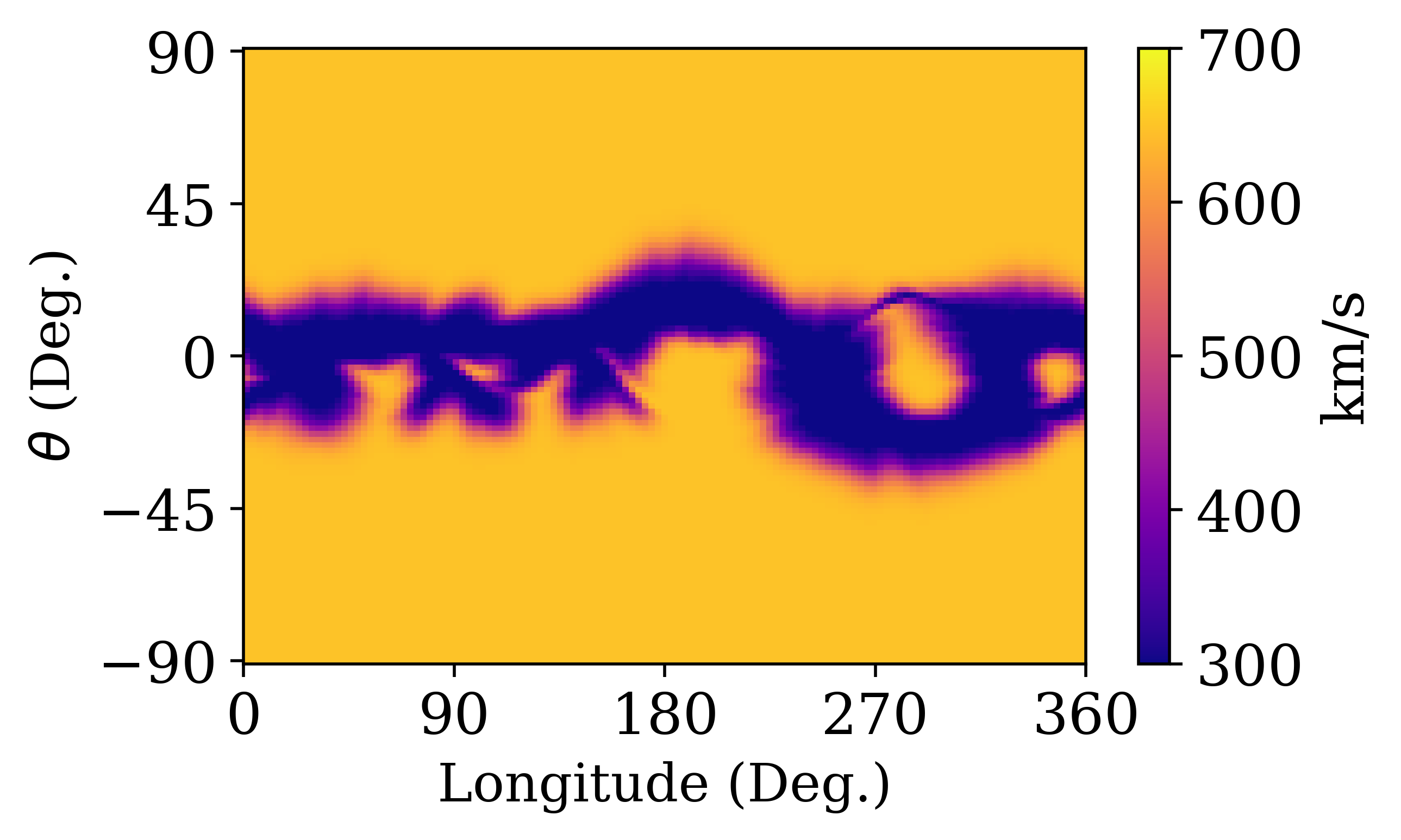}
        \label{fig:cc-30rs}
    \end{subfigure}
    \begin{subfigure}[b]{0.32\textwidth}
        \centering
        \caption{$V_{1 \text{AU}}$}
        \includegraphics[width=\textwidth]{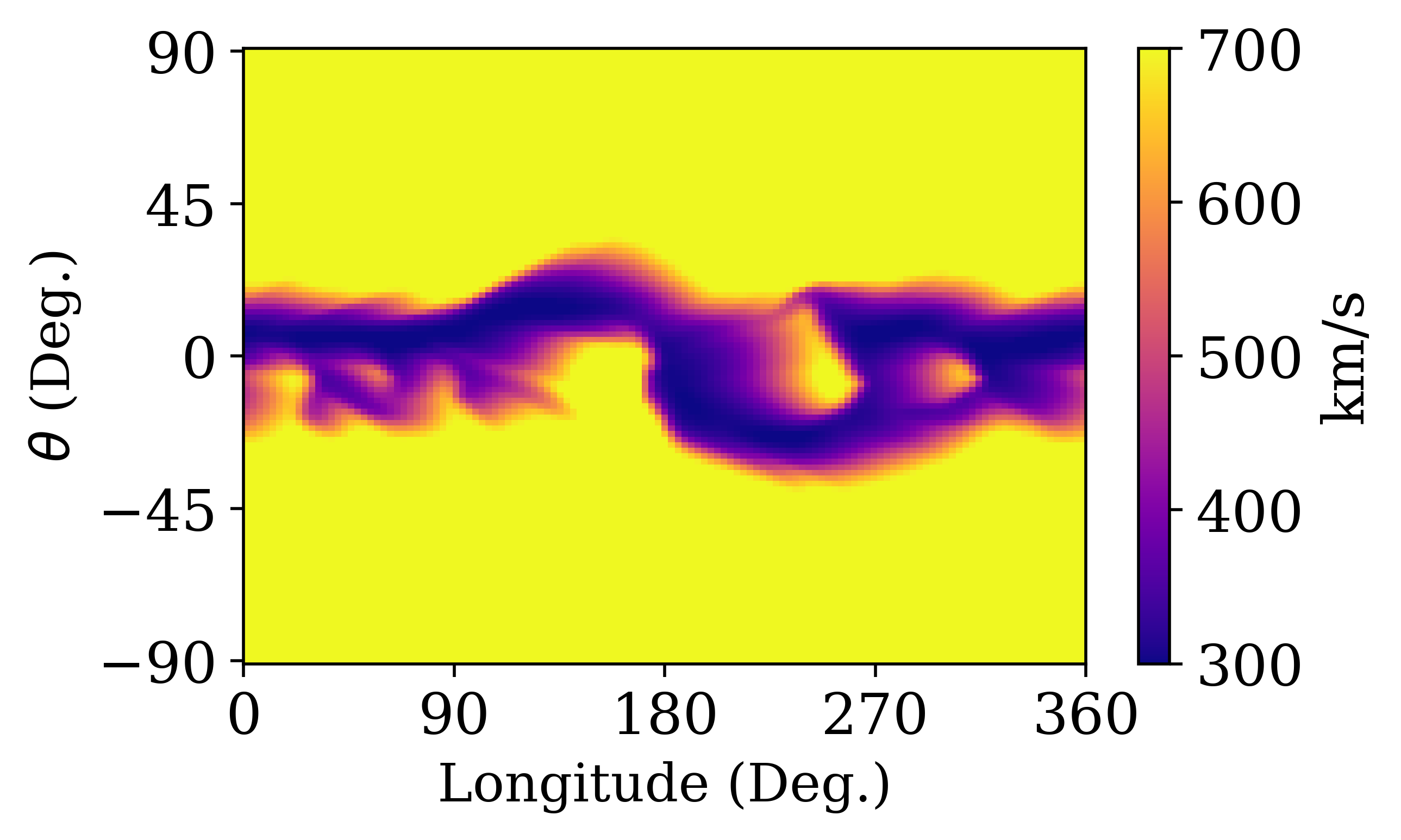}
        \label{fig:cc-1au}
    \end{subfigure}
    \begin{subfigure}[b]{0.32\textwidth}
        \centering
        \caption{$V_{30 R_{S}} \overset {2}{\star} V_{1 \text{AU}} $}
        \includegraphics[width=\textwidth]{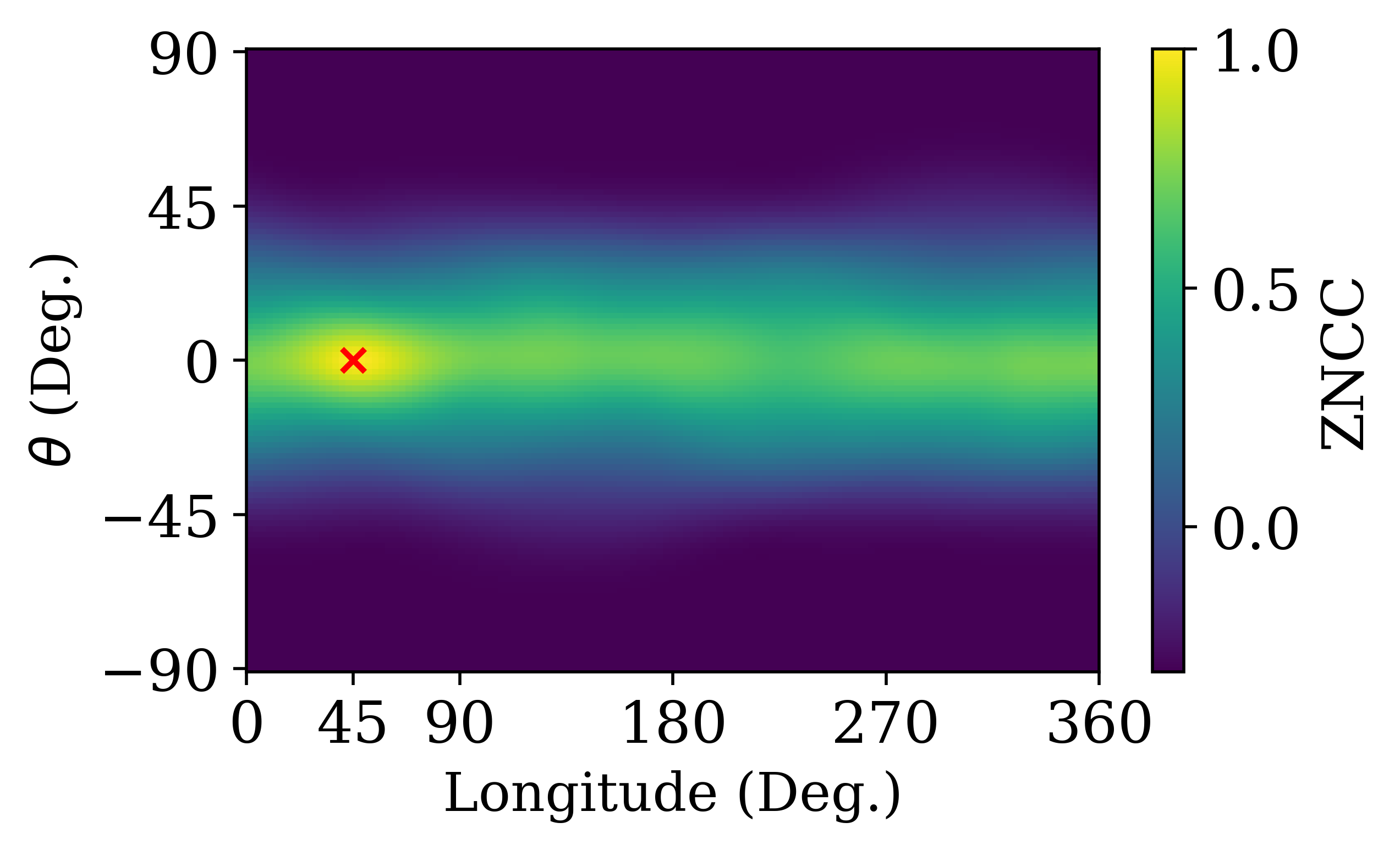}
        \label{fig:cc-1au-30rs}
    \end{subfigure}
    \caption{An illustration of discrete circular cross-correlation between the MAS CR2210 velocity results at (a)~the initial condition ($30 R_{S}$) and (b)~at Earth ($1 \text{AU}$). The bi-variate zero-normalized (circular) cross-correlation (ZNCC) between the two MAS velocity results at $30 R_{S}$ and $1 \text{AU}$ is shown in Graphic (c). The ZNCC is described by Eq.~\eqref{cc-dis-definition-multi-variate} along with normalizing the two signals by subtracting their mean and dividing by their standard deviations. The shift is found by the maximum of $V_{30R_{S}} \overset {2}{\star} V_{1\text{AU}}$, which is $45^{\circ}$ in longitude and $0^{\circ}$ in latitude (purely longitudinal translation).}
    \label{fig:cc-between-two-velocity-fields}
\end{figure}

We propose to find the optimal discrete displacement, $\boldsymbol{\tau}^{*} \in \mathbb{Z}^{k}$, by maximizing the cross-correlation between the two discrete signals (or snapshots), which amounts to solving
\begin{equation}\label{cc-max}
    \boldsymbol{\tau}^{*} \coloneqq {\argmax_{\boldsymbol{\tau} \in \mathbb{Z}^{k}}} {(\bf \star \overset {k}{\cdots } \star \bg)[\boldsymbol{\tau}]}.
\end{equation} 
Once the shift is computed for all training snapshots by applying the circular discrete cross-correlation between each snapshot $\bu_{i} \in \real^{n}$ and the initial condition $\bu_{0} \in \real^{n}$, we obtain the the multivariate discrete displacement for each time-step, i.e.,
\begin{equation}\label{cc-max-time-dependent}
    \boldsymbol{\tau}^{*}(t_{i}) \coloneqq {\argmax_{\boldsymbol{\tau} \in \mathbb{Z}^{k}}} {\{ \bu_{0} \star \overset {k}{\cdots } \star \bu_{i}\}[\boldsymbol{\tau}]}.
\end{equation}
Then, the shift function $c_{j}(t), \forall j \in \{1, 2, \ldots,  k\}$ is found via least squares polynomial curve fitting to the data points between the time increments and corresponding spatial location of the shift, such that the shift function is of the form
\begin{equation*} \label{polynomial-general-form}
    c_{j}(t) = \sum_{m= 0}^{d} a_{m} t^{m}
\end{equation*}
where $d$ is the degree of the polynomial approximation. 
To find the vector of real coefficients
$\ba=[a_{0}, a_{1}, \ldots, a_{d}]\in \real^{d+1}$, we solve the minimization problem 
\begin{equation*} \label{least-squares-minimization}
    \min_{\ba \in \real^{d}} \left \Vert \bT \ba - \bb \right \Vert^{2}_{2}
\end{equation*}
where $\bb= [x_{j, {\tau}^{\star}_{j}(t_{1})}, \ldots,  x_{j, {\tau}^{\star}_{j}(t_{K})}] \in~\real^{K}$ is the vector of corresponding spatial location of the shift, and the Vandermonde matrix $\bT$ is defined as
\begin{equation*}
    \bT = \begin{bmatrix}
    1 & t_{1} & \ldots & t_{1}^{d}\\
    1 & t_{2} & \ldots & t_{2}^{d}\\
    \vdots & \vdots &  & \vdots \\
    1 & t_{K} & \ldots & t_{K}^{d}\\
    \end{bmatrix} \in \real^{K\times (d + 1)}.
\end{equation*}
To make predictions outside the training interval, we approximate the shift by polynomial extrapolation of $\bc(t) = [c_{1}(t) , c_{2}(t), \ldots, c_{k}(t)]\in \real^{k}$. 
As an illustration, Figure~\ref{fig:cc-between-two-velocity-fields} shows the bi-variate discrete circular cross-correlation applied to the MAS CR2210 snapshot at $30R_{S}$ (the initial condition) and at $1\text{AU}$. Here, since the flow is steady, the independent variable  is the radial distance from the Sun, $r$, (instead of time $t$). For this case, the convective shift is $45^{\circ}$ in longitude and $0^{\circ}$ in latitude. Figure~\ref{fig:cc-between-two-velocity-fields} confirms that the translation in the solar wind is purely longitudinal due to the rotation of the Sun. 
The main idea behind the cross-correlation extrapolation technique is similar to the template-fitting technique studied in \cite{rowley_2000translation_symmetry,rowley_2003_selfsimilar} where the data is periodic and the template is set to be the initial condition.

%%%%%%%%%%%%%%%%%%%%%%%%%%%%%%%%%%%%%%%
%%%%%%%%%%%%%%%%%%%%%%%%%%%%%%%%%%%%%%%

%%%%%%%%%%%%%%%%%%%%%%%%%%%%%%%%%%%%%%%
\subsection{Illustrative Example: Shifted Operator Inference for the Inviscid Burgers' Equation} \label{sec:Burgers}
%%%%%%%%%%%%%%%%%%%%%%%%%%%%%%%%%%%%%%%
%
The one-dimensional inviscid Burgers' equation is of the form of Eq.~\eqref{general-form-pde-moc} describing the scalar quantity $u(x, t):~[a, b] \times [0, T] \mapsto \real^{+}$ with $f[u(x, t), t] = u(x, t)$ and $g[u(x, t), t] = 0$, such that
\begin{equation} \label{inviscid-burgers-equation}
    \frac{\partial u(x, t)}{\partial t} + u(x, t) \frac{\partial u(x, t)}{\partial x} = 0
\end{equation}
subject to the initial condition $u(x, t=0)=u_{0}(x)$ with appropriate boundary conditions at $x=a$ and $x=b$. In the moving coordinate frame defined by 
\begin{equation}
    \tilde{x}(x, t) = x + c(t) \qquad \text{and} \qquad u(x, t) = \tilde{u} (\tilde{x}(x, t), t)
\end{equation}
Burgers' equation~\eqref{inviscid-burgers-equation} becomes 
\begin{equation}\label{shifted-inviscid-burgers-equation}
\frac{\partial \tilde{u}(\tilde{x}, t)}{\partial t} + \left(\tilde{u}(\tilde{x}, t) + \frac{\text{d}c}{\text{d}t}\right) \frac{\partial \tilde{u}(\tilde{x}, t)}{\partial \tilde{x}} = 0.
\end{equation}
This can be written in conservative form as
\begin{equation}\label{shifted-burgers-conservative}
\frac{\partial \tilde{u}(\tilde{x}, t)}{\partial t} +  \frac{\partial}{\partial \tilde{x}}\left(\frac{1}{2}\tilde{u}(\tilde{x}, t)^2 + \frac{\text{d}c}{\text{d}t} \tilde{u}(\tilde{x}, t)\right)= 0.
\end{equation}
Then, by the conservative first-order upwind scheme, we approximate the spatial derivative by
\begin{equation} \label{conservative-upwind-scheme-burgers}
\frac{\partial}{\partial \tilde{x}}\left(\frac{1}{2}\tilde{u}(\tilde{x}^{(j)}, t)^2 + \frac{\text{d}c}{\text{d}t} \tilde{u}(\tilde{x}^{(j)}, t)\right) = 
\frac{1}{2\Delta \tilde{x}} \left[\tilde{u}(\tilde{x}^{{{(j)}}}, t)^2 - \tilde{u}(\tilde{x}^{{{(j-1)}}}, t)^2\right] + \frac{1}{\Delta \tilde{x}} \frac{\text{d}c}{\text{d}t} \left[\tilde{u}(\tilde{x}^{{(j)}}, t) - \tilde{u}(\tilde{x}^{{(j-1)}}, t)\right],
\end{equation}
where $j = 1, 2, \ldots, n$ denotes the grid index in $\tilde{x}$.
Based on the discretization scheme in Eq.~\eqref{conservative-upwind-scheme-burgers}, we can write the dynamics of Eq.~\eqref{shifted-burgers-conservative} in vector form as
\begin{equation} \label{shifted-burgers-vector-form-quad}
\frac{\text{d}\mathbf{\tilde{u}}(t)}{\text{d} t} = \bA(t) \mathbf{\tilde{u}(t)} + \bH \left[\mathbf{\tilde{u}}(t) \otimes \mathbf{\tilde{u}}(t)\right]
\end{equation}
where $\otimes$ denotes the Kronecker product and  $\mathbf{\tilde{u}}(t) = [\mathbf{\tilde{u}}(\tilde{x}^{(1)}, t), \mathbf{\tilde{u}}(\tilde{x}^{(2)}, t),\ldots, \mathbf{\tilde{u}}(\tilde{x}^{(n)}, t)]^{\top} \in \real^{n}$ denotes the state vector discretized over $n$ spatial points at time $t$. Here, $\bH \in \real^{n \times {n}^2}$ is the quadratic operator that corresponds to the discrete term $\frac{1}{2\Delta \tilde{x}} \left[\tilde{u}(\tilde{x}^{{(j)}}, t)^2 - \tilde{u}(\tilde{x}^{{(j-1)}}, t)^2\right]$ from Eq.~\eqref{conservative-upwind-scheme-burgers}.
Moreover, $\bA(t) \in \real^{n \times n}$ is the linear time-dependent operator corresponding to the discrete term $\frac{1}{\Delta \tilde{x}} \frac{\text{d}c}{\text{d}t} \left[\tilde{u}(\tilde{x}^{{(j)}}, t) - \tilde{u}(\tilde{x}^{{(j-1)}}, t)\right]$ in the moving coordinate frame. In the case where $c(t) \in \real$ is a linear function (so $\frac{\text{d}c}{\text{d}t} =$const.), meaning the wave is traveling at constant speed, the linear operator is time independent, i.e., $\bA(t) \equiv \bA$.

The traveling wave speed $c(t)$ can be estimated by the method of characteristics (see Section~\ref{sec:MethodOfCharacteristics}), where we can rewrite Eq.~\eqref{inviscid-burgers-equation} as the two coupled ODEs
\begin{equation} \label{burgers-characteristics}
    \frac{\text{d} x(t)}{\text{d} t} = u(x(t), t) \qquad\text{and}\qquad \frac{\text{d}u(x(t), t)}{\text{d}t} = 0.
\end{equation}
Hence, along the characteristic lines the quantity $u(x(t), t)$ remains constant, which can be verified by
\begin{equation} \label{verify-characteristics-burgers}
    \begin{aligned}
        \frac{\text{d}}{\text{d} t} u(x(t), t) & = \frac{\partial}{\partial t} u(x(t), t) +  \frac{\text{d} x(t)}{\text{d} t} \frac{\partial}{\partial x} u(x(t), t) =  \frac{\partial}{\partial t} u(x(t), t) + u(x(t), t) \frac{\partial}{\partial x} u(x(t), t) = 0.
    \end{aligned}
\end{equation}
Then, by integrating Eq. \eqref{burgers-characteristics} the characteristic curves are linear before shock formation. Let the spatial domain of $x$ be discretized on a uniform grid with $n$ points such that $\bx= [x^{(1)}, x^{(2)},\ldots, x^{(n)}] \in \real^{n}$. From here, we can approximate the shift function via Eq.~\eqref{shock-piecewise-c-of-t} as a piecewise continuous function:
\begin{equation}\label{burgers-c-of-t}
    c(t) = \begin{cases}
        \frac{1}{q-p}\sum_{j=p}^{q} {u(x^{(j)}, t=0)} t & \text{if } 0 < t < t_{s} \\
        s(t) - s(t_{s}) + a & \text{if } t > t_{s} 
    \end{cases}
\end{equation}
where $t_{s} \in \mathbb{R}$ is the time of shock formation, $a = \frac{1}{q-p}\sum_{j=p}^{q} {u(x^{(j)}, t=0)} t_{s}$, and $s(t)\in \mathbb{R}$ is the shock trajectory. 

To demonstrate the sOpInf method, we now consider a specific spatial domain $x \in [0,3]$ and time domain $t \in [0, 2]$ along  with a Gaussian initial condition $u_{0}(x) = 0.8 + 0.5 \exp{(-(x - 1)^2/0.1)}$ and periodic boundary conditions. Equation~\eqref{inviscid-burgers-equation} in the regular coordinates is solved numerically via the forward Euler method in time and the conservative upwind scheme in space on an equidistant computational grid with $500$ discretization points in space, and $1000$ points in time.
With this choice, the CFL condition $u(x, t) \frac{\Delta t}{\Delta x} \leq 1$ is satisfied, where $\Delta x$ and $\Delta t$ denote the grid spacing in $x$ and $t$, respectively. The training dataset consists of 80\% of the snapshots and the testing dataset consists of 20\% of the snapshots. The shock is formed at $t_{s}= \displaystyle \min_{x} \left(-{\text{d} u_{0}(x)/{\text{d} x}}\right)^{-1} \approx 0.737$ and the shock trajectory $s(t)$ is computed by Whitham's equal-area rule \cite{whitham_1976}, in which the area of each lobe in the mutli-valued solution is numerically approximated via the trapezoidal rule.

%%%%%%%%%%%%%%%%%%%%%%%%%%%%%%%%
\begin{figure}
   \centering
    \begin{subfigure}[b]{0.32\textwidth}
        \centering
        \caption{Original coordinates}
        \includegraphics[width=\textwidth]{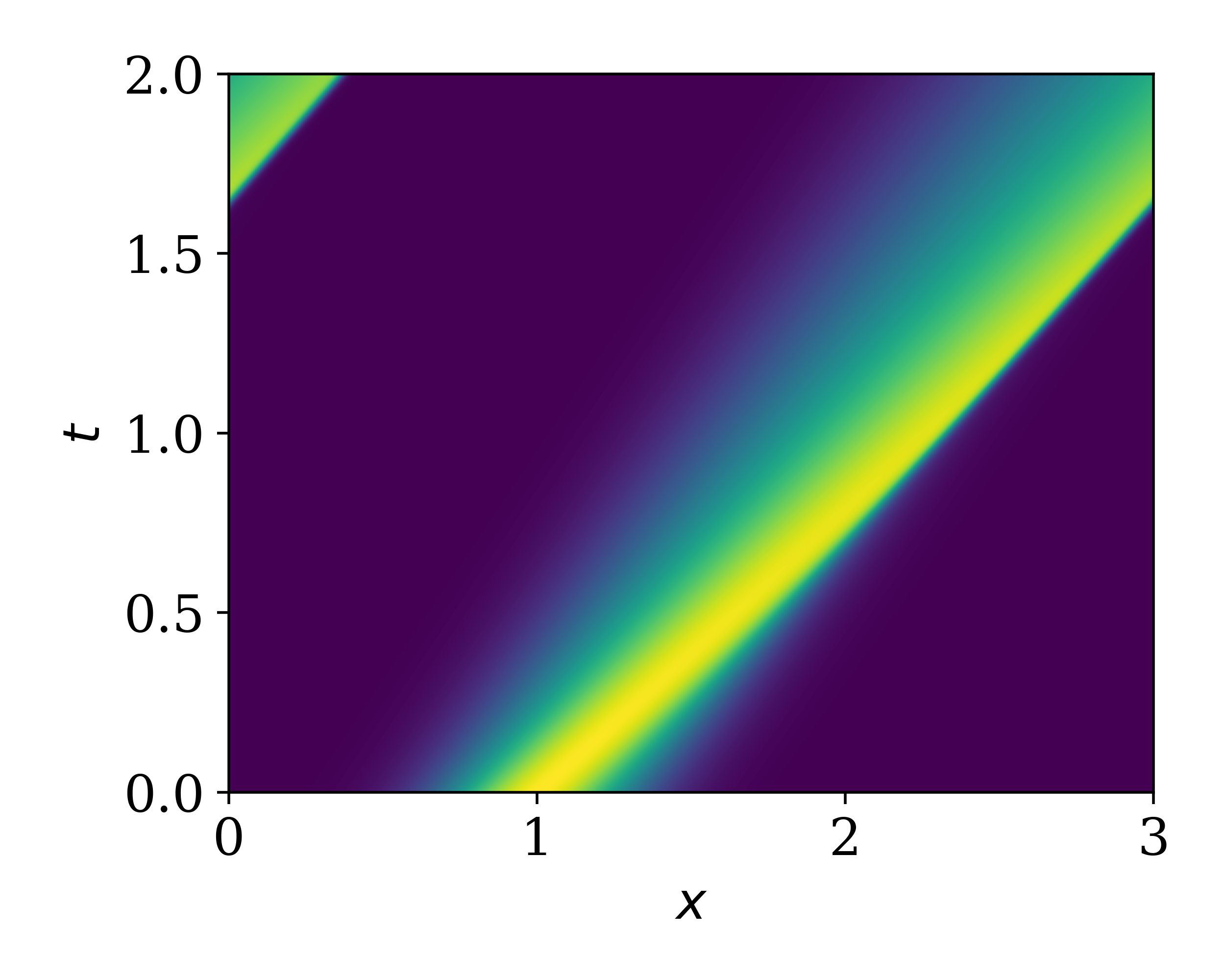}
        \label{fig:original-burgers}
    \end{subfigure}
    \begin{subfigure}[b]{0.32\textwidth}
        \centering
        \caption{Shifted coordinates}
        \includegraphics[width=\textwidth]{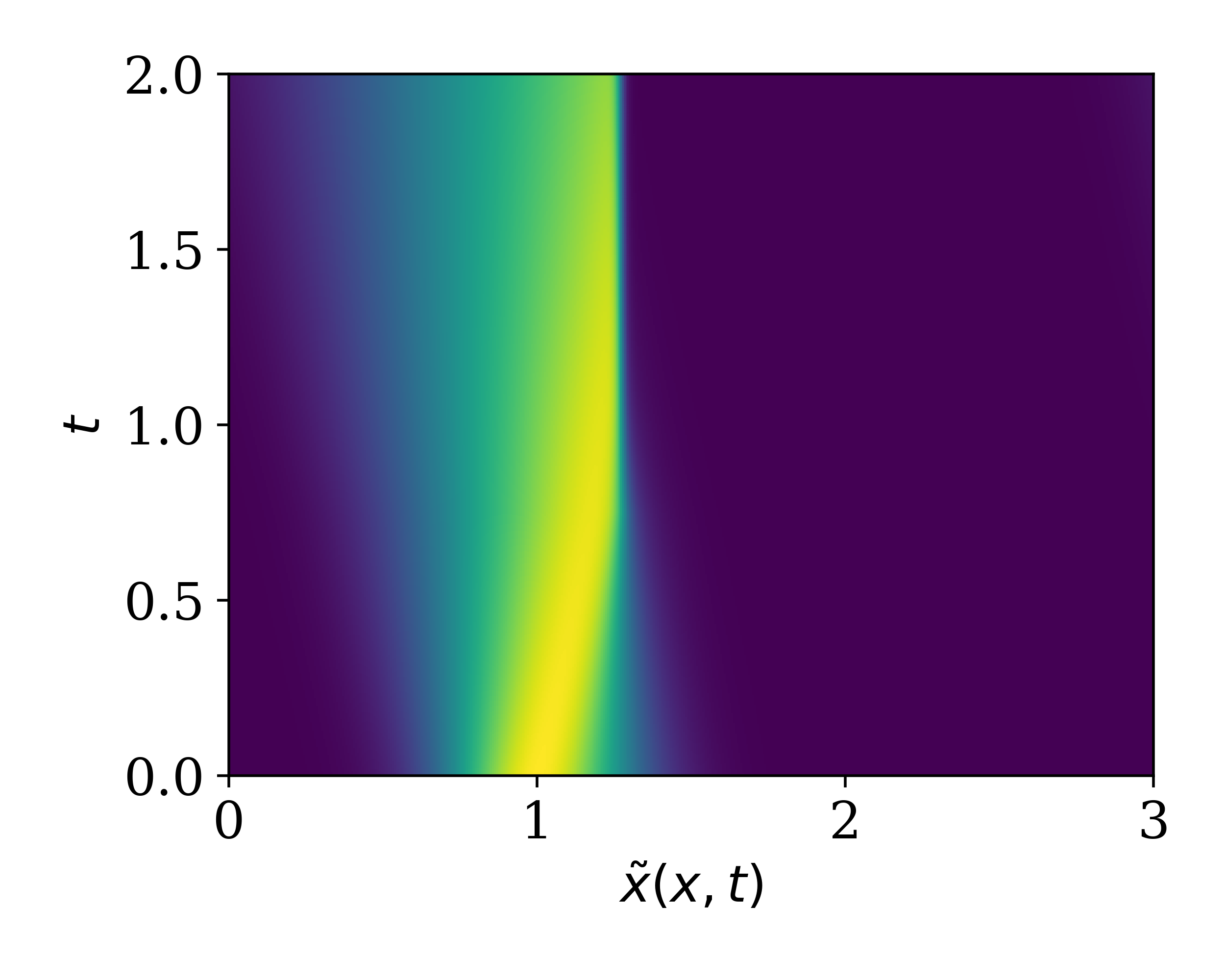}
        \label{fig:shifted-burgers}
    \end{subfigure}
    \begin{subfigure}[b]{0.32\textwidth}
        \centering
        \caption{Shift function $c(t)$}
        \includegraphics[width=\textwidth]{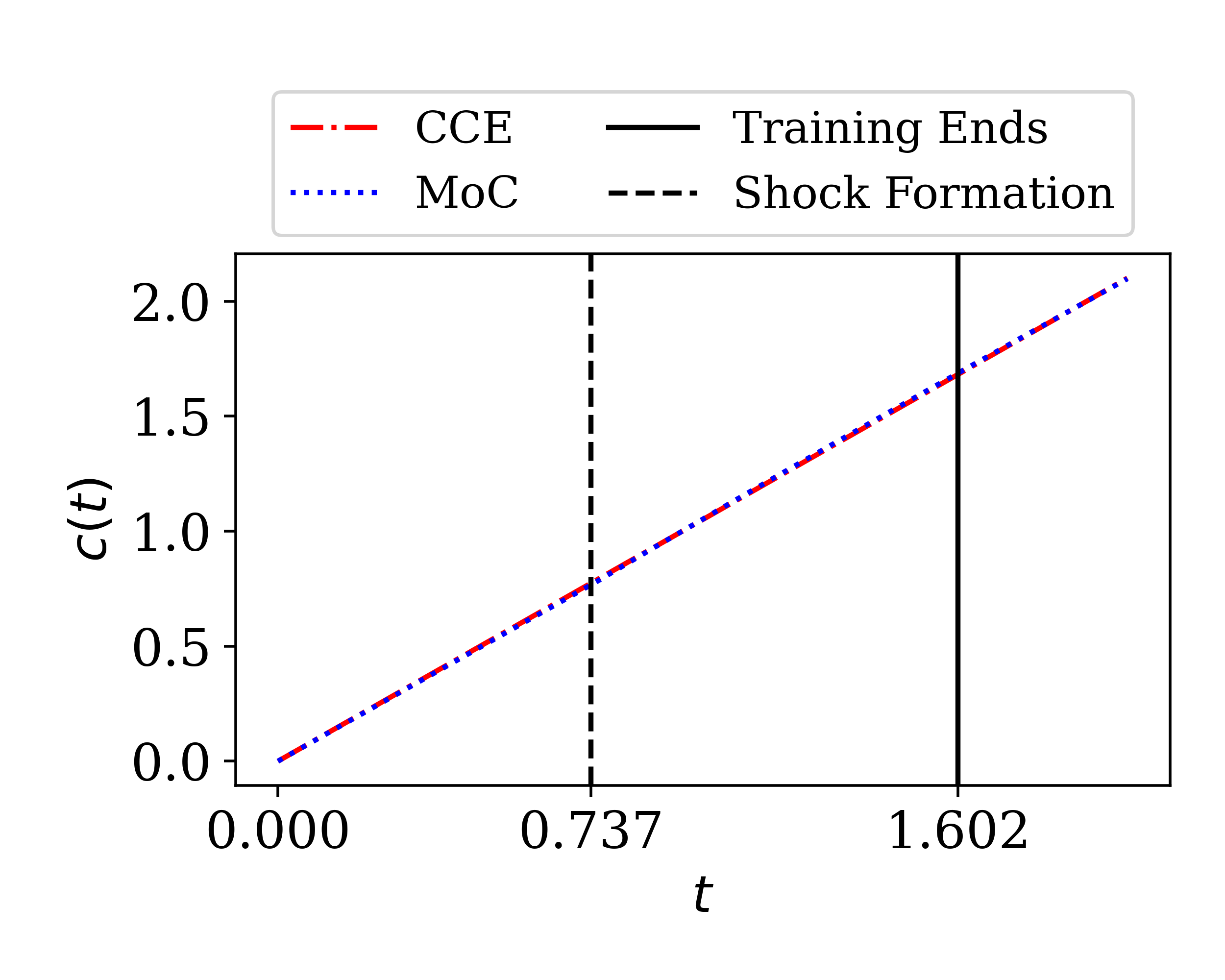}
        \label{fig:shift-function-burgers}
    \end{subfigure}
    \begin{subfigure}[b]{0.33\textwidth}
        \centering
        \caption{Cumulative Energy obtained by Eq.~\eqref{energy-criteria} }
        \includegraphics[width=\textwidth]{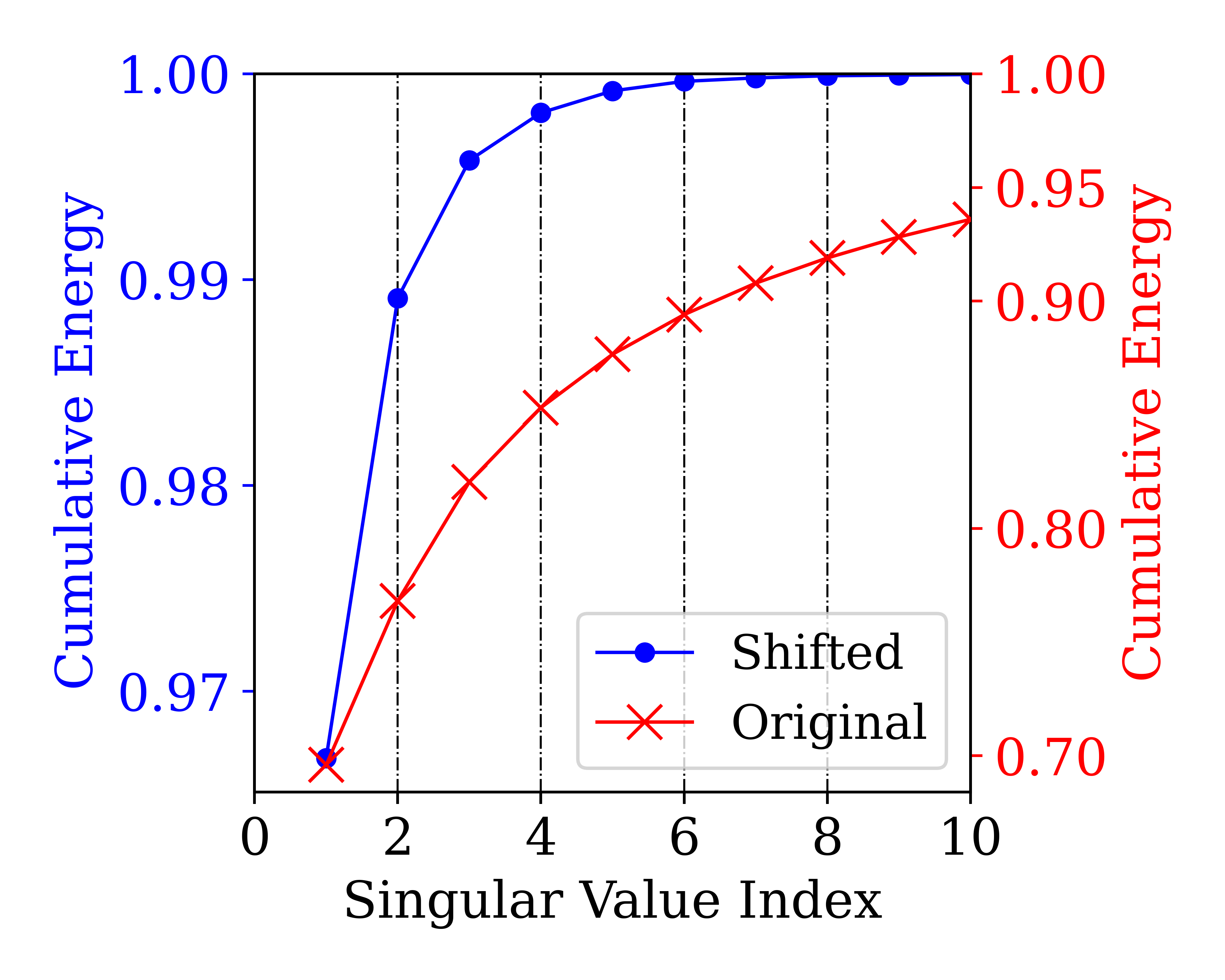}
        \label{fig:singular-value-decay-burgers}
    \end{subfigure}
        \begin{subfigure}[b]{0.52\textwidth}
        \centering
        \caption{Characteristic Curves}
        \includegraphics[width=\textwidth]{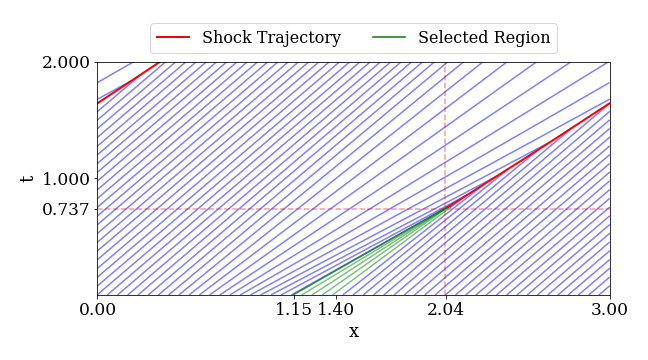}
        \label{fig:characteristic-curves}
    \end{subfigure}
    \caption{\footnotesize One-dimensional inviscid Burgers' equation with Gaussian initial condition and periodic boundary conditions. The numerical solutions are shown in the (a)~original coordinates $(x, t)$ and the (b)~shifted coordinates $\tilde{x}(x, t)$. Graphic~(c) presents a comparison between finding the wave speed function $c(t)$ using the method of characteristics (MoC) and the cross-correlation extrapolation method (CCE). The results are visually indistinguishable. Graphic~(d) shows the cumulative energy computed via the SVD of $\bU$ (original coordinates) vs. $\tilde{\bU}$ (shifted coordinates), illustrating that the data can be much better approximated in the shifted than in the original coordinates. The characteristic curves including the shock trajectory are shown in Graphic~(e).}
    \label{fig:Burgers-Eqn-Comparison-Of-Techniques}
\end{figure}
%%%%%%%%%%%%%%%%%%%%%%%%%%%%%%%%

The cross-correlation extrapolation method (Section~\ref{sec:cross-correlation}) approximated $c(t)$ by least-squares linear fitting to the cross-correlation between the training snapshots and the initial condition $u_{0}(x)$, resulting in $c(t) = 1.05 t$. Whereas, via the method of characteristics (Section~\ref{sec:MethodOfCharacteristics}) the shift function, $c(t)$ described in Eq.~\eqref{burgers-c-of-t}, is linear only before shock formation, where we compute the mean of characteristics emanating from the spatial interval $[1.15, 1.4]$ before shock formation, see the green characteristic curves in Figure~\ref{fig:characteristic-curves}. By the Rankine-Hugoniot jump condition \cite{whitham_1976}, the shock curve $s(t)$, and hence the shift function after shock formation, is non-linear, see the red curve in Figure~\ref{fig:characteristic-curves}.
However, the absolute difference between the shift function $c(t)$ obtained by the method of characteristics vs. the cross-correlation extrapolation is less than $10^{-2}$ for all $t \in [0, 2]$, indicating there is not a significant difference between the two methods, see Figure~\ref{fig:shift-function-burgers} for a visual comparison. 
With that, we found that using the method of characteristics to find~$c(t)$ resulted in up to $2\%$ more accurate ROM results in comparison to the results obtained by the cross-correlation extrapolation linear shift function. By aligning the wave discontinuity at one spatial point, the ROM modes can better approximate the shock. We continue the Burgers' ROM analysis using the shift function computed by the method of characteristics.

Figure~\ref{fig:original-burgers} shows the inviscid Burgers' equation results on the original coordinate system $(x, t)$ while Figure~\ref{fig:shifted-burgers} shows the inviscid Burgers' equation results on the shifted coordinates $(\tilde{x}, t)$. It is apparent from Figure~\ref{fig:shifted-burgers} that the dynamics are largely absent of translation in the shifted coordinate frame. The evolution of the initial wave depends solely on shock formation as the discontinuity in the wave sharpens. 
Figure~\ref{fig:singular-value-decay-burgers} compares the singular value decay of the solution data matrix $\bU$ on the original coordinates and of $\tilde{\bU}$ on the shifted coordinates. For the former, the singular value decay is very slow due to the translation properties in the system; a ROM in those coordinates would require a large number of basis functions. In contrast, once the dynamics are absent of translation on the shifted coordinates the system can be approximated using only a few modes. 

\begin{figure}
    \centering
    \includegraphics[width=\textwidth]{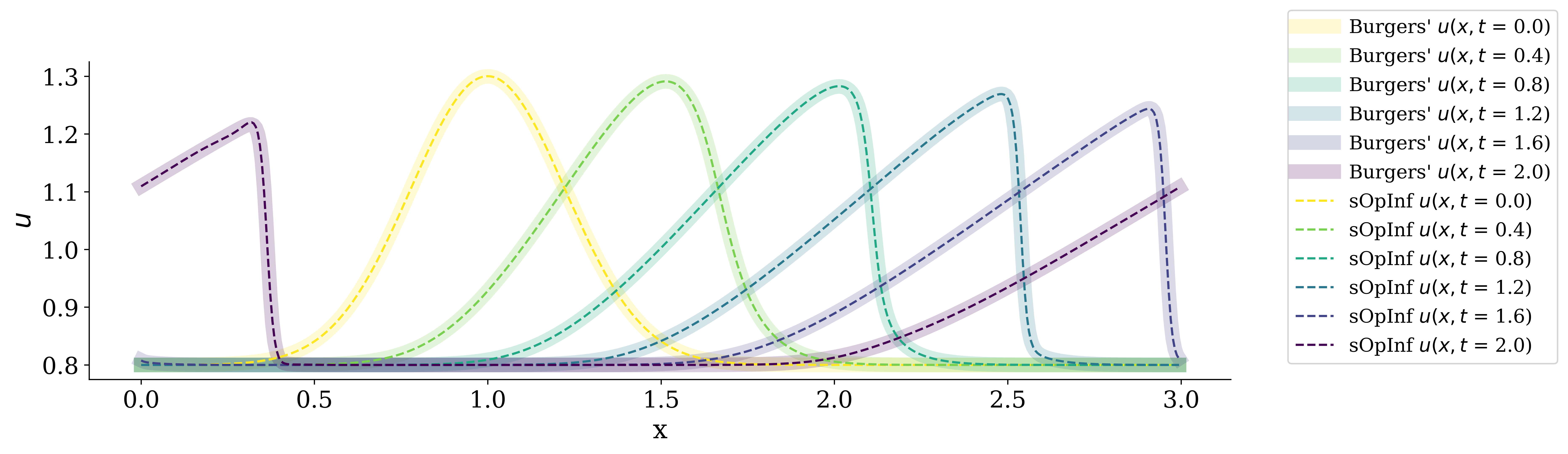}
    \caption{\footnotesize Solutions from the sOpInf model of the form $\dot{\hat{\bu}} = \hat{\bA} \hat{\bu} + \hat{\bH} (\hat{\bu} \otimes^{\prime} \hat{\bu})$ with $\ell = 9$ modes show very good agreement with the high-fidelity solutions for the one-dimensional inviscid Burgers' equation with Gaussian initial condition and periodic boundary conditions.}
    \label{fig:Burgers-Eqn-ROM-Results}
\end{figure}

Figure~\ref{fig:Burgers-Eqn-ROM-Results} illustrates the predicted solutions from sOpInf compared to the FOM results for the inviscid Burgers' equation. The sOpInf model is of the form $\dot{\hat{\bu}} = \hat{\bA} \hat{\bu} + \hat{\bH} (\hat{\bu} \otimes^{\prime} \hat{\bu})$ with $\ell = 9$ modes. The regularization coefficients of the operators $\hat{\bA} \in \real^{\ell \times \ell}$ and $\hat{\bH} \in \real^{\ell \times {\frac{1}{2}\ell(\ell+1)}}$ are $\lambda_{1} = 1$ and $\lambda_{2} = 1$, respectively. The sOpInf results in 0.9999958 Pearson correlation coefficient (PCC) and $1.204\times 10^{-4}$ mean relative error (MRE) in comparison to the inviscid Burgers' results, indicating that the sOpInf model is able to capture well the evolution of the high-fidelity Burgers' simulation with only a few modes. On the shifted coordinates we are able to construct a ROM with only 9 modes, meanwhile, in the original coordinates, we would need 89 modes to achieve the same projection error. This demonstrates that the shifting procedure produces significant computational speedups.

\begin{figure}
    \centering
    \includegraphics[width=\textwidth]{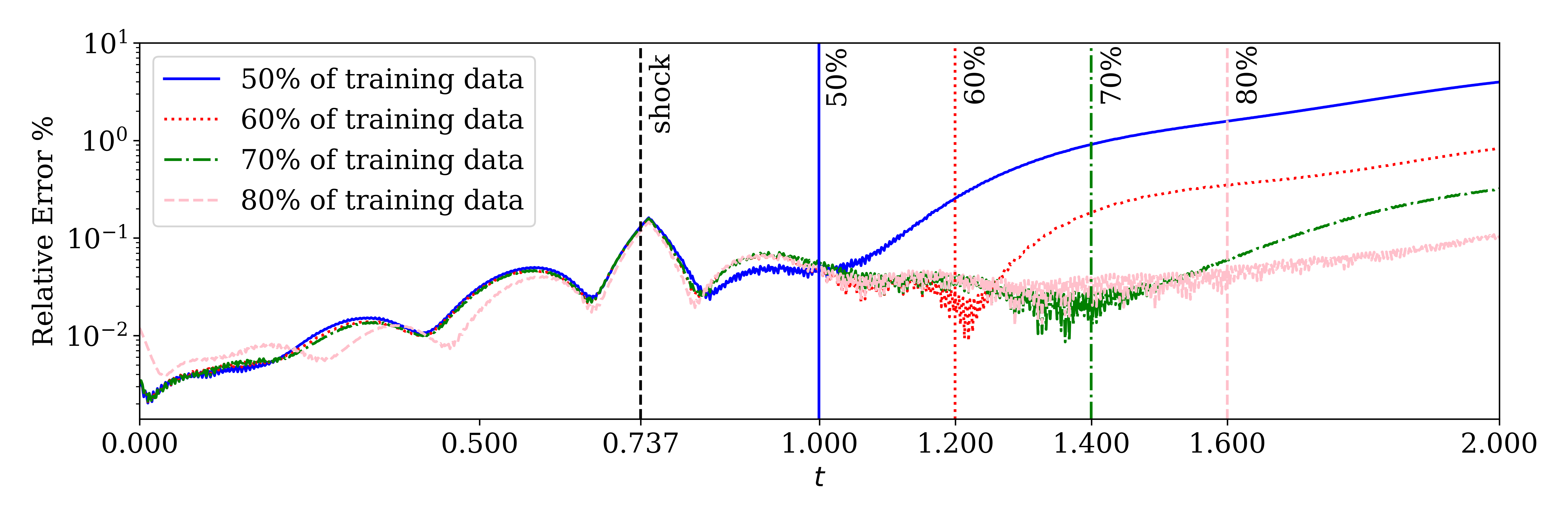}
    \caption{\footnotesize The sensitivity of sOpInf ROM to the amount of training data with 50\%, 60\%, 70\%, and 80\% of the total snapshots used for training. The relative error in the $L_{2}$-norm is shown as a function of time. The numerical results show that the sOpInf methodology is able to learn adequate ROMs with as low as 50\% of the total snapshots utilized for training.}
    \label{fig:Burgers-Eqn-Sensitivity-to-training}
\end{figure}

We next assess how the amount of training data affects the accuracy of the resulting ROM by training sOpInf on 50\%, 60\%, 70\%, and 80\% of the total snapshots, see Figure~\ref{fig:Burgers-Eqn-Sensitivity-to-training}. The numerical results show that---as anticipated---increasing the amount of training data improves the ROMs accuracy. Additionally, the relative error measured in the $L_{2}$-norm is bounded by $10\%$ for all four simulations, resulting in an adequate ROM using as low as 50\% of the total snapshots for training. To analyze the framework's robustness to noise, we tested the sOpInf methodology on the inviscid Burgers' simulated data with added noise.
The results are presented in \ref{appA} and demonstrate that the method seems to be robust to the addition of moderate levels of Gaussian noise. 

%%%%%%%%%%%%%%%%%%%%%%%%%%%%%%%%
%%%%%%%%%%%%%%%%%%%%%%%%%%%%%%%%
%%%%%%%%%%%%%%%%%%%%%%%%%%%%%%%%
%%%%%%%%%%%%%%%%%%%%%%%%%%%%%%%%

%%%%%%%%%%%%%%%%%%%%%%%%%%%%%%%%%%%%%%%
\section{Numerical Results for MAS and HUX Solar Wind Models} \label{sec:Numerical-Results}
%%%%%%%%%%%%%%%%%%%%%%%%%%%%%%%%%%%%%%%
We apply the sOpInf methodology to learn low-dimensional models for the ambient solar wind radial velocity predicted by the HUX and MAS models described in Section~\ref{sec:sw-model}. We consider a specific event of interest and give some background on that event in Section~\ref{sec:physical-relevance}. In Section~\ref{sec:data-computational-enviroment-etc}, we give details on the data and implementation. Section~\ref{sec:hux-numerical-results} and Section~\ref{sec:mas-equator-results} present the sOpInf two-dimensional steady results trained on HUX and MAS equatorial plane data, respectively.  Section \ref{sec:mhd-full-sun-results} presents sOpInf three-dimensional steady results trained on the full-Sun MAS results. Section~\ref{sec:comparison-of-streamlines} compares sOpInf and HUX reconstruction of MAS equatorial plane streamlines. Lastly, Section~\ref{sec:model-form} discusses the choice of ROM model form.
The public repository \url{https://github.com/opaliss/Space-Weather-ROM} contains a collection of Jupyter notebooks in Python~3.9 containing the code and data used in this study.

%%%%%%%%%%%%%%%%%%%%%%%%%%%%%%%%%%%%%%%%%%%%%%%%%%%%%%%%%%%
\subsection{Physical Relevance of Data} \label{sec:physical-relevance}
\begin{figure}
    \centering
     \centering
    \begin{subfigure}[b]{0.52\textwidth}
        \centering
        \caption{MAS radial velocity}
        \includegraphics[width=\textwidth]{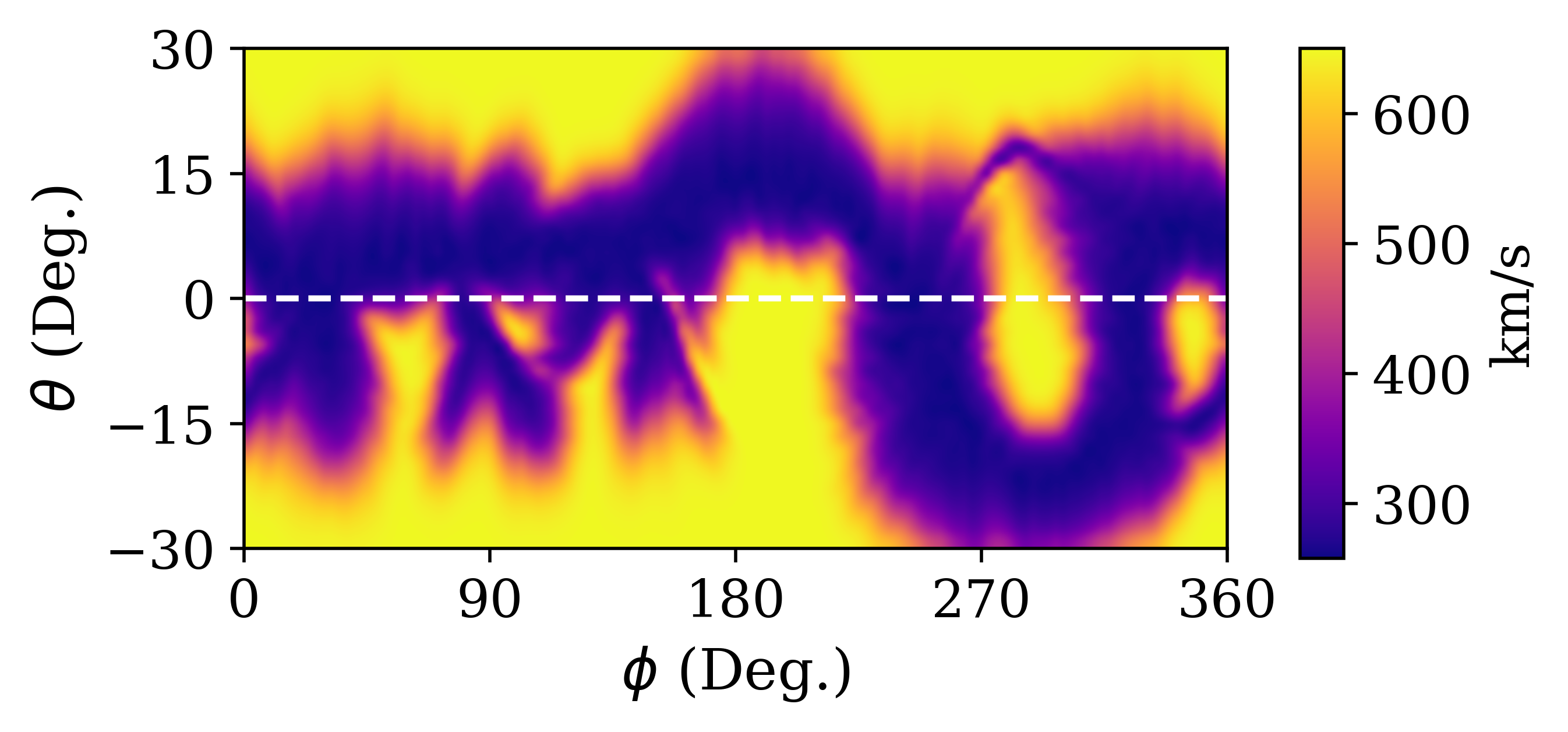}
        \label{fig:60-lat-30rs}
    \end{subfigure}
      \hfill
    \begin{subfigure}[b]{0.4\textwidth}
        \centering
        \caption{MAS equatorial plane radial velocity}
        \includegraphics[width=\textwidth]{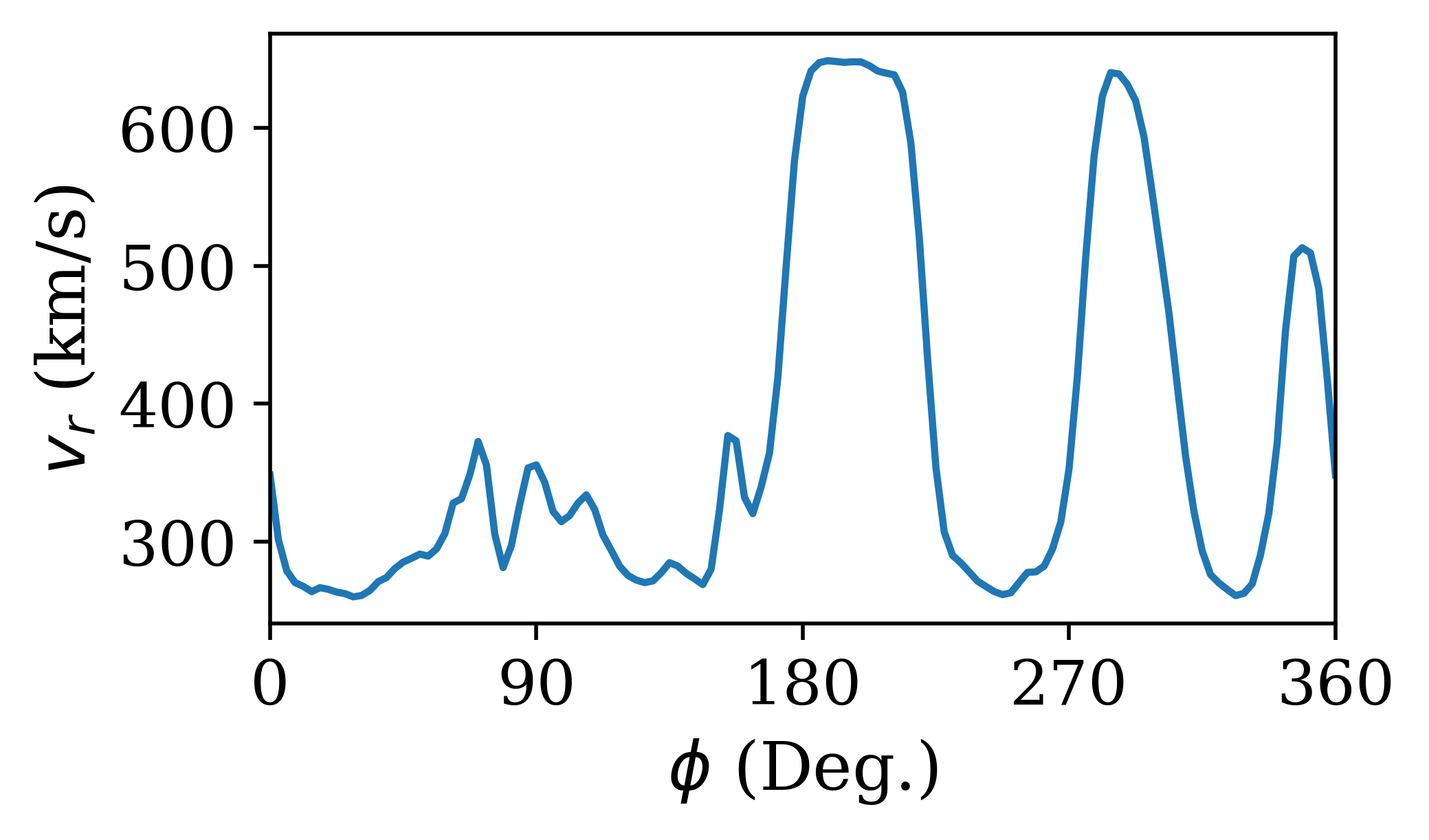}
        \label{fig:equator-30rs}
    \end{subfigure}
    \caption{\footnotesize The MAS solar wind radial velocity solution at heliocentric distance $r=30R_{S}$ during CR2210 (a) ranging from $\theta \in [-30^{\circ},30^{\circ}]$ in latitude and (b) the equatorial ($\theta_{e.p.} = 0^{\circ}$) velocity profile. The solar wind radial velocity at the equator has three main peaks, originating from coronal holes, which make this period a great candidate to study large-scale structure in the solar wind.}
    \label{fig:MAS-30RS-60deg-span-in-lat}
\end{figure}

%%%%%%%%%%%%%%%%%%%%%%%%%%%%%%%%%%%%%%%%%%%%%%%%%%%%%%%%%%%
We focus on Carrington Rotation (CR) 2210, which occurred from 26 October to 23 November 2018, during a solar minimum. Figure~\ref{fig:60-lat-30rs} shows the heliospheric solar wind radial velocity MAS results for CR2210 in the latitude region of $\theta \in [-30^{\circ},30^{\circ}]$ and Figure~\ref{fig:equator-30rs} presents the equatorial solar wind radial velocity profile. 
As observed in Figure~\ref{fig:MAS-30RS-60deg-span-in-lat}, during CR2210, the solar wind radial velocity at the equator has three main peaks, essentially making this period a great candidate to study large-scale structure in the solar wind. Although not shown here, the origin of the fast wind (ranging from $500$ to $700$ km/s) is from equatorial coronal holes located at approximately $180^{\circ}-210^{\circ}$, $280^{\circ}-300^{\circ}$, and $330^{\circ}-360^{\circ}$ in longitude at the source surface (see \cite[Fig. 8(a)]{badman_2020_psp} for a synoptic view at $2R_{S}$ of the coronal hole regions). Although fast streams commonly originate in large coronal holes, slow streams come from various coronal sources (e.g., coronal hole boundaries, coronal loops, etc.). Another reason to consider this data is that this Carrington period is well-studied in literature due to Parker Solar Probe (PSP) reaching its first perihelion pass of $35.7R_{S}$ after its first Venus gravity assist on 6 November 2018, as it broke records by becoming the closest spacecraft to the Sun. Moreover, the authors in \cite{riley2021_psp_mas} showed that the thermospheric MAS solar-wind speed results highly match the observations made by PSP during CR2210. Thus, while we do not present a comparison with \textit{in-situ} solar-wind observations we can treat the MAS results as a physically meaningful representation of the solar wind.

%%%%%%%%%%%%%%%%%%%%%%%%%%%%%%%%
%%%%%%%%%%%%%%%%%%%%%%%%%%%%%%%%

%%%%%%%%%%%%%%%%%%%%%%%%%%%%%%%%%%%%%%%
\subsection{Data and Implementation Details} \label{sec:data-computational-enviroment-etc}
%%%%%%%%%%%%%%%%%%%%%%%%%%%%%%%%%%%%%%%

%%%%%%%%%%%%%%%%%%%%%%%%%%%%%%%%%%%%%%%
\subsubsection{MAS Data} \label{MAS-Data}
%%%%%%%%%%%%%%%%%%%%%%%%%%%%%%%%%%%%%%%
The MAS model is described in detail in Section~\ref{sec:MAS-Model}. The MAS model solar wind velocity results are time-stationary in spherical coordinates. The data covers the entire domain in longitude $0^\circ\leq\phi\leq360^\circ$, latitude $-90^\circ \leq \theta \leq 90^\circ$, and radial axis (a.k.a. heliocentric distance) $0.14\text{AU}\leq r \leq1.1\text{AU}$. The MAS simulation results are on a rectangular grid with $n_{\phi}=128$  uniformly spaced points in Carrington longitude, $n_{\theta} = 111$  uniformly spaced points in heliographic latitude, and $n_{r} = 140$ points on a non-uniformly spaced grid in the radial axis. Hence, the snapshot data dimension is $n_x  = n_{\phi} \times n_{\theta} = 14,208$ where we treat $r$ as the independent variable. A convergence check with a higher-resolution grid found that this resolution is sufficient for physical accuracy. 
The MAS coronal and heliospheric models (implemented in FORTRAN) were run with medium and high resolution by the authors of \cite{MAS}, which took approximately four and 28 hours of wall-clock time using four NVIDIA RTX 2080Ti GPUs, respectively \cite{Caplan_2019_GPU}. We derive data-driven ROMs from the medium-resolution MAS heliospheric simulation which is publicly available at PSI's web page \cite{MHDweb}. The MAS training datasets contain 70\% of the snapshots, with $98$ and $42$ snapshots for training and testing, respectively. The training domain is from $0.14$ to $0.82\text{AU}$ and the testing domain is from $0.82$ to $1.1\text{AU}$.

%%%%%%%%%%%%%%%%%%%%%%%%%%%%%%%%%%%%%%%
\subsubsection{HUX Data} \label{HUX-Data}
%%%%%%%%%%%%%%%%%%%%%%%%%%%%%%%%%%%%%%%
The HUX model is described in detail in Section~\ref{sec:HUX}. To simulate the HUX model at the heliographic solar equatorial plane we numerically solve the ODE in  Eq.~\eqref{ode-form-hux} via the forward explicit Euler's method in time and setting the initial condition to the MAS velocity profile results at $30R_{S}\approx 0.14 \text{AU}$. The HUX dataset is two-dimensional with $n_{\phi} = 128$ and $n_{r} = 140$ (satisfying the CFL condition) on the same domain as listed above for the MAS results. On the equatorial plane ($\theta_{\text{e.p}} = 0$), the angular frequency of the Sun is $\Omega_{\text{rot}}(\theta_{\text{e.p}}) = \frac{2 \pi}{25.38} \text{1/days}$.
The HUX dataset took 0.03s to simulate on a MacBook Pro 2.3 GHz Quad-Core Intel Core i7 processor with 16 GB RAM. We note that the HUX model is already computationally efficient. We train a ROM for HUX merely as a introductory example since it produces the same physics (advection-dominated solutions) that we expect in much more expensive solvers, and not to show any computational improvements.
The HUX training and testing domains are identical to the MAS equatorial training domain, see details in Section~\ref{MAS-Data}. 

%%%%%%%%%%%%%%%%%%%%%%%%%%%%%%%%%%%%%%%
\subsubsection{ROM Implementation} \label{ROM-Setup}
%%%%%%%%%%%%%%%%%%%%%%%%%%%%%%%%%%%%%%%
The MAS and HUX numerical results are on a non-uniformly spaced grid in the radial axis, consequently, we approximate the derivatives of the training data with respect to $r$, i.e. $\frac{\text{d}}{\text{d}r} \hat{\bu}$ in Eq.~\eqref{data-matrix-and-derivative}, via a second-order accurate central difference in the interior points and second-order accurate one-sided (forward and backward) difference at the boundaries. For the case of uniform grid meshing, e.g. Burgers' equation ROM presented in Section~\ref{sec:Burgers}, we used a sixth-order finite difference scheme.
We simulate the ROM via an implicit multi-step variable method based on a backward differentiation formula using the \texttt{scipy.integrate.solve\_ivp()} Python function.
The ROM regularization coefficients, $\lambda_{1}$ and $\lambda_{2}$, are chosen from the logarithmically spaced set, i.e. $\{10^{0}, 10^{1}, 10^{2}, \ldots, 10^{10}\}$, such that the best coefficients minimizes the relative error measured via the $L_{\infty}$-norm over the training regime. 

%%%%%%%%%%%%%%%%%%%%%%%%%%%%%%%%%%%%%%%
\subsection{HUX Equatorial Plane Numerical Results} \label{sec:hux-numerical-results}
\begin{figure}
\centering
    \includegraphics[width=1\linewidth]{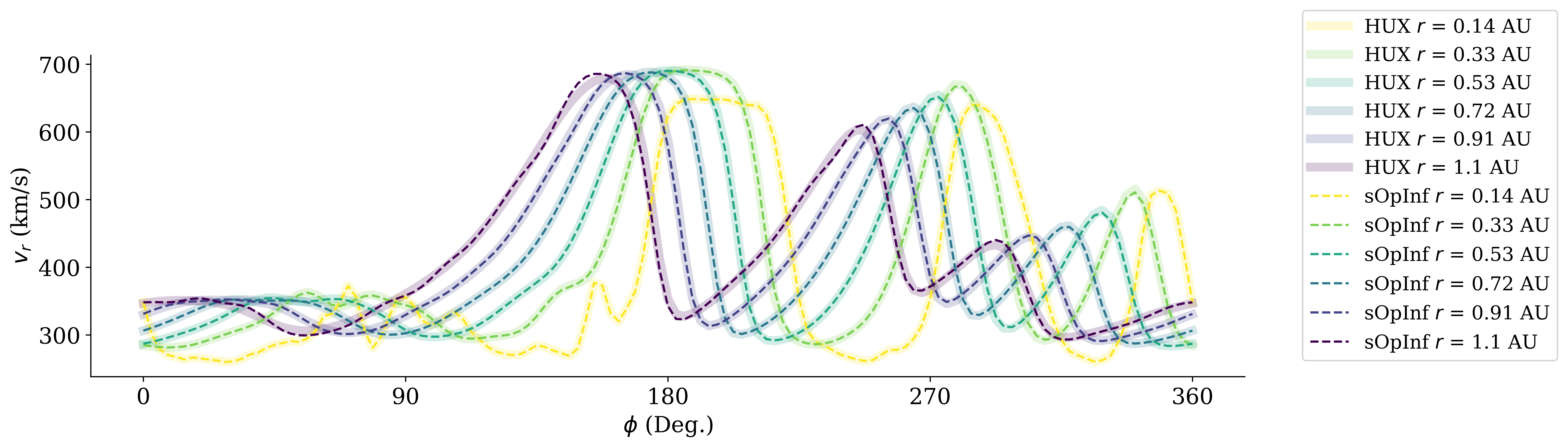}
    \caption{The HUX solar wind radial velocity results at the heliographic equatorial plane for CR2210 along with sOpInf quadratic ROM results. The sOpInf ROM aligns very well with the data past the training interval; the snapshots at $r=0.91$AU and $r=1.1$AU are testing data where the ROM is fully predictive.}
    \label{fig:sOpInf-HUX-1D}
\end{figure}

%%%%%%%%%%%%%%%%%%%%%%%%%%%%%%%%%%%%%%%
We apply the sOpInf framework to learn a ROM of the HUX CR2210 equatorial plane radial velocity. As derived in Section~\ref{sec:hux-pde-section}, the HUX underlying equation is 
\begin{equation*} 
-\Omega_{\text{rot}}\left(0\right ) \frac{\partial v_{r}(r, \phi)}{\partial \phi} + v_{r}(r, \phi)\frac{ \partial v_{r}(r, \phi)}{\partial r}=0,
\end{equation*}
where $r, \phi$ are the independent variables.
To begin, we shift the HUX dynamics to a moving coordinate frame defined by 
\begin{equation*}
    \tilde{\phi}(r, \phi) = \phi + c(r) \qquad \text{and} \qquad v_{r}(r, \phi) = \tilde{v}_{r} (\tilde{\phi}(r, \phi), r).
\end{equation*}
The shift function $c(r)$ can be learned via either the method of characteristics (Section~\ref{sec:MethodOfCharacteristics}) or the cross-correlation extrapolation method (Section~\ref{sec:cross-correlation}), in particular, the circular uni-variate cross-correlation method described in Eq.~\eqref{cc-dis-circular-definition}. The linear-fit cross-correlation shift function resulted in $c(r) = -51.711^{\circ}r +7.032^{\circ}$.
Our numerical studies found that there is no substantial differences between the numerical results of the two methods.
Following the steps described in Section~\ref{sec:MethodOfCharacteristics}, the HUX characteristic curves are derived by the following two coupled ODEs:
\begin{equation} \label{hux-characteristics}
\frac{\text{d}}{\text{d} r} v_{r}(\phi(r), r)= 0 \quad\text{and}\quad \frac{\text{d} \phi(r)}{\text{d} r} = -\frac{\Omega_{\text{rot}}(0)}{v_{r}(\phi(r), r)}.
\end{equation}
Then, by integration of Eq.~\eqref{hux-characteristics}, the characteristics before shock formation are straight lines described by 
\begin{equation*}
\phi(r) = \phi -\frac{\Omega_{\text{rot}}(0)}{v_{r_{0}}(\phi)}  (r - r_{0}),
\end{equation*}
where $r_{0} = 0.14 \text{AU}$. The HUX characteristic curves are also called the \textit{ballistic} approximation, which assumes that each spiral field line of plasma continues at a constant speed throughout the heliosphere~\cite{riley_HUXP2_2021}. After obtaining the characteristic curves, we are able to approximate the shift function by 
\begin{equation*}
    c(r) = \begin{cases}
        \frac{1}{q-p}\sum_{j=p}^{q}\frac{-\Omega_{\text{rot}}(0)}{\mathbf{v}(\phi_{j}, r_{0})} (r-r_{0}) & \text{if } r_{0} < r < r_{s} \\
        s(r) - s(r_{s}) + a & \text{if } r > r_{s} 
        \end{cases}
\end{equation*}
so that $r_{s}$ is the radial position where the characteristics first intersect, $s(r)$ is the shock trajectory, and $ a = \frac{1}{q-p}\sum_{j=p}^{q}\frac{-\Omega_{\text{rot}}(0)}{\mathbf{v}(\phi_{j}, r_{0})} (r_{s}-r_{0})$. The indices $p, q$ can include the whole spatial domain or instead a bounded spatial interval to track specific regions of the initial wave. For CR2210, we limit $p, q$ to include the characteristics emanating from a main equatorial high solar wind peak, originating from an equatorial coronal hole, on the longitudinal interval $[180^{\circ}, 260^{\circ}]$. There are three main shock curves, and we choose to follow the first shock curve that emerged at $r_{s} = 0.344 \text{AU}$ and $\phi_{s} = 211.737^\circ$. 
The learned sOpInf ROM is of the form 
\begin{equation*}
    \dot{\hat{\bv}} = \hat{\bH} (\hat{\bv} \otimes^{\prime} \hat{\bv}), \qquad \hat{\bH} \in \real^{\ell \times \frac{1}{2} \ell (\ell +1)}
\end{equation*}
and it is able to sufficiently model the dynamics of the HUX model with only $\ell= 4$ modes. 
A practical implementation of operator inference requires regularization, and for the least-squares fitting we found $\lambda = 10^{3}$ to give good results. 
The comparison between sOpInf and HUX velocity profiles is provided in Figure~\ref{fig:sOpInf-HUX-1D}. The figure shows that the advective solutions are well approximated both in the training regime until $r=0.82$AU and in the testing regime, where the ROM is fully predictive. This conclusion is also supported by Table~\ref{tab:CR2210} where we provide the mean/median/maximum relative error and the Pearson correlation coefficient comparing the HUX solutions with the ROM solutions, both in the testing and training regime. The error measures show that a sOpInf ROM can sufficiently predict the HUX dynamics while reducing the dimensionality of the problem from $n_{\phi}=128$ to $\ell = 4$, i.e., a factor of 32 reduction of state-space dimension. 

%%%%%%%%%%%%%%%%%%%%%%%%%%%%%%%%
\noindent 
\begin{table}
\caption{Comparison of the sOpInf ROM performance for the test case CR2210. Given are the mean/median/maximum relative error (RE) measured in percent and the Pearson correlation coefficient (PCC) comparing the ROM with the respective high-fidelity models (HUX, MAS-2D, MAS-3D) for both training and testing datasets. }
\label{tab:CR2210}
\begin{tabular}{|p{4.5cm}||p{1.5cm}||p{1.5cm}|p{1.8cm}|p{1.5cm}|p{1.5cm}|}
\hline
Model & Regime & RE mean & RE median & RE max. & PCC  \\
\hline
HUX Equatorial Plane (2D) & \makecell{Training \\ Testing} & \makecell{0.286 \\  0.526}  & \makecell{0.143 \\  0.385} & \makecell{3.379 \\  4.114} & \makecell{0.99991 \\  0.99956}\\
\hline
MAS Equatorial Plane (2D)  & \makecell{Training \\ Testing} & \makecell{0.449 \\ 1.264} & \makecell{0.229 \\  0.849} & \makecell{5.564 \\  8.235} &  \makecell{0.99980 \\ 0.99901} \\
\hline
MAS Full Sun (3D) & \makecell{Training \\ Testing} & \makecell{0.451 \\  0.539}  & \makecell{0.334 \\  0.353} & \makecell{7.731 \\ 20.492} & \makecell{0.99969 \\ 0.99964} \\
\hline
\end{tabular} 
\end{table}
%%%%%%%%%%%%%%%%%%%%%%%%%%%%%%%%

%%%%%%%%%%%%%%%%%%%%%%%%%%%%%%%%
%%%%%%%%%%%%%%%%%%%%%%%%%%%%%%%%

%%%%%%%%%%%%%%%%%%%%%%%%%%%%%%%%%%%%%%%
\subsection{MAS Equatorial Plane Numerical Results} \label{sec:mas-equator-results}
\begin{figure}
    \centering
    \begin{subfigure}[b]{0.3\textwidth}
        \centering
        \caption{Original Coordinates}
        \includegraphics[width=\textwidth]{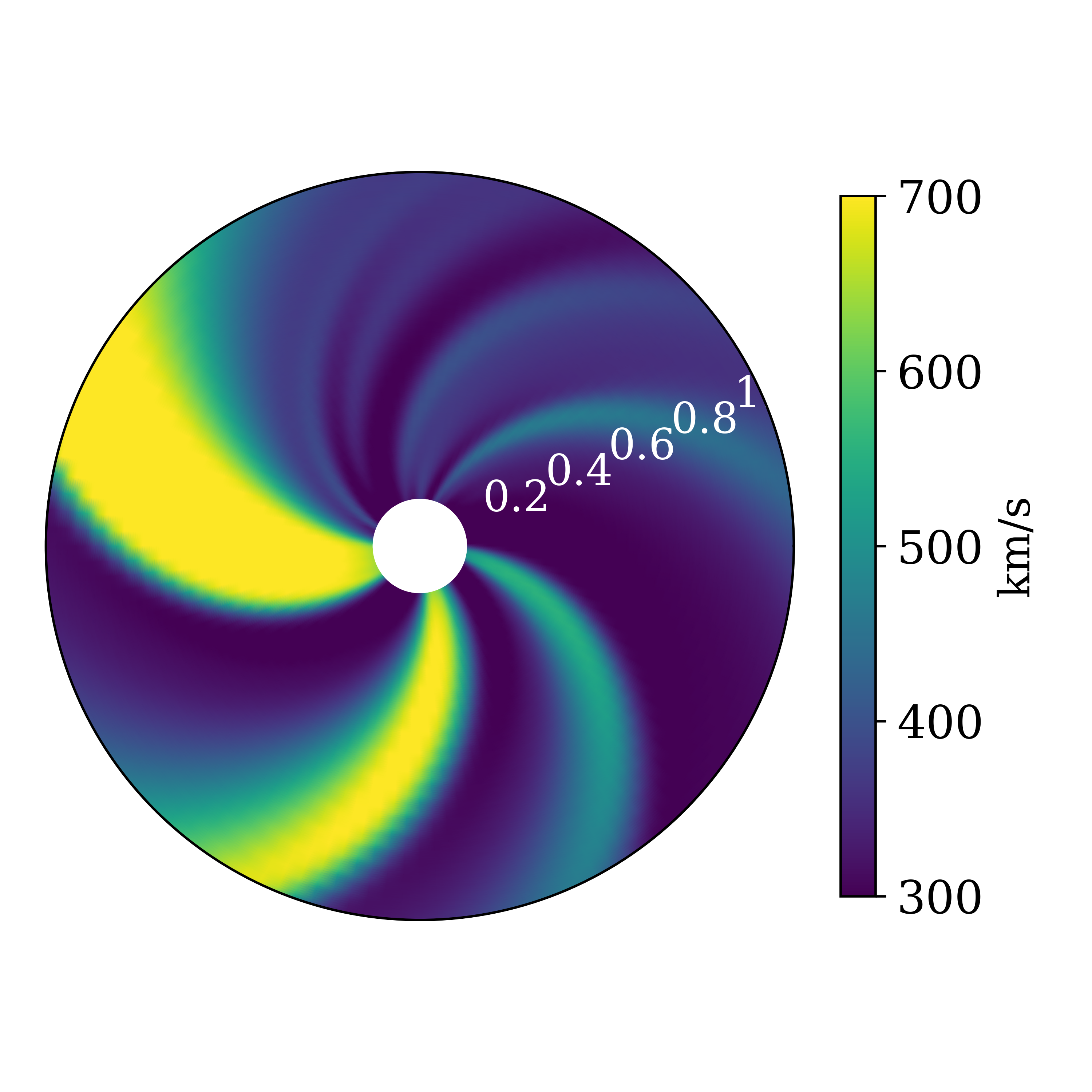}
        \label{fig:polar-MAS-original}
    \end{subfigure}
    \begin{subfigure}[b]{0.3\textwidth}
        \centering
        \caption{Shifted Coordinates}
        \includegraphics[width=\textwidth]{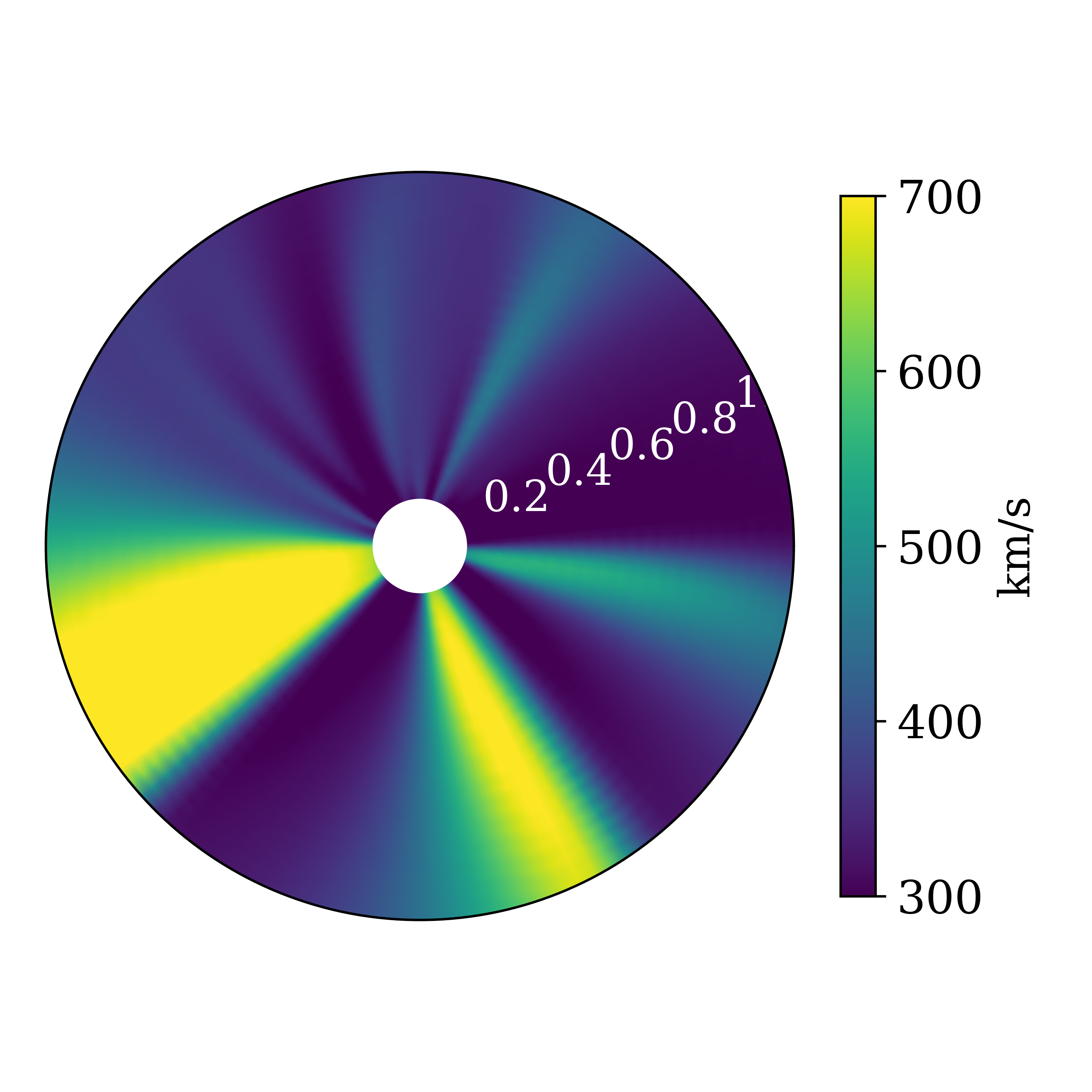}
        \label{fig:polar-MAS-shifted}
    \end{subfigure}
    \begin{subfigure}[b]{0.3\textwidth}
        \centering
        \caption{Cumulative Energy obtained by Eq.~\eqref{energy-criteria}}
        \includegraphics[width=\textwidth]{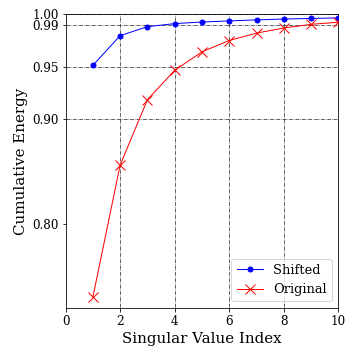}
        \label{fig:mas-equatorial-singular-value-decay}
    \end{subfigure}
    \caption{\footnotesize (a)~The MAS solar wind radial velocity results at the equatorial plane for CR2210 from $0.14 \text{AU}$ to $1.1\text{AU}$. (b)~The solar wind data in shifted coordinates eliminating the translational properties caused by the Sun's rotation. (c)~The Singular value cumulative energy of the data in the original and shifted coordinates. The singular values decay more rapidly in the shifted coordinates, indicating that the ROM will require less modes in the shifted coordinates. }
    \label{fig:Shifted-MHD-SVD-Decay}
\end{figure}

%%%%%%%%%%%%%%%%%%%%%%%%%%%%%%%%%%%%%%%
We apply the sOpInf framework to learn a ROM for the MAS CR2210 equatorial plane velocity field. Since the MAS Eqs.~\eqref{MAS-EQ1}--\eqref{MAS-EQ6-Energy} are not in the form of Eq.~\eqref{general-form-pde-moc}, the method of characteristics can not be applied to approximate the shift function $c(r)$; instead, we use the cross-correlation extrapolation method, which resulted in $c(r) = -54.98^{\circ} r +7.39^{\circ}$, where $r$ is measured in AU.
Figure~\ref{fig:Shifted-MHD-SVD-Decay} shows the MAS equatorial plane heliospheric results on the original and shifted polar coordinates along with the cumulative singular value energy criteria described in Eq.~\eqref{energy-criteria} in each coordinate system.  Evidently, we see that shifting the snapshots to a moving coordinate frame creates a faster singular value decay, leading to an accurate representation of the shifted data with far fewer modes. This observation is in line with the results we showed for Burgers' equation in Figure~\ref{fig:singular-value-decay-burgers} above.

\begin{figure}
\centering
    \includegraphics[width=1\linewidth]{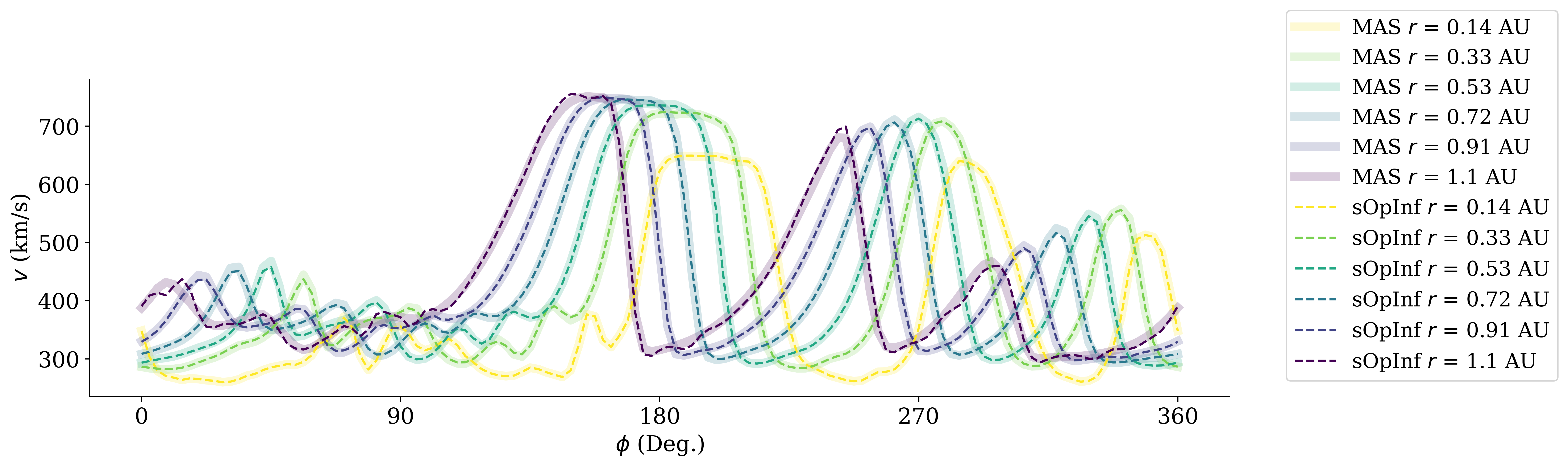}
    \caption{The MAS solar wind radial velocity results at the heliographic equatorial plane for CR2210 along with sOpInf quadratic ROM results. The sOpInf ROM aligns very well with the data past the training interval; the snapshots at $r=0.91$AU and $r=1.1$AU are testing data where the ROM is fully predictive.}
    \label{fig:sOpInf-MAS-1D} 
\end{figure}

\begin{figure}
    \centering
    \begin{subfigure}[b]{0.32\textwidth}
        \centering
        \caption{MAS}
        \includegraphics[width=\textwidth]{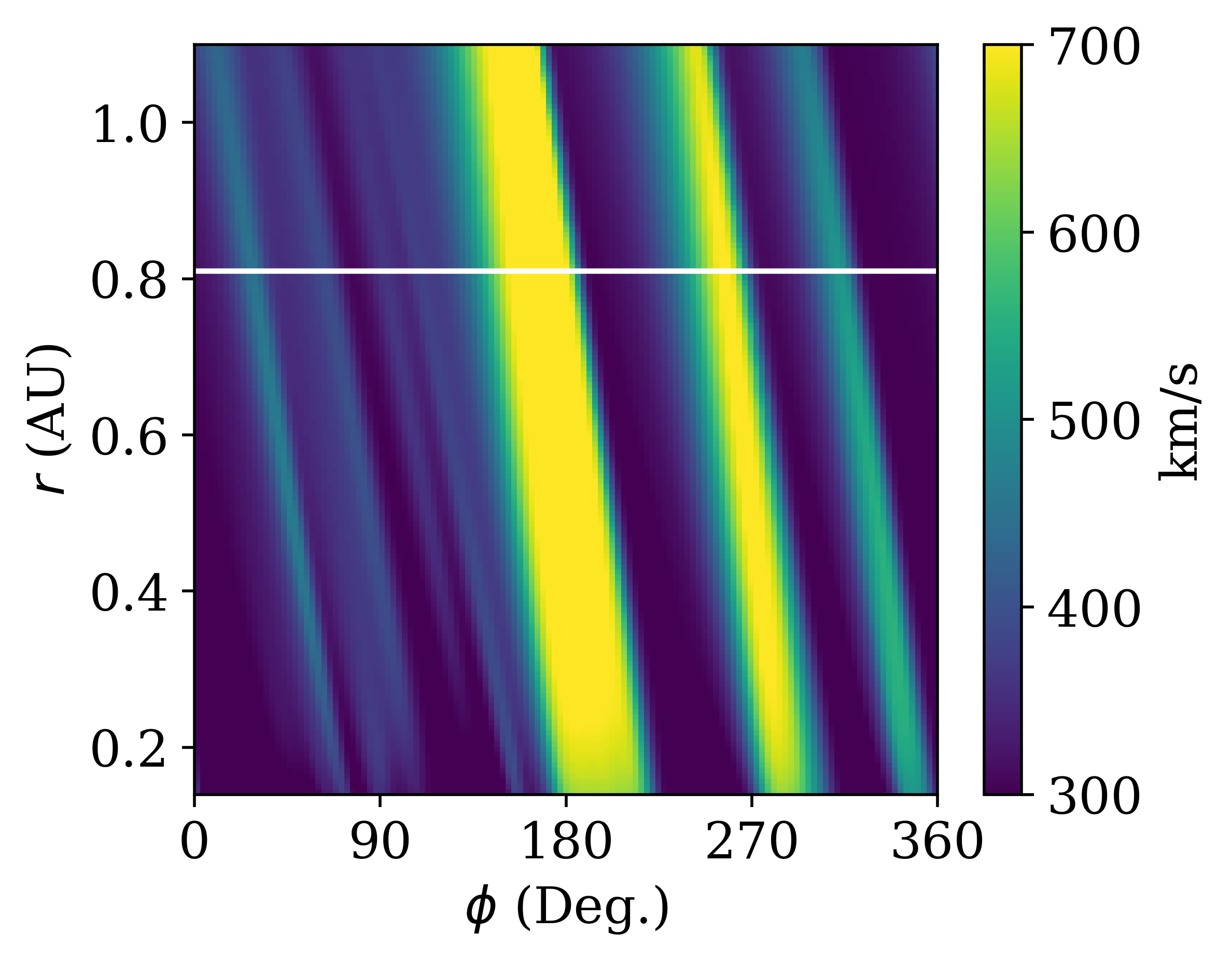}
        \label{fig:equator-MAS}
    \end{subfigure}
    \begin{subfigure}[b]{0.32\textwidth}
        \centering
        \caption{sOpInf}
        \includegraphics[width=\textwidth]{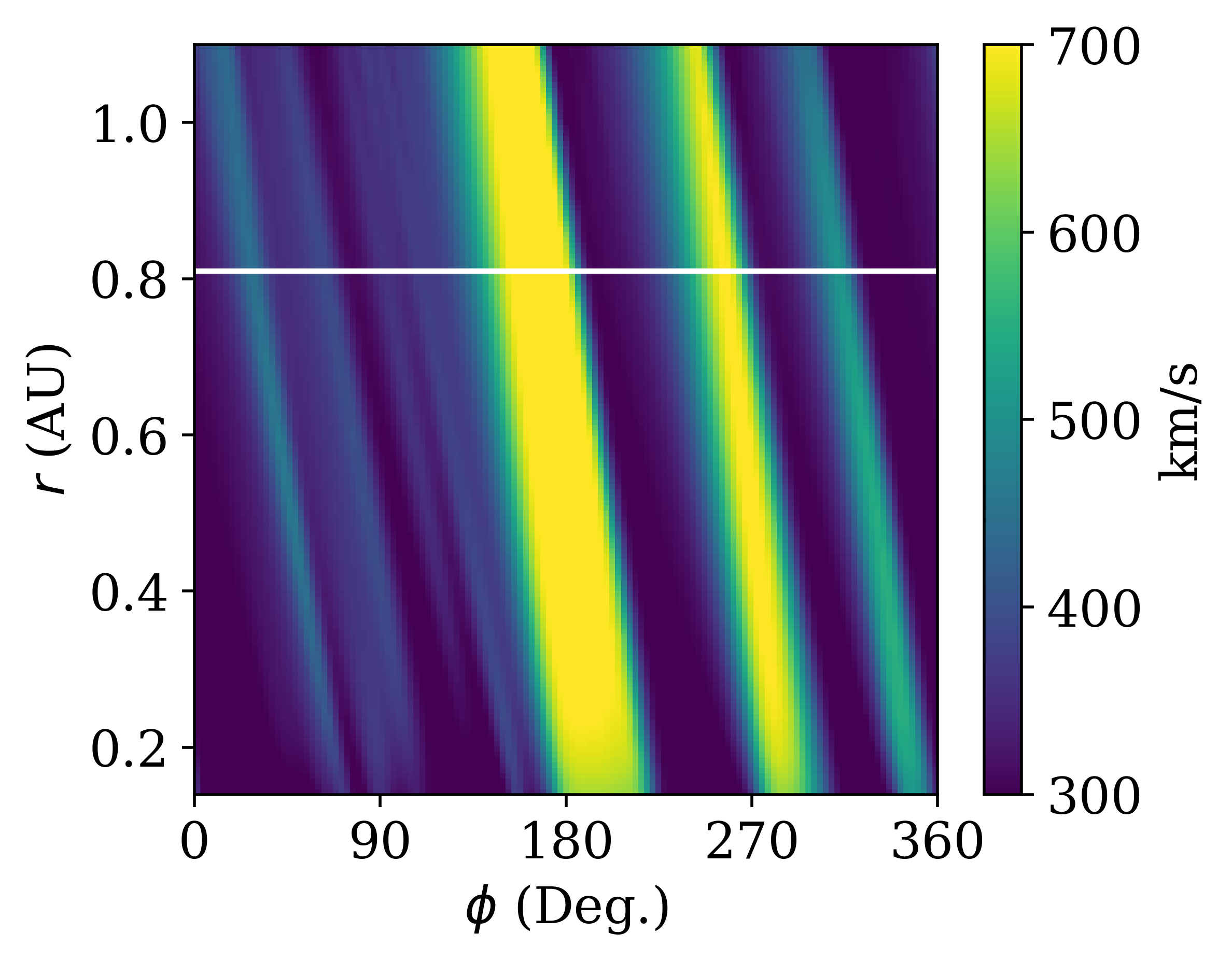}
        \label{fig:equator-sOpInf}
    \end{subfigure}
    \begin{subfigure}[b]{0.32\textwidth}
        \centering
        \caption{Relative Error}
        \includegraphics[width=\textwidth]{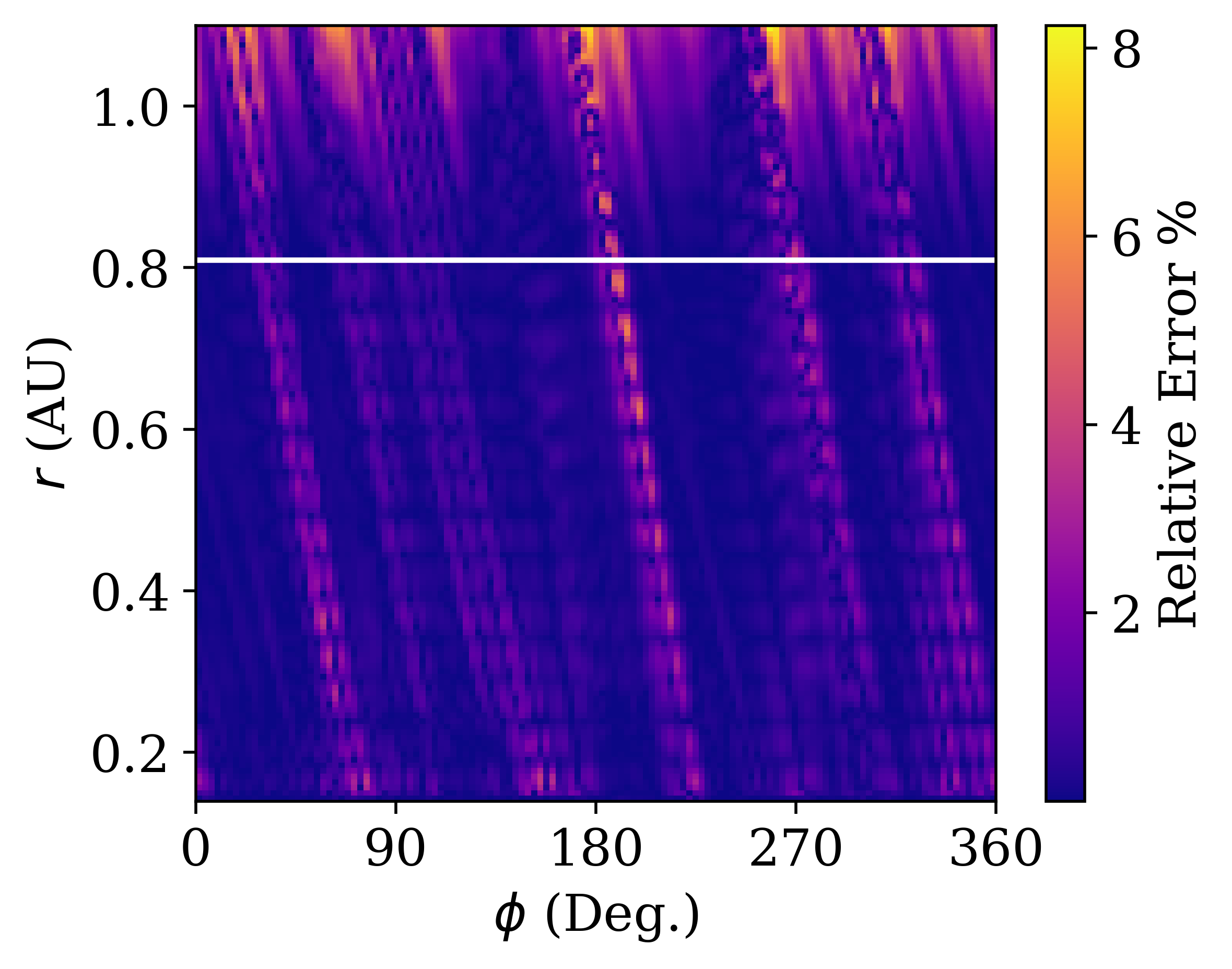}
        \label{fig:equator-MAS-relative-error}
    \end{subfigure}

    \caption{A comparison between (a)~the MAS solar wind radial velocity solution at the heliographic equatorial plane for CR2210 and (b)~the learned quadratic sOpInf ROM results with $\ell = 9$ basis modes. The relative error between MAS and sOpInf results is illustrated in sub-figure (c). }
    \label{fig:MHD-1D-Relative-Error}
\end{figure}

The results shown in this section are for the ROM model form
\begin{equation*}
    \dot{\hat{\bv}} = \hat{\bH} (\hat{\bv} \otimes^{\prime} \hat{\bv}), \qquad \hat{\bH} \in \real^{\ell \times \frac{1}{2} \ell (\ell +1)}
\end{equation*}
with $\ell = 9$ modes. This model form provided the best overall results in training and extrapolation, compared to other model combinations including linear and constant terms, see Section~\ref{sec:model-form} for further discussion.
The regularization coefficient for computing the operator $\hat{\bH}$ is $\lambda = 10^{5}$.
Figure~\ref{fig:sOpInf-MAS-1D} visually demonstrates that sOpInf is capable of accurately approximating the MAS equatorial results, where the snapshots at $r=0.91$AU and $r=1.1$AU are in the fully predictive regime of the ROM. We highlight that the MAS data contains more complex features than the HUX data, specifically in the region $0^{\circ}\leq \phi\lesssim 120^{\circ}$ where small localized wave structures exist. Nevertheless, sOpInf covers those equally well as the HUX data in the previous section. 
Figure~\ref{fig:MHD-1D-Relative-Error} presents are more qualitative assessment of the relative error between the sOpInf ROM and MAS velocity fields, which shows that the relative error is less than 8.3\% in the entire domain. The mean/median/maximum relative error and PCC in the testing and training regime are again provided in Table~\ref{tab:CR2210} above. 

%%%%%%%%%%%%%%%%%%%%%%%%%%%%%%%%
%%%%%%%%%%%%%%%%%%%%%%%%%%%%%%%%

%%%%%%%%%%%%%%%%%%%%%%%%%%%%%%%%
%%%%%%%%%%%%%%%%%%%%%%%%%%%%%%%%

%%%%%%%%%%%%%%%%%%%%%%%%%%%%%%%%
%%%%%%%%%%%%%%%%%%%%%%%%%%%%%%%%

%%%%%%%%%%%%%%%%%%%%%%%%%%%%%%%%%%%%%%%
\subsection{3-D Full-Sun MAS Numerical Results} \label{sec:mhd-full-sun-results}
\begin{figure}
    \centering
  \begin{tabular}[c]{cc}
    \begin{subfigure}[c]{0.495\textwidth}
        \centering
        \caption{sOpInf vs. MAS}
        \includegraphics[width=\textwidth]{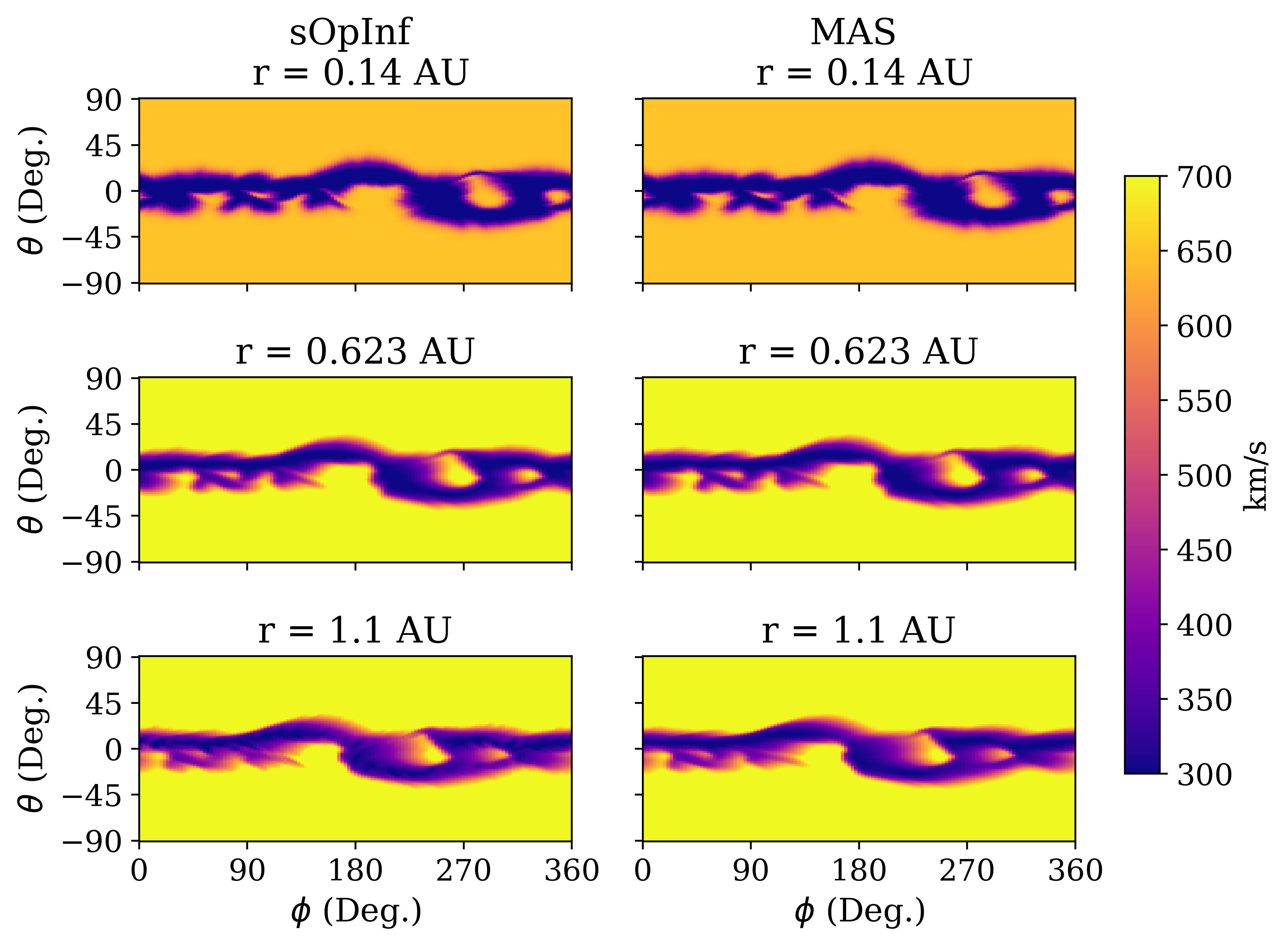}
        \label{fig:MHD-2D-Relative-Error}
    \end{subfigure}
    \begin{subfigure}[c]{0.495\textwidth}
        \centering
        \caption{Relative Error}
        \includegraphics[width=\textwidth]{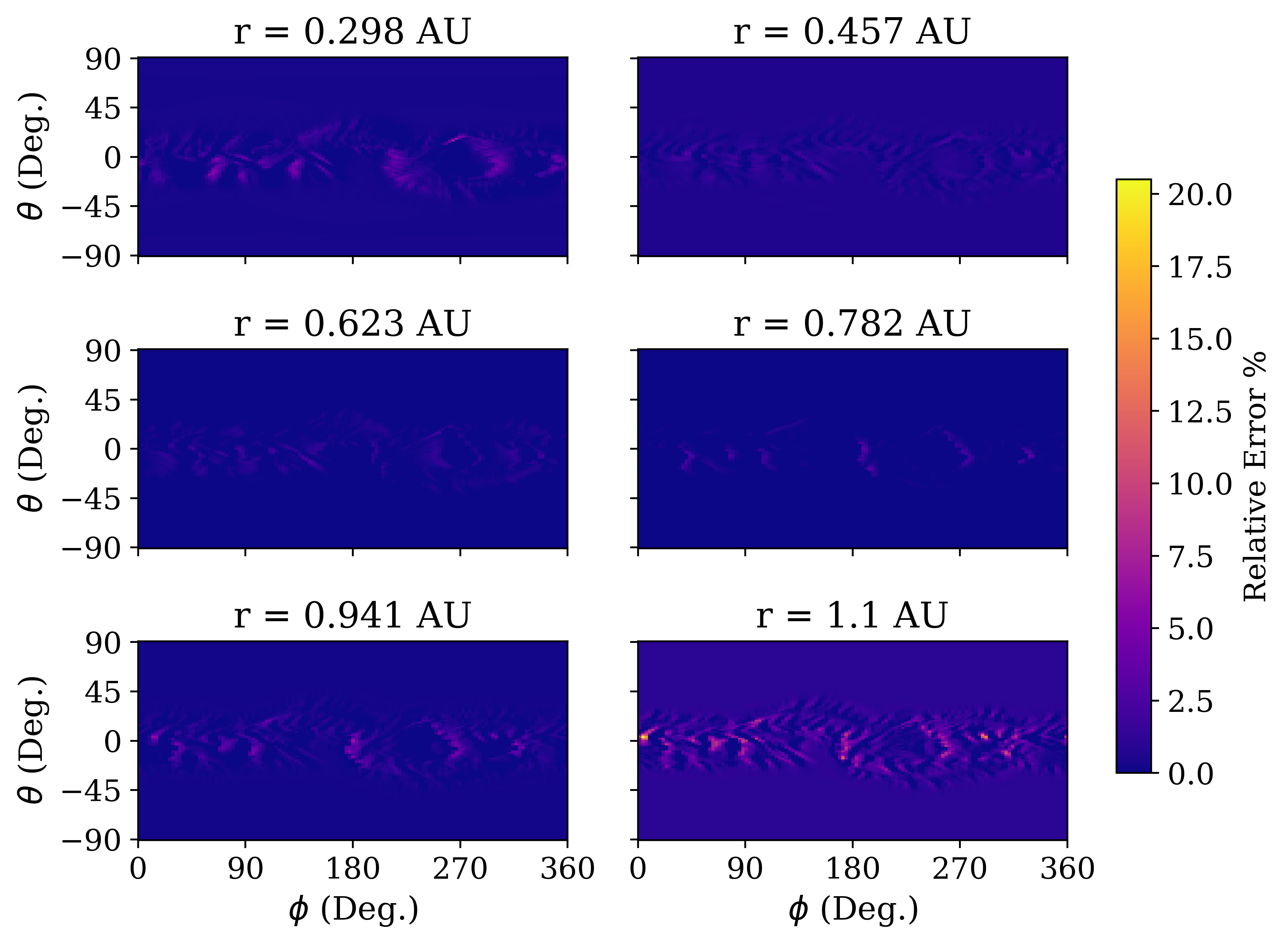}
        \label{fig:MHD-2D-Snapshot-comparison}
    \end{subfigure}
    \end{tabular}
    \caption{Graphic (a) shows the MAS solar wind velocity results (right column) and sOpInf model of the form $\dot{\hat{\bv}} = \hat{\bA} \hat{\bv} + \hat{\bH} (\hat{\bv} \otimes^{\prime} \hat{\bv}) + \hat{\bB}$ results (left column) with $\ell=8$ modes. The training ends at $0.82 \text{AU}$, hence, the velocity profile at $r=1.1 \text{AU}$ (last row) is in the purely predictive regime. Graphic (b) presents the relative error of learned full-Sun sOpInf ROM vs. MAS. The sOpInf ROM results at $r=0.968 \text{AU}$ and $r=1.1 \text{AU}$ (last row) are in the purely predictive regime. The ROM shows good qualitative and quantitative agreement with the MAS solution, yet can be evaluated at a much lower cost.}
\end{figure}

%%%%%%%%%%%%%%%%%%%%%%%%%%%%%%%%%%%%%%%
We showcase sOpInf trained on MAS three-dimensional steady-state velocity results. As mentioned previously, the MAS Eqs.~\eqref{MAS-EQ1}--\eqref{MAS-EQ6-Energy} are not in the form of Eq.~\eqref{general-form-pde-moc}, hence, the method of characteristics is not a valid choice in approximating the shift function $\bc(r)\in \real^{2}$. Therefore, we turn to the cross-correlation extrapolation method described in Section~\ref{sec:cross-correlation}. More specifically, the full-Sun MAS snapshots in matrix form, $\bv(r) \in \real^{n_{\phi} \times n_{\theta}}$, are bi-variate in Carrington longitude ($\phi$) and latitude ($\theta$), with $n_{\phi}, n_{\theta}$ mesh points in longitude and latitude, respectively. The shift function $\bc(r) \in \real^{2}$ is found via the bi-variate circular cross-correlation defined by Eqs.~\eqref{cc-dis-definition-multi-variate}--\eqref{cc-max} with $k=2$. The bi-variate circular cross-correlation is applied between each snapshot and the velocity profile at the initial condition ($30R_{S}$). Since the translation in the MAS solar wind results is due to the Sun's rotation, the shift function can be expressed as $\bc(r) = c(r) \hat{\be}_{\phi}$, where $\hat{\be}_{\phi}$ is the unit vector in longitude direction, and $c(r) \in \real$ is the shift in longitude, where we computed $c(r) = -52.44^\circ r +6.94^\circ$ with $r$ measured in AU.
The results in this section are for a ROM of the form 
\begin{equation*}
    \dot{\hat{\bv}} = \hat{\bA} \hat{\bv} + \hat{\bH} (\hat{\bv} \otimes^{\prime} \hat{\bv}) + \hat{\bB}
\end{equation*}
where $\hat{\bB} \in \real^{\ell}$, $\hat{\bA} \in \real^{\ell \times \ell}$, $\hat{\bH} \in \real^{\ell \times \frac{1}{2} \ell (\ell +1)}$ with only $\ell = 8$ modes. 
The regularization coefficient for computing the operator $\hat{\bB}$ and $\hat{\bA}$ is $\lambda_{1} = 10^{4}$ and the regularization coefficient for $\hat{\bH}$ is $\lambda_{2} = 10^{8}$. Figure~\ref{fig:MHD-2D-Relative-Error} shows the relative error between the two velocity fields and the mean/median/max relative error in the testing and training regime is shown in Table~\ref{tab:CR2210}. For a visual comparison, Figure~\ref{fig:MHD-2D-Snapshot-comparison} shows a comparison between the sOpInf reconstructed and predicted two-dimensional full-Sun snapshots; the visual comparison and error estimates indicate that sOpInf can successfully reproduce the high fidelity full-Sun MAS dataset, where $n_{x} = n_{\phi} \times n_{\theta} = 14,208$ with only $\ell = 8$ modes, leading to a substantial reduction in the model's dimensionality. 

In practice, the sOpInf framework can be employed to speed up the MAS computational time by setting the MAS heliospheric outer boundary condition to $0.82\text{AU}$ (which is 70\% of the current computational domain) and run the time-dependent MAS simulation until it relaxes to steady-state. Then, the steady-state MAS snapshots are used as training data for sOpInf, which takes 0.355 seconds to simulate from $30 R_{s}$ up to $1.1\text{AU}$ on a MacBook Pro 2.3 GHz Quad-Core Intel Core i7 processor with 16 GB RAM (for 3-D full-Sun simulation). If we assume that running MAS on 70\% of the computational domain would take 70\% of the MAS current computational time, then we would be able to speed up the MAS computational time by approximately 1.2 hours and 8.4 hours for the medium and high resolution runs, respectively (i.e. save 30\% of the MAS computational time).  Another important sOpInf speed-up contribution can be in the case when one is interested in studying the solar wind dynamics much further in the heliosphere (e.g. conditions in the vicinity of Mars or Jupiter). In this case, one can use the MAS simulation up to 1.1\text{AU} for training sOpInf and use the reduced model to predict up to 5\text{AU}, resulting in a more significant speed up. It is important to mention that the MAS model solves for several plasma flow quantities, i.e. Eqs.~\eqref{MAS-EQ1}--\eqref{MAS-EQ6-Energy}, and sOpInf is currently only solving for the radial velocity component. Thus, a direct comparison of their run-time can be equivocal.

%%%%%%%%%%%%%%%%%%%%%%%%%%%%%%%%
%%%%%%%%%%%%%%%%%%%%%%%%%%%%%%%%

%%%%%%%%%%%%%%%%%%%%%%%%%%%%%%%%%%%%%%%
\subsection{A Comparison of Surrogate Model Accuracy via Equatorial Plane Streamlines} \label{sec:comparison-of-streamlines}
%%%%%%%%%%%%%%%%%%%%%%%%%%%%%%%%%%%%%%%
%%%%%%%%%%%%%%%%%%%%%%%%%%%%%%%%
\begin{figure}[t]
  \centering
  \begin{tabular}[c]{ccc}
    \begin{subfigure}[c]{0.31\textwidth}
        \centering
        \caption{MAS}
        \includegraphics[width=\textwidth]{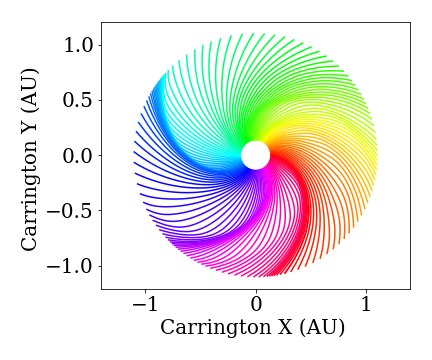}
        \label{fig:MAS-streamlines}
    \end{subfigure}
    \begin{subfigure}[c]{0.31\textwidth}
        \centering
        \caption{HUX}
        \includegraphics[width=\textwidth]{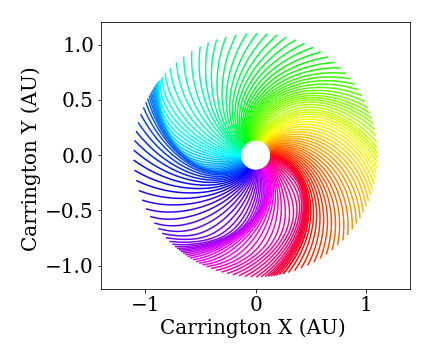}
        \label{fig:HUX-streamlines}
    \end{subfigure}
    \begin{subfigure}[c]{0.31\textwidth}
        \centering
        \caption{sOpInf}
        \includegraphics[width=\textwidth]{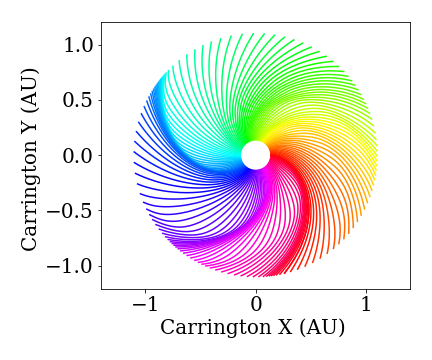}
        \label{fig:sOpInf-streamlines}
    \end{subfigure}
    \\
    \begin{subfigure}[c]{0.31\textwidth}
        \centering
        \caption{CDF}
        \includegraphics[width=\textwidth]{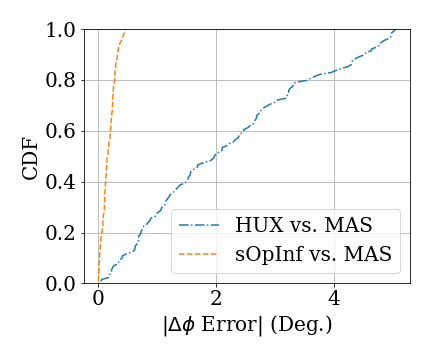}
        \label{fig:CDF-streamlines}
    \end{subfigure}
    \begin{subfigure}[c]{0.31\textwidth}
        \centering
        \caption{HUX vs. MAS}
        \includegraphics[width=\textwidth]{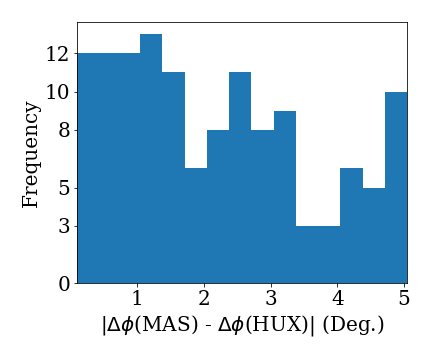}
        \label{fig:HUX-vs-MAS-streamlines}
    \end{subfigure}
    \begin{subfigure}[c]{0.31\textwidth}
        \centering
        \caption{sOpInf vs. MAS}
        \includegraphics[width=\textwidth]{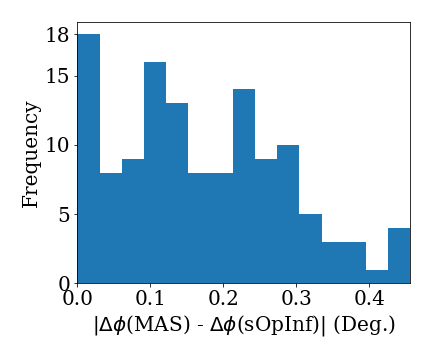}
        \label{fig:sOpInf-vs-MAS-streamlines}
    \end{subfigure}
    \end{tabular}
    \caption{The solar wind streamlines (or Parker spiral) for the CR2210 equatorial plane is shown using the velocity results of (a)~MAS, (b)~HUX, and (c)~sOpInf trained on MAS. Sub-figure (d)~shows the cumulative distribution function (CDF) of the streamlines longitude absolute difference at $1.1$AU, where sOpInf outperformed HUX in approximating the MAS solar wind streamlines. The longitude absolute difference between the streamlines are shown for (e)~HUX vs. MAS and (f)~sOpInf vs. MAS. 
    }
    \label{fig:MHD-HUX-ROM-Streamlines}
\end{figure}
%%%%%%%%%%%%%%%%%%%%%%%%%%%%%%%%
Given the two surrogate models of MAS, namely HUX (a reduced-physics approximation) and sOpInf (a data-driven ROM trained on MAS), we are interested in comparing their accuracy as surrogates of MAS. We do so by drawing the equatorial streamlines for each model. The streamlines of a flow field are curves that are tangential to the local velocity vector. 
In Figure~\ref{fig:MHD-HUX-ROM-Streamlines}(\subref{fig:MAS-streamlines}-\subref{fig:sOpInf-streamlines}), the streamlines are mapped from the inner-heliosphere at $0.14 \text{AU}$ to $1.1 \text{AU}$. The shape of the streamlines depends on the velocity field, such that the fast solar wind results in less tightly wound lines than the slow solar wind. At regions where the streamlines interact, a \textit{compression wave} is formed, whereas, at regions where the streamlines are distant there is a \textit{rarefaction wave}. At these regions of compression and rarefaction, the solar wind streams go through substantial changes in density and flow speed~\cite{kivelson_russell_1995}. For a better understanding of the mapped streamline accuracy, Figure~\ref{fig:MHD-HUX-ROM-Streamlines}(\subref{fig:HUX-vs-MAS-streamlines}-\subref{fig:sOpInf-vs-MAS-streamlines}) presents a histogram of the mapped streamlines longitude difference at $1.1 \text{AU}$ along with plotting the cumulative distribution function (CDF) in Figure~\ref{fig:CDF-streamlines} of the streamlines longitude difference at $1.1 \text{AU}$ for each surrogate model: HUX and sOpInf. The MAS in comparison to HUX and sOpInf trained on MAS mean/median/maximum and standard deviation (SD) of the streamline longitude error are presented in Table~\ref{tab:mapped-streamline-error}. The streamline mean longitude absolute error at $1.1\text{AU}$ of the HUX model is a factor of $12$ larger than the ones associated with sOpInf. The numerical results show that the sOpInf model is a more accurate approximation of the MAS model in comparison to the reduced-physics HUX model. Moreover, the sOpInf framework can account for the solar wind dynamics in three-dimensional space, whereas HUX is strictly two-dimensional.
%%%%%%%%%%%%%%%%%%%%%%%%%%%%%%%%
\noindent 
\begin{table} 
\caption{Streamline longitude absolute error (AE), measured in degrees,  at $1.1\text{AU}$ of the surrogate models HUX and sOpInf in comparison to MAS. That snapshot is in the fully predictive regime of the ROM, showing that the sOpInf ROM can predict well outside the training interval and provides a better surrogate model than the reduced-physics HUX model.}
\label{tab:mapped-streamline-error}
\begin{tabular}{|p{4cm}||p{2cm}|p{3cm}|p{2cm}|p{2cm}|}
 \hline
Model Comparison & AE mean  & AE median  & AE max.  & AE SD \\
\hline
HUX vs. MAS & 
2.186$^{\circ}$& 
1.977$^{\circ}$& 
5.047$^{\circ}$&  
1.463$^{\circ}$\\
\hline
\\[-2em]
{\color{black} {sOpInf vs. MAS}} & 
{{0.172$^{\circ}$}}&  
{{0.155$^{\circ}$}}&  
{{0.457$^{\circ}$}}&  
{{0.115$^{\circ}$}}\\
\hline
\end{tabular} 
\end{table}
%%%%%%%%%%%%%%%%%%%%%%%%%%%%%%%%

%%%%%%%%%%%%%%%%%%%%%%%%%%%%%%%%%%%%%%%
\subsection{Additional Considerations when Choosing the Operator Model Form} \label{sec:model-form}
%%%%%%%%%%%%%%%%%%%%%%%%%%%%%%%%%%%%%%%
%%%%%%%%%%%%%%%%%%%%%%%%%%%%%%%%
\subsubsection{Comparing Models With the Same Number of Modes}
\begin{figure}
    \centering
    \begin{subfigure}[b]{0.26\textwidth}
        \centering
        \caption{HUX-2D ROM Form Comparison}
        \includegraphics[width=\textwidth]{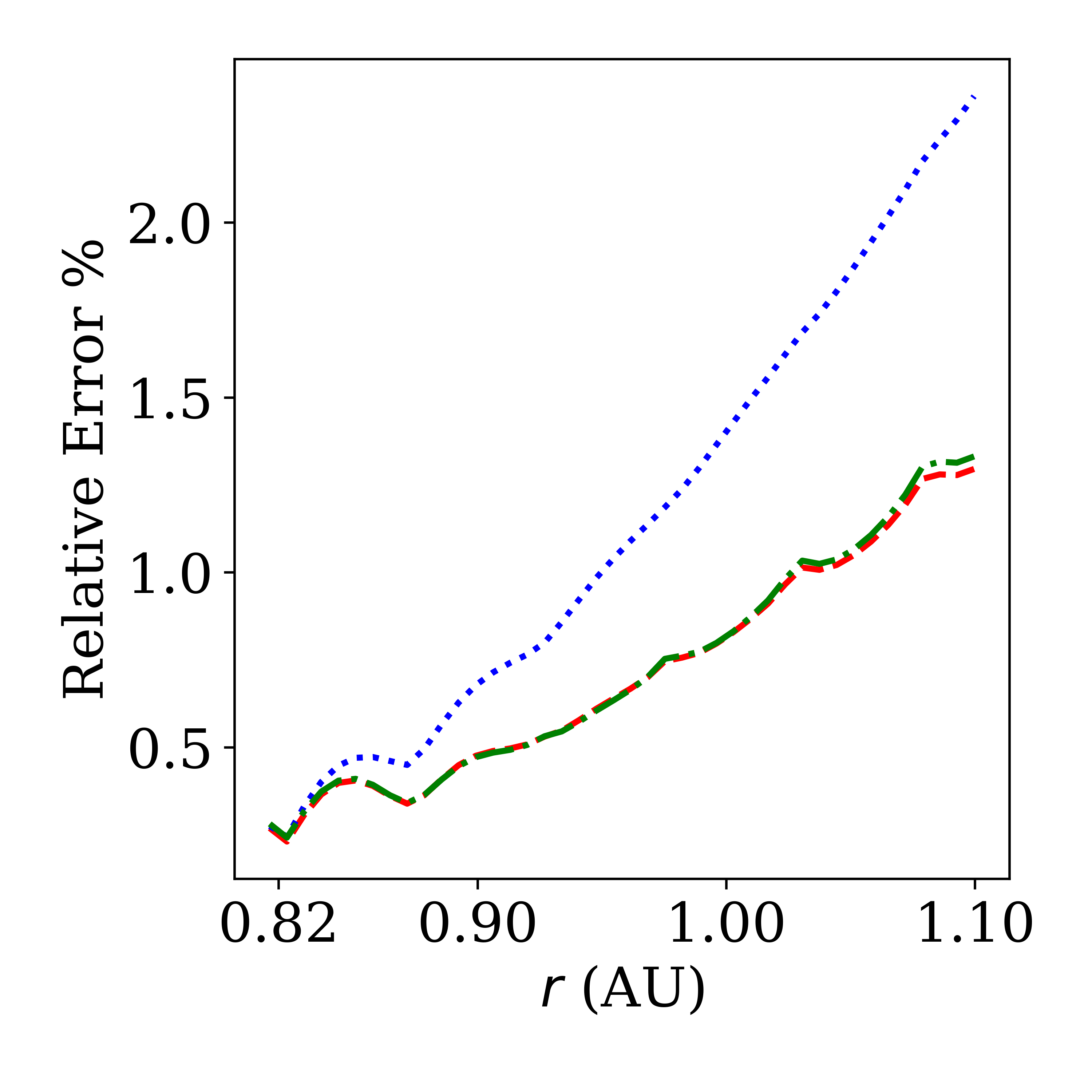}
        \label{fig:hux-model-comparison}
    \end{subfigure}
      \hfill
    \begin{subfigure}[b]{0.26\textwidth}
       % \centering
        \caption{MAS-2D ROM Form Comparison}
        \includegraphics[width=\textwidth]{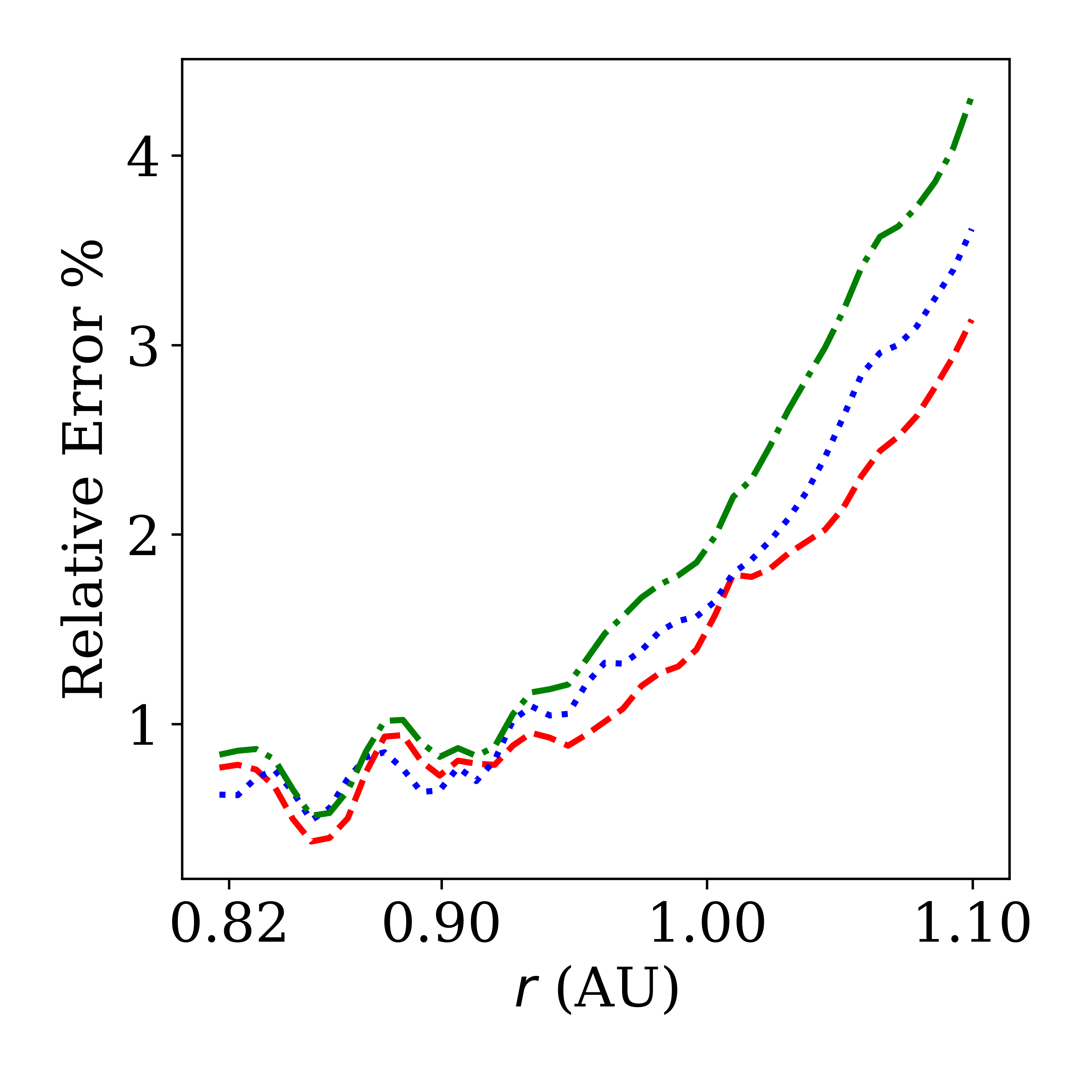}
        \label{fig:mas-2d-model-comparison}
    \end{subfigure}
    \hfill
    \begin{subfigure}[b]{0.45\textwidth}
        %\centering
        \caption{MAS-3D Full-Sun ROM Form Comparison}
        \includegraphics[width=\textwidth]{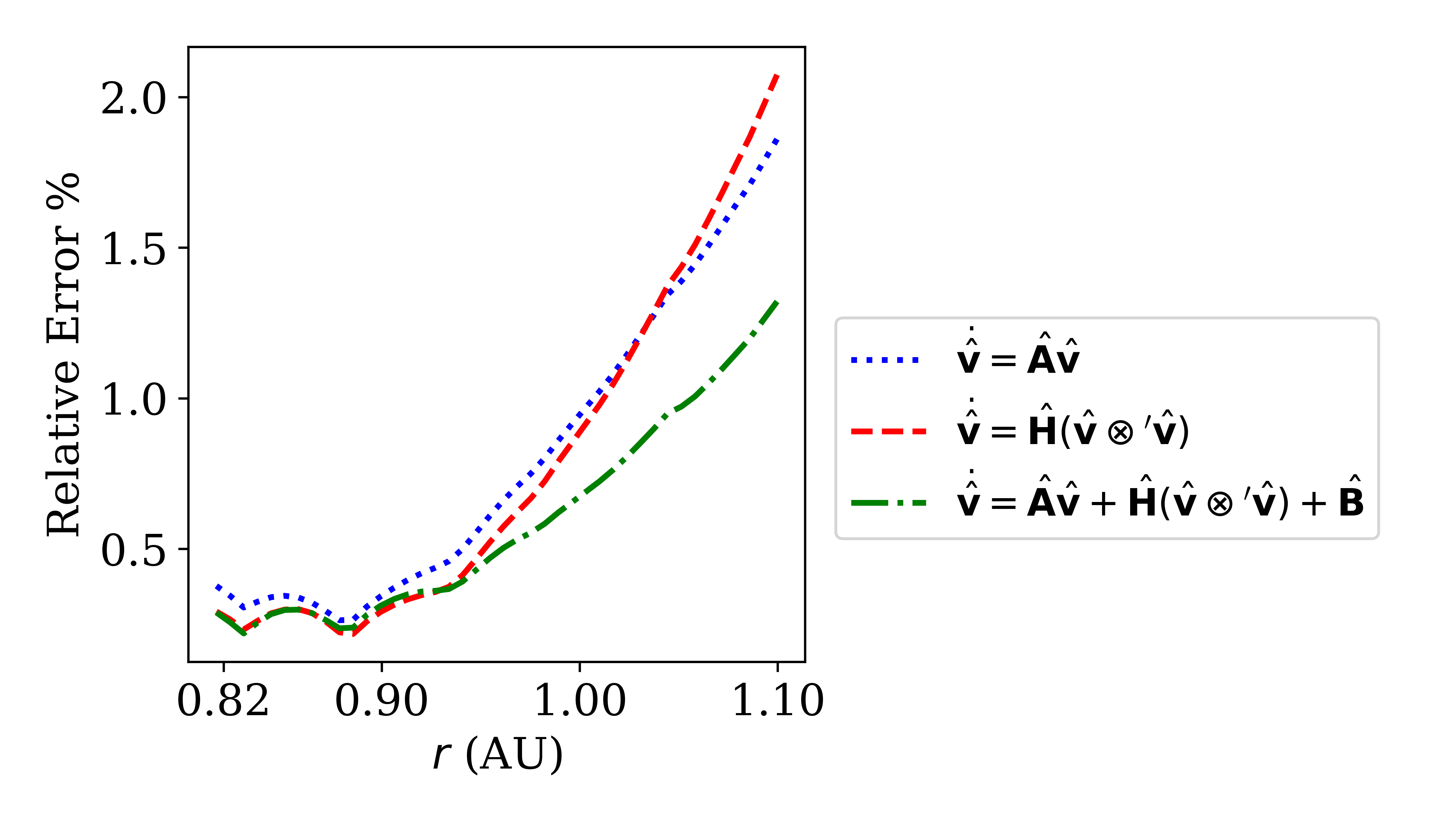}
        \label{fig:mas-3d-model-comparison}
    \end{subfigure}
    \caption{\footnotesize A comparison of three ROM model forms, i.e. purely linear, purely quadratic, and quadratic with linear and constant term, relative error measured via the $L_{2}$-norm in the testing regime. The three models are trained on 70\% of the (a) HUX-2D equatorial data, (b) MAS-2D equatorial data, and (c) MAS-3D full-Sun data.}
    \label{fig:model-form-comparison}
\end{figure}

%%%%%%%%%%%%%%%%%%%%%%%%%%%%%%%%
The choice of polynomial ROM form for the HUX and MAS dataset is an approximation of the governing equations (unlike the inviscid Burgers' example in Section~\ref{sec:Burgers}) since both models have nonpolynomial terms. 
The above Sections~\ref{sec:hux-numerical-results}--\ref{sec:mhd-full-sun-results} showcase the ROM model form that performed the best in the testing regime, i.e. purely-linear, purely-quadratic, or a combination thereof. 
Here, we present the results of a detailed investigation of how other polynomial ROM forms performed on each dataset. Figure~\ref{fig:model-form-comparison} compares the state error in the fully predictive (testing) regime for each model: (\ref{fig:hux-model-comparison}) shows the HUX-2D equatorial plane results, (\ref{fig:mas-2d-model-comparison}) shows the MAS-2D equatorial plane results, and (\ref{fig:mas-3d-model-comparison}) shows the MAS-3D full-Sun results. 
In each case, we consider strictly-linear, strictly-quadratic, and linear-quadratic plus a constant term model forms in our analysis. 
Figure~\ref{fig:hux-model-comparison} shows a comparison of three different model forms for HUX dataset, in which strictly-quadratic and linear-quadratic plus constant term ROMs perform better than the strictly-linear ROM. There is not a significant difference between the two quadratic forms, yet since the strictly-quadratic ROM has fewer model parameters and resulted in a slightly better relative error in the testing regime, we choose to employ a strictly-quadratic ROM form.
For the MAS-2D dataset, the results in Figure~\ref{fig:mas-2d-model-comparison} show that the strictly-quadratic model outperformed the other two model forms in the testing regime. 
Lastly, for the MAS-3D dataset, Figure~\ref{fig:mas-3d-model-comparison} shows that the quadratic model with linear and constant terms outperforms the strictly-linear and strictly-quadratic ROMs with the same amount of modes. 

%%%%%%%%%%%%%%%%%%%%%%%%%%%%%%%%

%%%%%%%%%%%%%%%%%%%%%%%%%%%%%%%%

%%%%%%%%%%%%%%%%%%%%%%%%%%%%%%%%
\subsubsection{Comparing Models with Similar Computational Cost}
\begin{figure}
    \centering
    \begin{subfigure}[b]{0.325\textwidth}
        \centering
        \caption{HUX-2D ROM Form}
        \includegraphics[width=\textwidth]{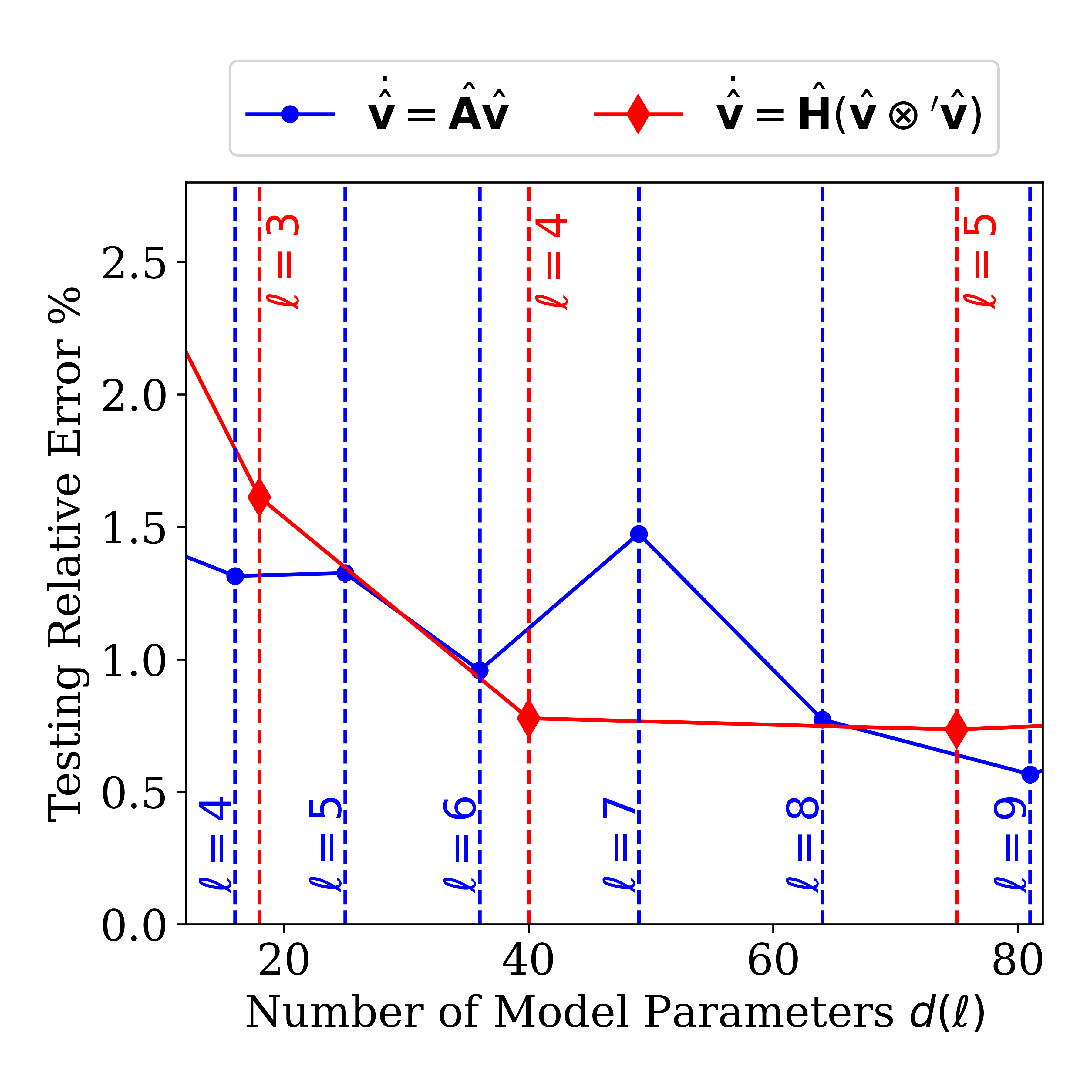}
        \label{fig:hux-model-comparison-dof}
    \end{subfigure}
      \hfill
    \begin{subfigure}[b]{0.325\textwidth}
       % \centering
        \caption{MAS-2D ROM Form}
        \includegraphics[width=\textwidth]{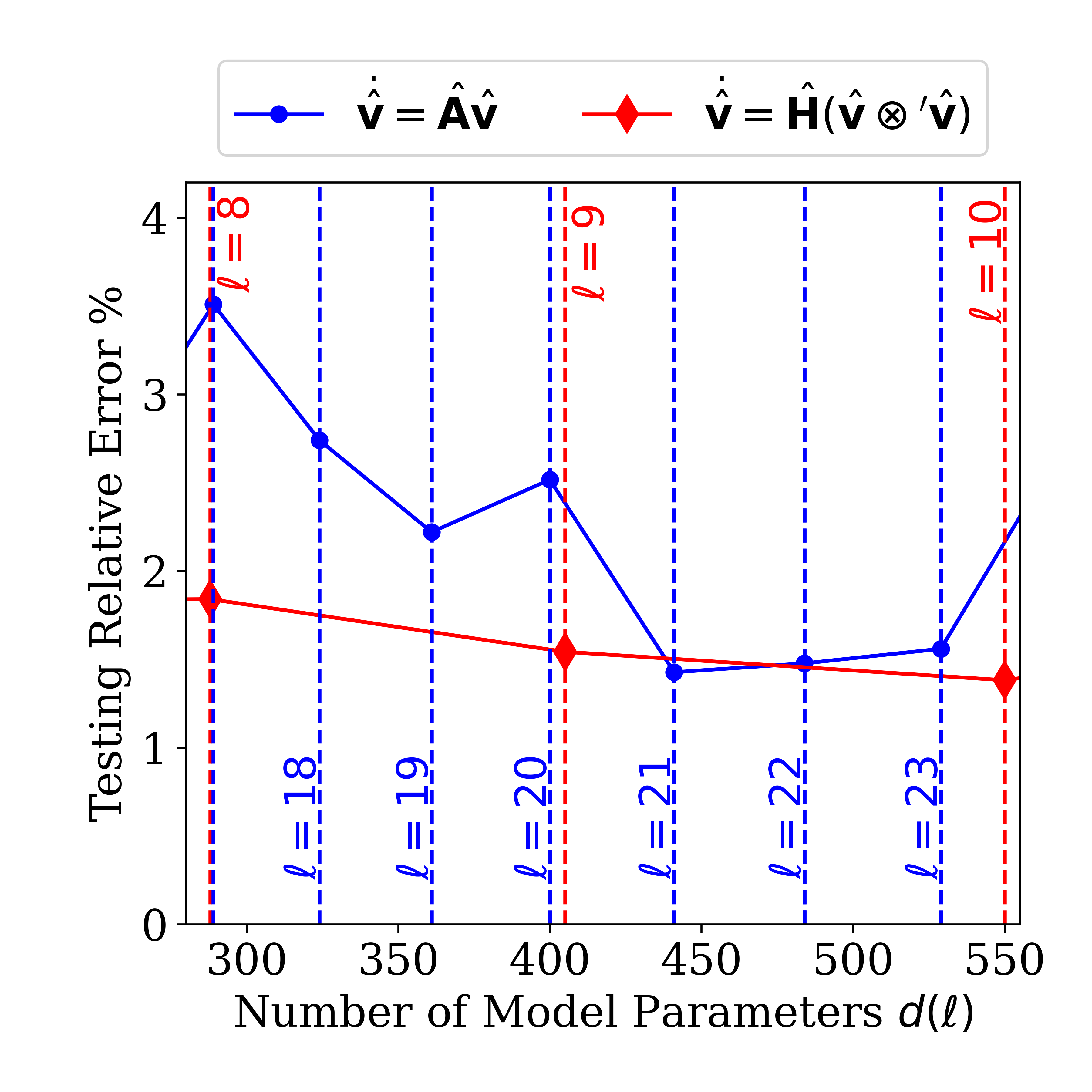}
        \label{fig:mas-2d-model-comparison-dof}
    \end{subfigure}
    \hfill
    \begin{subfigure}[b]{0.325\textwidth}
        %\centering
        \caption{MAS-3D Full-Sun ROM Form}
        \includegraphics[width=\textwidth]{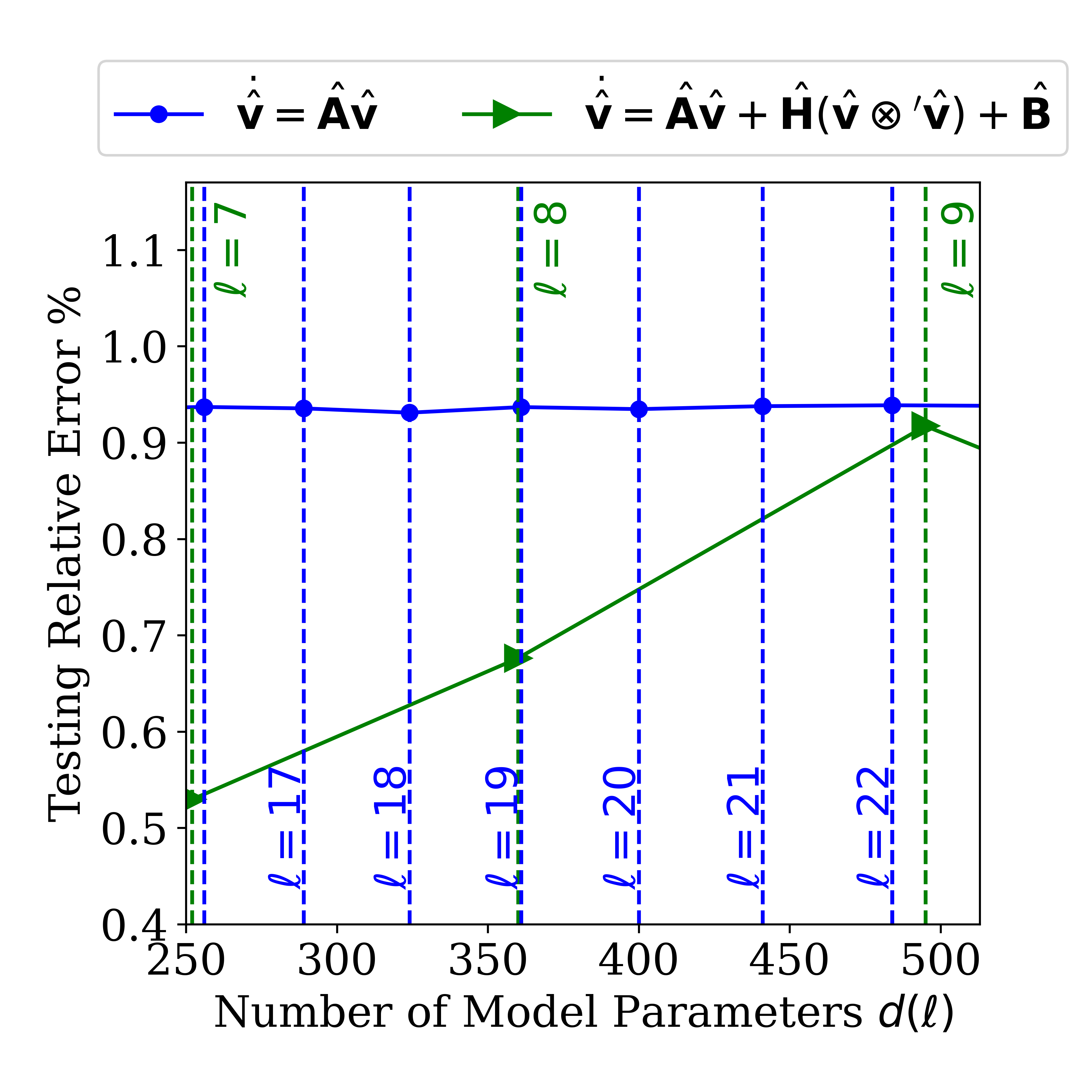}
        \label{fig:mas-3d-model-comparison-dof}
    \end{subfigure}
    \caption{\footnotesize A comparison of quadratic ROM forms, i.e. purely-quadratic and linear-quadratic plus constant term, to purely-linear ROM form via the testing relative error measured by the Frobenius norm. The numerical results for (a) HUX-2D, (b) MAS-2D, and (c) MAS-3D, show that the quadratic ROM forms are generally more accurate than the strictly-linear ROMs learned with a comparable number of model parameters.}
    \label{fig:model-form-comparison-dof}
\end{figure}

%%%%%%%%%%%%%%%%%%%%%%%%%%%%%%%%
The main question we seek to answer: \textit{Is it better to have a linear ROM with larger $\ell$ or a quadratic ROM with smaller $\ell$?}
The cost of simulating the sOpInf ROM depends on the number of ROM parameters in the matrices on the right-hand side of Eq.~\eqref{eq:poly_ROM}. That number of parameters is determined by both the model form and the reduced basis dimension $\ell$ and determines the models computational cost. 
As mentioned in Section~\ref{sec:ShiftedOperatorInference}(III), we use the compact Kronecker product in the sOpInf ROM. Consequently, the number of model parameters~$d(\ell)$ (i.e., parameters in the system matrices) is 
\begin{equation}\label{number-of-rom-params}
d(\ell) = \begin{cases}
\ell \times \ell & \text{if } \dot{\hat{\mathbf{v}}} = \hat{\mathbf{A}} \hat{\mathbf{v}} \\
\ell \times \frac{1}{2}\ell (\ell + 1) & \text{if } \dot{\hat{\mathbf{v}}} = \hat{\mathbf{H}} (\hat{\mathbf{v}} \otimes^{\prime} \hat{\mathbf{v}})\\
 \ell \times (\ell + \frac{1}{2}\ell (\ell + 1) + 1) & \text{if } \dot{\hat{\mathbf{v}}} =  \hat{\mathbf{A}} \hat{\mathbf{v}} + \hat{\mathbf{H}} (\hat{\mathbf{v}} \otimes^{\prime} \hat{\mathbf{v}}) + \hat{\mathbf{B}} \\
\end{cases}
\end{equation}
for the different ROM forms we investigated.
We thus compare the quadratic model forms chosen in Section~\ref{sec:hux-numerical-results}--\ref{sec:mhd-full-sun-results} to a linear ROM with a comparable number of model parameters $d(\ell)$ in  Figure~\ref{fig:model-form-comparison-dof}. That figure shows results for (\ref{fig:hux-model-comparison-dof}) HUX-2D ROM presented in Section~\ref{sec:hux-numerical-results}, (\ref{fig:mas-2d-model-comparison-dof}) MAS-2D ROM presented in Section~\ref{sec:mas-equator-results}, and  (\ref{fig:mas-3d-model-comparison-dof}) MAS-3D ROM presented in Section~\ref{sec:mhd-full-sun-results}.
As seen in all three examples, the chosen quadratic model forms perform better than the purely-linear model form in the testing regime using a comparable number of model parameters (and hence comparable computational cost), more specifically:
\begin{itemize}[leftmargin=*]
    \item[--] For the HUX-2D example, we set $\ell = 4$ with a strictly-quadratic model, so a linear model with $\ell \approx 6$ would have a comparable number of model parameters. The quadratic model performs better with approximately the same number of model parameters. 
    \item[--]  For the MAS-2D example, we set $\ell = 9$ with a strictly-quadratic model, so a linear model with $\ell \approx 20$ would have a comparable number of model parameters. The quadratic model is more accurate with approximately the same number of model parameters.
    \item[--] For the MAS-3D example, we set $\ell = 8$ with a linear-quadratic plus constant term model, so a linear model with $\ell \approx 19$ would have a comparable number of model parameters. The quadratic model performs better with approximately the same number of model parameters.
\end{itemize}
The above numerical evidence highlights the importance of including quadratic nonlinearity in the sOpInf ROM. From another perspective, adding a quadratic term to the model allows us to have a lower ROM dimension $\ell$ than if only linear terms were present.

%%%%%%%%%%%%%%%%%%%%%%%%%%%%%%%%
%%%%%%%%%%%%%%%%%%%%%%%%%%%%%%%%

%%%%%%%%%%%%%%%%%%%%%%%%%%%%%%%%%%%%%%%
\section{Conclusion} \label{sec:conclusion}
%%%%%%%%%%%%%%%%%%%%%%%%%%%%%%%%%%%%%%%
We proposed a reduced-order modeling strategy that uses simulated data to learn low-dimensional models for efficient solar wind predictions. The method leverages physical knowledge in that it first seeks to detect a spatial shift in the data/model (arising from advection) either through the method of characteristics or the fully data-based cross-correlation method. Given that shift, the system is then transformed into a moving coordinate frame, where a ROM can efficiently be learned via operator inference. 
The numerical results showed that for the full-Sun MAS simulations, a ROM with $\ell=8$ modes was sufficient to accurately predict the solar winds, yet produced significant computational speedup compared to the full-order model.
From a surrogate modeling perspective, we investigated and compared the accuracy of two surrogates for the MAS model: a reduced-physics approximation (HUX) and the proposed sOpInf ROM approximation. We found that the latter is a much more accurate model than HUX; therefore, it is worth investigating ROM approaches for solar physics applications. 
Although we developed the sOpInf methodology to learn ROMs from simulated solar wind data, we found that the sOpInf framework is robust to moderate levels of noise, which is promising as one hopes to use sOpInf for noisy observational data. Additionally, when considering which sOpInf ROM model form should be chosen, we found that smaller quadratic ROMs perform better than larger, purely-linear ROMs in the testing regime (with comparable number of model parameters), highlighting the importance of quadratic terms in the ROM.
While applied to use cases where solar wind velocities are most relevant, our methodology is applicable to forecasting additional solar wind quantities such as the density, pressure, etc. These models are our focus of future work. Moreover, while most quadratic forms of the ROM were sufficient to represent the physics with good accuracy, we expect more nonlinearly behaving systems to benefit from additional variable transformations (and lifting approaches similar to \cite{KW18nonlinearMORliftingPOD,QKPW2020_lift_and_learn}). 

A long-term goal of our project is to solve uncertainty quantification problems, by efficiently assessing the impact of model input
uncertainties, such as uncertain model coefficients, boundary conditions, and initial conditions, for which we anticipate our ROMs to be very useful. The ROMs can substantially accelerate ensemble methods, e.g. the most direct approach of Monte Carlo simulations and Bayesian inference, which are highly valuable in space weather operational forecasting.

%%%%%%%%%%%%%%%%%%%%%%%%%%%%%%%%%%%%%%%
\section{Acknowledgement} \label{sec:acknowledgement}
%%%%%%%%%%%%%%%%%%%%%%%%%%%%%%%%%%%%%%%
This research was partially supported by the National Science Foundation under Award 2028125 for ``SWQU: Composable Next Generation Software Framework for Space Weather Data Assimilation and Uncertainty Quantification".

%%%%%%%%%%%%%%%%%%%%%%%%%%%%%%%%%%%%%%%%%%%%%%%%%%%%
\appendix
\section{Extending Shifted Operator Inference to Applications with Noisy Data} \label{appA}
%%%%%%%%%%%%%%%%%%%%%%%%%%%%%%%%%%%%%%%%%%%%%%%%%%%%
%
We developed the sOpInf methodology to predict the ambient solar wind from simulated data, yet the sOpInf framework extends to a wide class of advection-dominated systems described on a periodic domain and can be trained directly from observational data which is always noise-corrupted.
We thus tested the sOpInf sensitivity to noise on the inviscid Burgers' equation example described in Section~\ref{sec:Burgers} with Gaussian noise added to each snapshot entry. 
Following the work by~\cite{Guo_2022}, the noise is drawn from a $\mathcal{N}(0, \nu^2)$ with $\nu = \zeta\cdot(\max_{x} u(x, 0) - \min_{x} u(x, 0)) = \zeta \cdot (1.3 - 0.8) = \zeta/2$, where the coefficient $\zeta$ is the noise level. 
The same Gaussian initial condition, periodic boundary conditions, and first-order finite difference numerical solver described in Section~\ref{sec:Burgers} are used to generate the training and testing snapshots, where $80\%$ of the total snapshots are used for training. 

\begin{figure}[p]
    \centering
        \begin{subfigure}[b]{0.38\textwidth}
        \centering
        \caption{Sensitivity of $c(t)$ to Noise}
        \includegraphics[width=\textwidth]{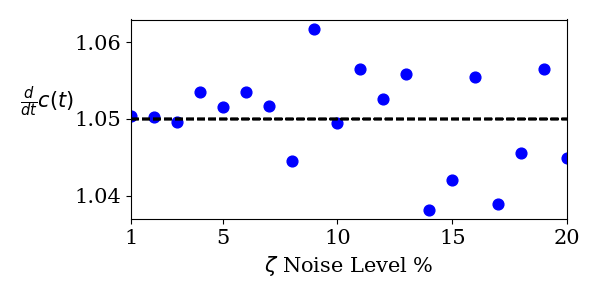}
        \label{fig:robust-cce-noise}
    \end{subfigure}
    \begin{subfigure}[b]{0.61\textwidth}
        \centering
        \caption{Singular Value Decay}
        \includegraphics[width=\textwidth]{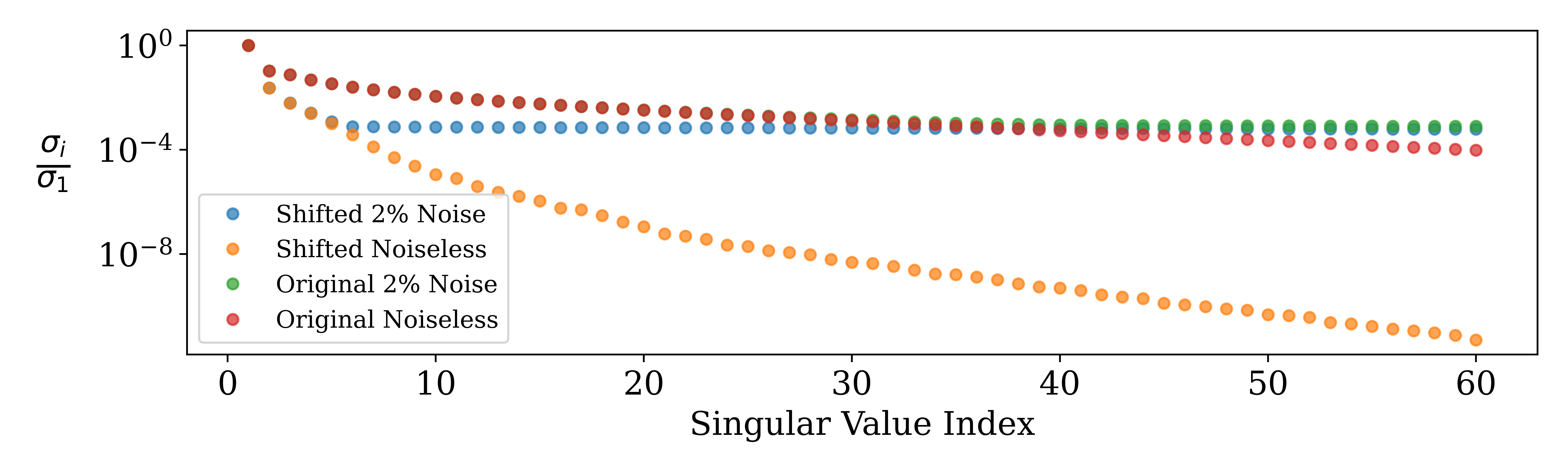}
        \label{fig:singular-value-decay-noise}
    \end{subfigure}
        \centering
    \begin{subfigure}[b]{\textwidth}
        \centering
        \caption{POD Reduced Basis}
        \includegraphics[width=\textwidth]{pod_modes_noise.png}
        \label{fig:noisy-pod-modes}
    \end{subfigure}
    \begin{subfigure}[b]{\textwidth}
        \centering
        \caption{Reduced State}
        \includegraphics[width=\textwidth]{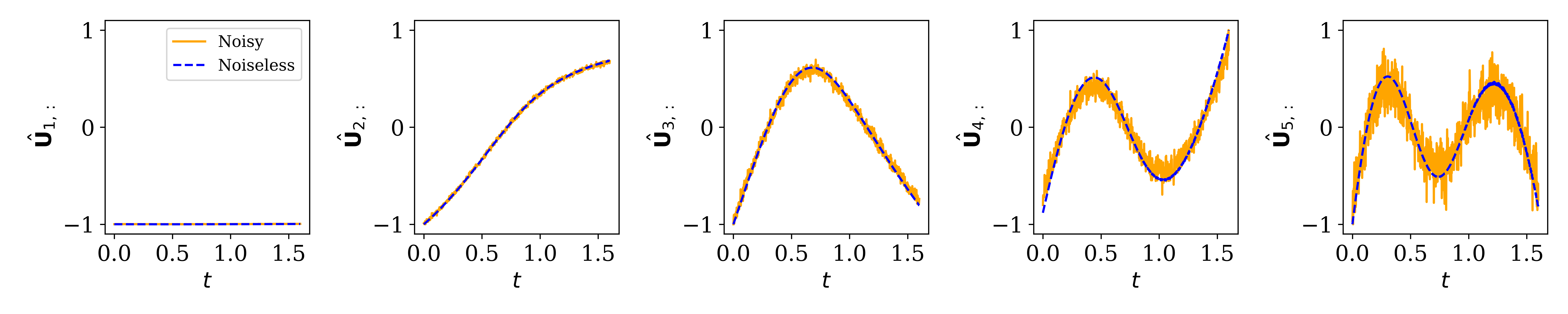}
        \label{fig:noisy-reduced-state}
    \end{subfigure}
    \caption{Graphic (a) shows the slope $\frac{\text{d}}{\text{d}t}c(t)$ of the linear shift function as the noise level $\zeta = 1, 2, \dots, 20\%$ varies. The results show that the cross-correlation extrapolation technique is robust to noise as the shift function slope remains relatively close to $1.05$ as we increase the noise level. Graphic (b) shows the singular value decay of the noisy (with $\zeta = 2\%$) and noiseless Burgers' equation snapshots on the original and shifted coordinates. Graphic~(c) shows the first eight POD modes normalized between $[-1, 1]$ of the noisy (noise level $\zeta=2\%$) and noiseless inviscid Burgers' training snapshots. Of those, the first five POD modes of the noisy dataset are able to filter most of the noise, suggesting to learn a ROM with $\ell=5$. Graphic~(d) shows the first five temporal ROM coefficients (normalized to $[-1, 1]$) in the training regime for noisy and noiseless snapshots.}
\end{figure}

We generate noisy snapshots with various levels of noise $\zeta= 1,2, \ldots, 20 \%$, and for each noise level compute the shift function $c(t)$ via the cross-correlation extrapolation technique fitting a linear polynomial (see Section~\ref{sec:cross-correlation}). Figure~\ref{fig:robust-cce-noise} shows the cross-correlation extrapolation linear shift function $c(t)$ slope as we vary the level of noise $\zeta= 1,2, \ldots, 20 \%$. We found that the cross-correlation technique is robust to noise as the shift function remained within $c(t) = (1.05 \pm 0.012) t$ for $\zeta= 1,2, \ldots, 20 \%$. Note that $c(t) = 1.05 t$ is the linear shift function for the noiseless data as mentioned in Section~\ref{sec:Burgers}. 

Figure~\ref{fig:singular-value-decay-noise} shows the singular values of the noisy and noiseless Burgers' training snapshots on the shifted and original coordinate frame for $\zeta = 2\%$. In agreement with Figure~\ref{fig:singular-value-decay-burgers}, we see in Figure~\ref{fig:singular-value-decay-noise} that the singular values decay faster in the shifted coordinates (in comparison to the original coordinates).
Next, we compute the POD basis $\mathbf{V}_{\ell} = [\mathbf{v}_{1}, \ldots, \mathbf{v}_{\ell}] \in \real^{n \times \ell}$, shown in Figure~\ref{fig:noisy-pod-modes}. Inspecting the POD modes, we choose to keep $\ell=5$ POD basis functions, since after the fifth mode the POD modes are polluted with noise.
This choice of ROM dimension $\ell=5$ is consistent with the singular values of the noisy data in the shifted coordinates,  which plateau after $\ell=5$.
The projected snapshots $\hat{\mathbf{U}} = \mathbf{V}_{\ell}^{\top} \tilde{\mathbf{U}}$, where the rows of $\hat{\bU} \in \real^{\ell \times K}$ are denoted by $\hat{\bU}_{i, :} \in \real^{K}$ for $i= 1, 2, \ldots, \ell$ are the temporal ROM coefficients and are shown in Figure~\ref{fig:noisy-reduced-state}. 
Since the temporal coefficients are polluted with noise, using a uniform sixth-order finite difference scheme as we did in Section~\ref{sec:Burgers} will be highly inaccurate. Instead, the time derivative $\dot{\hat{\mathbf{U}}}$ of the reduced states is computed via a simple factor method based on bimodal kernels of \cite{brabanter_2013}. 

\begin{figure}
    \centering
    \includegraphics[width=\textwidth]{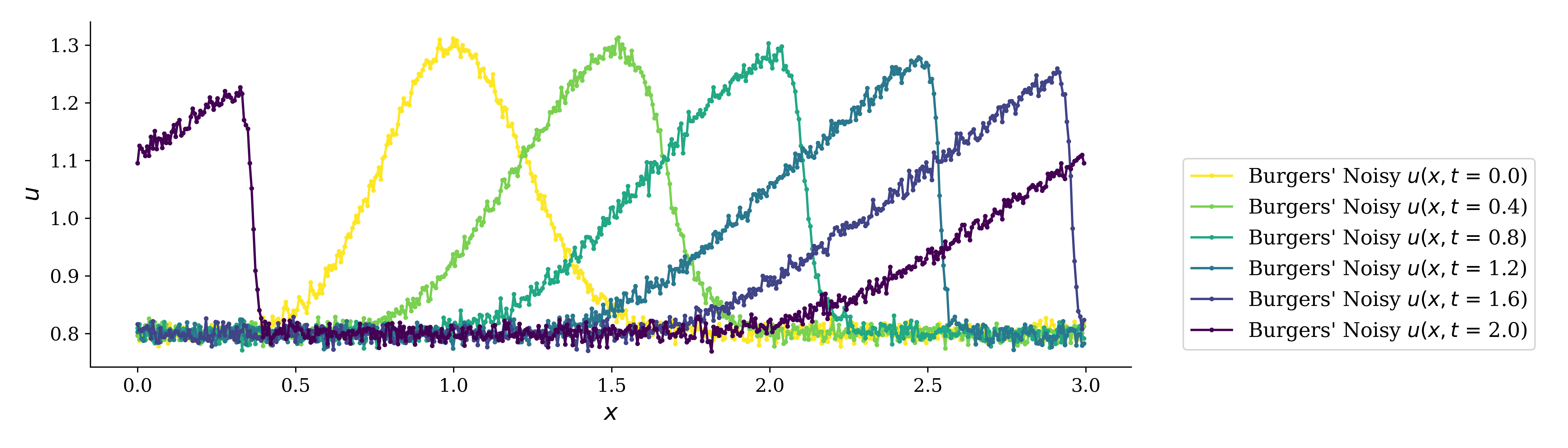}
    \caption{The inviscid Burgers' equation simulated snapshots with added Gaussian noise (noise level $\zeta = 2\%$).}
    \label{fig:burgers-data-noise}
\end{figure}

\begin{figure}
    \centering
    \includegraphics[width=\textwidth]{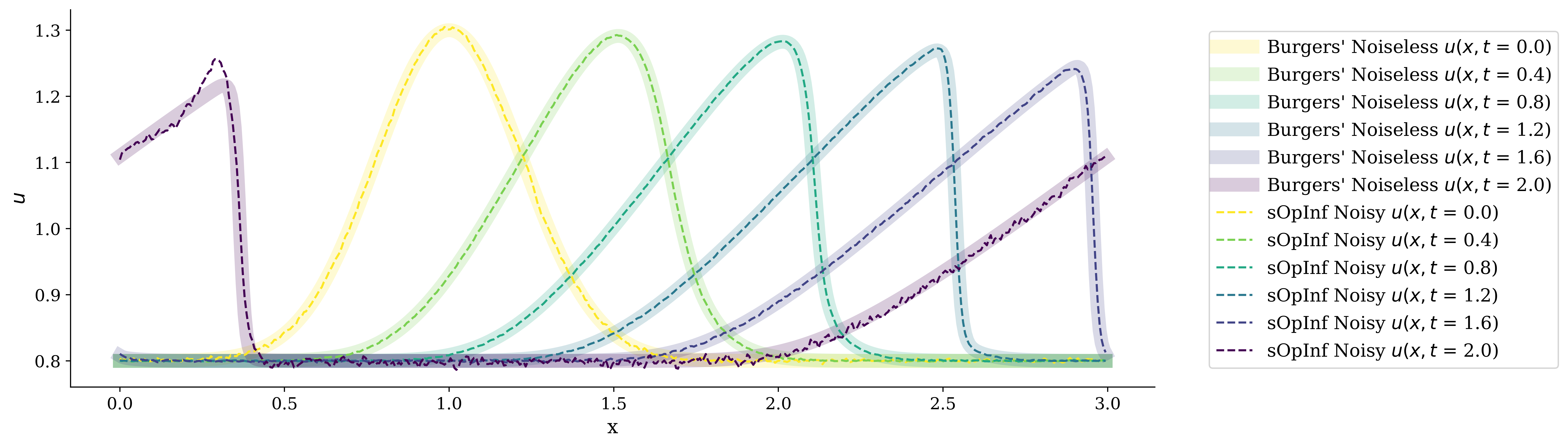}
    \caption{Solutions from the sOpInf model of the form $\dot{\hat{\bu}} = \hat{\bA} \hat{\bu} + \hat{\bH} (\hat{\bu} \otimes^{\prime} \hat{\bu})$ with $\ell = 5$ modes trained on noisy (noise level $\zeta = 2\%$) inviscid Burgers' snapshots. The results show good agreement with the noiseless snapshots. }
    \label{fig:burgers-noise-results}
\end{figure}

Figure~\ref{fig:burgers-data-noise} shows the noisy snapshots for noise level $\zeta = 2\%$. Correspondingly, the sOpInf ROM results with $\ell = 5$, $\lambda_{1} = 1$ and $\lambda_{2} = 10^{4}$ trained on noisy (noise level $\zeta = 2\%$) snapshots are shown in Figure~\ref{fig:burgers-noise-results}. The numerical results illustrate that the sOpInf framework successfully reconstructs and predicts the dynamics of the inviscid Burgers' equation from noisy snapshots. The results show small oscillations near the shock in the testing regime. We suspect this behavior is due to the low number of modes ($\ell=5$). 
The mean relative error is $1.1416 \times 10^{-3}$ and the Pearson correlation coefficient is $0.99987$ in comparison to the noiseless snapshots.
The numerical results show that the sOpInf methodology is robust to noise and can be potentially extended to applications with noisy observational data. 

%%%%%%%%%%%%%%%%%%%%%%%%%%%%%%%%%%%%%%%
%%%%%%%%%%%%%%%%%%%%%%%%%%%%%%%%%%%%%%%
%

%%%%%%%%%%%%%%%%%%%%%%%%%%%%%%%%%%%%%%%
%%%%%%%%%%%%%%%%%%%%%%%%%%%%%%%%%%%%%%%

%%%%%%%%%%%%%%%%%%%%%%%%%%%%%%%%%%%%%%%
%%%%%%%%%%%%%%%%%%%%%%%%%%%%%%%%%%%%%%%

%%%%%%%%%%%%%%%%%%%%%%%%%%%%%%%%%%%%%%%
% BIBLIOGRAPHY
%%%%%%%%%%%%%%%%%%%%%%%%%%%%%%%%%%%%%%%
\clearpage 
\newpage
\bibliography{references}
\bibliographystyle{abbrv}

\clearpage 
\end{document}

%% file: ListPackages.tex
\usepackage[margin = 1.2in]{geometry}
\usepackage[T1]{fontenc}
\usepackage[english]{babel}
\usepackage[latin1]{inputenc}
\usepackage{makecell}

\usepackage{mathtools}
\usepackage{amsmath}
\usepackage{amsfonts}
\usepackage{amsthm}
\usepackage{amssymb}
\usepackage{array}
\usepackage{algorithm,algorithmicx,algpseudocode}
\usepackage{subcaption}
\usepackage{siunitx}

\usepackage{color}
\usepackage{url}
\usepackage{cleveref}
\usepackage{graphicx}
\usepackage{breakcites}
\usepackage{soul} % for using the \ul command with linebreak
\usepackage{enumitem}
\usepackage[title,titletoc,toc]{appendix}

%% file: mymacros.tex
{\noindent {\textbf{Proof}:} }%
{\hfill $\Box$ \\[1ex] }

\newcommand{\bit}{\begin{itemize}}
\newcommand{\eit}{\end{itemize}}
\newcommand{\ben}{\begin{enumerate}}
\newcommand{\een}{\end{enumerate}}

\newcommand {\real} {\mathbb{R}}

\newcommand {\prhodt} {\frac{\partial \rho}{\partial t}}

\newcommand {\pTdt} {\frac{\partial T}{\partial t}}

\DeclareMathOperator*{\argmin}{arg\,min}%
\DeclareMathOperator*{\argmax}{arg\,max}%

\newcommand{\diverg}{\nabla \cdot}

\newcommand{\bA}{\ensuremath{\mathbf{A}}}
\newcommand{\bB}{\ensuremath{\mathbf{B}}}
\newcommand{\bC}{\ensuremath{\mathbf{C}}}
\newcommand{\bD}{\ensuremath{\mathbf{D}}}
\newcommand{\bE}{\ensuremath{\mathbf{E}}}

\newcommand{\bH}{\ensuremath{\mathbf{H}}}
\newcommand{\bI}{\ensuremath{\mathbf{I}}}
\newcommand{\bJ}{\ensuremath{\mathbf{J}}}

\newcommand{\bT}{\ensuremath{\mathbf{T}}}
\newcommand{\bU}{\ensuremath{\mathbf{U}}}
\newcommand{\bV}{\ensuremath{\mathbf{V}}}

\newcommand{\bX}{\ensuremath{\mathbf{X}}}

\newcommand{\ba}{\ensuremath{\mathbf{a}}}
\newcommand{\bb}{\ensuremath{\mathbf{b}}}
\newcommand{\bc}{\ensuremath{\mathbf{c}}}

\newcommand{\be}{\ensuremath{\mathbf{e}}}
\renewcommand{\bf}{\ensuremath{\mathbf{f}}}
\newcommand{\bg}{\ensuremath{\mathbf{g}}}

\newcommand{\bu}{\ensuremath{\mathbf{u}}}
\newcommand{\bv}{\ensuremath{\mathbf{v}}}

\newcommand{\bx}{\ensuremath{\mathbf{x}}}